\documentclass{iopconfser}
\usepackage{amssymb}
\usepackage{amsfonts}
\usepackage{amsmath}

\begin{document}

\title{Towards a General Theory of GPS Inter-Satellite Communications
Between Moving Satellites on Arbitrary Space Trajectories, Accounting for
General Relativity Effects}

\author{Bogdan G. Dimitrov}

\affil{ Institute for Advanced Physical Studies , Sofia Tech Park, 111
Tzarigradsko shaussee, 1784 Sofia, Bulgaria}

\email{dimitrov.bogdan.bogdan@gmail.com}
\email{bogdan.dimitrov@iaps.institute}

\begin{abstract}
In previous publications it has been shown that 1. It is possible to
calculate (in terms of zero and higher order elliptic integrals) the
propagation time of a signal, emitted by a satellite, moving along a plane
(or space-determined) Kepler orbit. 2. By means of a method of intersecting
null cones, together with the algebraic equation for the differential of the
Euclidean distance, the propagation time was found for the case, when the
signal is intercepted by a second satellite. In both cases, the curving of
the signal trajectory has been accounted due to the action of the
gravitational field.

\ Since the real motion of satellites is highly perturbed and beyond the
idealized approximation of Kepler orbits, the present paper has
the purpose to generalize the above approach for satellites, exchanging
signals and moving along arbitrary space trajectories. Previously, the
important physical notions of space-time
distance (negative, equal to zero or positive), condition for inter-satellite
communications and geodesic distance (only positive and greater than the Euclidean
distance, since is related to the real distance, at which the signal is
propagating) have been introduced. 

\ It will be shown that these notions can
be introduced also for the case of general motion of the satellites, and
more importantly, they have the same physical meaning.

\ \ \ The s.c. "ranging equation" will be
briefly discussed, introduced in A. Einstein's known monograph. In the
partial cases and in the general case, the equation establishes the relation
between the propagation times, the Euclidean distance and the geodesic
distance R. A generalization of the range equation for the case of a curved 
space-time has been proposed, using the introduced notion for "geodesic distance" for the 
signal. A detailed calculation of the geodesic distance has been performed for the general 
case of satellites, exchanging signals in the "forward"and "backward" directions. These geodesic 
distances are not equal and the physical reason is hidden in the non-inertial reference systems of 
motion of the two satellites. Such a situation is not considered in the classical monographs in 
gravitational physics (Hermann Bondi, Peter Bergmann, Wolfgang Pauli), where observers in two reference systems
exchange signals, but these systems are only inertial.  
     
\ Also it has been proved that the signal propagation is irreversible, 
meaning that the propagation time in the "forward direction" is different from the 
propagation time in the "backward direction". The differential of the "forward" (second) 
propagation time of the second satellite has been found to depend as a square root of a 
quadratic polynomial of the differential of the first propagation time. This proves the 
irreversibility of the propagation of the signal in the two directions. 

\ For the first time a mathematical model has been created for the "forward" and subsequent 
"backward" propagation of the signal between two moving satellites in a gravitational field. 
The model is based on $42$ nonlinear differential equations.    

\ \ \ \ Such a general theory of inter-satellite communications is the first
step towards constructing a modern theory for optimal propagation (i.e. for
the least time) of signals in a set of satellites in the near-Earth space,
thus achieving high-rate data transfer by means of laser communications
between the satellites. Creating a dense network of satellites, moving not only on orbits
but also on arbitrary space trajectories is crucial for the current $6G$ and 
future modern $7G$ satellite network communication systems. 
\end{abstract}

\section{Introduction}
\bigskip In the last 10-15 years there has been a rapid development of
multiple-satellite cooperative information and communication technologies,
designed to provide ubiquitous communication services across the 
globe~\cite{GenNetworks} . According to this new concept, satellites are considered
to be relay stations in space for the transmission of voice, video and data
communications. This means that large constellations of satellites in the LEO
(Low Earth Orbit) such as Starlink and the development of multi-layer
satellite systems, such as the European Infrastructure for Resilience,
Inter-connectivity and Security by Satellite (IRIS) will provide broadband
connectivity to critical users by integrating LEO satellite constellations
with Inter-Satellite Links (ISL) to Geostationary satellites \cite{Hoyhtya}.
\subsection{Basic facts about the propagation of signals in the
gravitational field around the Earth with account of General Relativity
Theory}
\bigskip The connectivity between the satellites in a large constellation depends on
the propagation delay of the signal between the satellites and also on the
fast movement of the satellites. This means that an accurate theoretical
model of the time of arrival (TA) of the signal is needed, which should
account for:

1. The action of the Earth gravitational field on the propagation of the
signal, which is to be accounted in the framework of the General Relativity
Theory. Since the publication of the paper by Shapiro \cite{Shapiro} it has been known that a signal in
the gravitational is travelling a longer time, compared to the time in a
flat background field. Mathematically, this is expressed by the known
Shapiro delay formulae 
\begin{equation}
T_{AB}=\frac{R_{AB}}{c}+\frac{2GM_{E}}{c^{3}}\ln \left( \frac{%
r_{A}+r_{B}+R_{AB}}{r_{A}+r_{B}-R_{AB}}\right) \text{,}  \label{AA19}
\end{equation}%
where the coordinates of the emitting and of the receiving satellite are
correspondingly $\mid x_{A}(t_{A})\mid =r_{A}$ and $\mid x_{B}(t_{B})\mid
=r_{B}$, and $R_{AB}=$ $\mid x_{A}(t_{A})-x_{B}(t_{B})\mid $ is the
Euclidean distance between the signal-emitting satellite and the
signal-receiving satellite. This formulae states that in the gravitational
field the propagation time acquires an additional logarithmic term due to
the curving of the trajectory of the signal. Note that the space points of
emission $x_{A}(t_{A})$ and arrival $x_{B}(t_{B})$ of the signal are taken
at the time $t_{A}$ of emission and $t_{B}$ of arrival correspondingly.

In order to calculate the propagation time $T_{AB}$ of the signal, according
to general relativity (see the book \cite{Straumann}), the emitter of
the first satellite emits a signal propagating on the gravitational null
cone after setting up to zero the infinitesimal metric element $ds^{2}=0$: 
\begin{equation}
ds^{2}=0=g_{00}c^{2}dT^{2}+2g_{oj}cdTdx^{j}+g_{ij}dx^{i}dx^{j}\text{.}
\label{AAF3}
\end{equation}%
If one has to find the propagation time, this null cone equation should be
solved as a quadratic algebraic equation and then the obtained expression
for $dT$ should be solved as a differential equation. The solution
essentially will depend on the parametrization of the space coordinates.
\subsection{Standard parametrizations of plane elliptic and space-oriented
orbits}
\bigskip The propagation time will depend on the parametrization of the orbit of the
signal-emitting satellite. If the simplest parametrization of a Kepler orbit
is taken in the standard form 
\begin{equation}
\ x=a(\cos E-e)\text{,}\ y=a\sqrt[.]{1-e^{2}}\sin E\text{, }  \label{Kep}
\end{equation}%
where $E$ is the eccentric anomaly angle, associated with the motion of the
satellite, then in \cite{Dimitrov2022} and later in the review paper \cite%
{Dimitrov2025} it has been proved that the propagation time is represented
by a sum of elliptic integrals of the first, second and the third kind, all
of which are of zero order. For the classification of elliptic integrals of
different kinds, one may consult the known mathematical monographs \cite%
{PrasSol}, \cite{Smirnov3} and \cite{Fichten2}.
\bigskip The second, more complicated case is when the signal emitting satellite is
moving along a space-oriented orbit, parametrized $\ $by the full set of six
Kepler parameters $(f,a,e,\Omega ,\omega ,I)$ \cite{KopEfroimKaplan}
\begin{equation}
x=\frac{a(1-e^{2})}{1+e\cos f}\left[ \cos \Omega \cos (\omega +f)-\sin
\Omega \sin (\omega +f)\cos I\right] \text{,}  \label{K11}
\end{equation}%
\begin{equation}
y=\frac{a(1-e^{2})}{1+e\cos f}\left[ \sin \Omega \cos (\omega +f)+\cos
\Omega \sin (\omega +f)\cos I\right] \text{,}  \label{K22}
\end{equation}%
\begin{equation}
z=\frac{a(1-e^{2})}{1+e\cos f}\sin (\omega +f)\sin I\text{,}  \label{K33}
\end{equation}%
where $r=\frac{a(1-e^{2})}{1+e\cos f}$ \ is the radius-vector in the orbital
plane, $f$ is the dynamical parameter, characterizing the motion of the
satellite on the elliptical orbit. For this, second case again in the papers 
\cite{Dimitrov2022} and \cite{Dimitrov2025} it has been proved that the
solution for $T$ is represented by a complicated combination of higher-order
elliptic integrals (second and fourth). The final formulae for the
propagation time is rather complicated and is given in paragraph $7.5$
(formulae $(113)$) of the review paper \cite{Dimitrov2025}.
\subsection{The problem about emission of a signal and its subsequent
perception and its relation to the modern concept about satellite networks
and high rate data transfer}
\bigskip Here it is important to stress that these two cases give an answer to the
problem about the magnitude of the propagation time, without presupposing
that the signal is to be received by another satellite. This means that in
the case of just emitting a signal by an emitter, the integration in the
calculated formulae for the propagation time is from the initial moment of
time to some final moment of time. This final time can be chosen arbitrary.
\bigskip The problem will be much more complicated, when the signal has to be
intercepted by a second, also moving satellite because the propagation time
has to reflect the fact not only of curving the trajectory, but also the
reception of the signal at some proper position of the second satellite. Of
course, the initial position of the first satellite is also important for
this process of transmission of signals.
\bigskip In principle, the collocation of satellites at certain points on an orbit
might allow the delivery of greater quantities of communication channels 
\cite{Elbert}. The actual transfer rate of data in the literature is
referred to as bandwidth and is measured in  bits of seconds. The NASA
Space Communications and Navigation program (SCaN) has demonstrated the
power of laser communications by using invincible infrared light, which
allows to send and receive information at higher data rates of $1.2$
gigabits-per-second \cite{LaserCommun}. Another example, related to the
contemporary concept about satellite networks is the s.c. cross-vendor laser
communication between satellites in low-Earth orbit, a part of the Proliferated Warfighter 
Space Architecture (PWSA) network. It is designed to enhance
communications between users on the Earth, making possible missile tracking
and early missile launching warnings \cite{CrosVendor}. The PWSA satellite
networks comprises $27$ satellites, operating at approximately $1000$ $km$
altitude above the Earth. In principle, satellite networks consist of
hundreds of satellites, which have the purpose not only to track targets on
land, but also to ensure a global coverage of the Earth by the satellites on
low-Earth orbit. This also means that satellites should be able to "mesh"
and relay data around the globe at high speed \cite{SmallSatel}, thus
ensuring seamless data exchange between all the satellites in the network by
using laser communications.
\bigskip From a theoretical point of view, it is interesting to trace how the
transfer rate of information depends on the dynamical change of the position
of the satellites and more concretely - how satellites on different orbits
(plane elliptic or space-determined orbits) can communicate. The importance
of this problem has been mentioned for example in the literature on
satellite networks \cite{Hoyhtya}, but the formulation does not refer to the
case of any curved background. In the monograph \cite{Hoyhtya} the important
notion of "delay" or "latency" has been defined, which is the measure of the
time it takes for the data to reach the destination and then to get back
again to its source. Consequently, "latency" is related to two-way
communication between satellites in permanent motion.
\bigskip Also, it has been mentioned that "each satellite is moving along its orbital
plane and can connect to adjacent satellites in the same plane and in the
neighboring planes, requiring four antennas for ISL (Inter-Satellite Links)
connectivity". Consequently, from a theoretical point of view the first
problem which can be formulated and solved is how a signal propagates
between a signal-emitting satellite and a signal-receiving satellite, both
of which move along definite orbits. Later on, in this paper and also in 
\cite{Dimitrov2026} it shall be shown that this problem about exchange of
signals can be extended to the case, when the satellites move along
arbitrary space trajectories. 
\subsection{The requirement for ultra-low latency for present 6G and future
7G mobile networks}
\bigskip The requirements for ultra-low latency, ultra-high reliability and
ultra-massive connectivity are the three basic requirements, which the
existing 5G technologies do not meet. This insufficiency of 5G wireless
networks is due to the postulation that the propagation environment of
wireless signals is predetermined and uncontrollable \cite{Wu6GMobile}.
Thus, the future 6G wireless networks (expected to appear around 2030) are
expected to provide a shift for future communications to "connected
intelligence", where the mobile devices are combined as a distributed
intelligent platform, integrating the capabilities of sensing,
communications and computations. This means also that wireless channels are
treated as controllable and programmable entities, which can be configured
with the transmitters and the receivers.
\bigskip All this requires also a new theoretical setting for the propagation between
signals between the satellites in a network, taking into account the
gravitational background and the requirement for the increased and optimal
connectivity between the satellites. This advanced connectivity is related
also to the expectation for achieving sub-millisecond latency, leveraging
Terahertz frequencies and AI (Artificial Intelligence). For example, while
for the 4G networks the latency is $50$ $ms$ (milliseconds), for 5G \ it is $%
5$ $ms$, for 6G it is $0.1$ $ms$ (sub-millisecond latency), while for the
future 7G networks (expected to appear around the year 2040) it will be
considerably smaller \cite{4G7G}, but with anticipated data transfer rate $%
5000$ $Gbps^{+}$ (gigabits-per-second). It is interesting also that future
6G networks and wireless applications will operate in the $100$ $GHz$ to $3$ 
$THz$ (terahertz) frequency band \cite{6GBeyond}, while the 7G $THz$
communications can achieve high data rates also via visible light
communication \cite{SharifPhD}. From a physical point of view, it is
interesting that in the Terahertz band the extremely small wavelength will
enable extremely high gain antennas to be made in extremely small physical
dimensions. If there is a relation between small wavelength-high
frequency-propagation time, then the problem is how such an electromagnetic
signal will propagate between satellites in a network in a gravitational
background? In principle, from a theoretical point of view, it is not clear
electromagnetic signals of different frequencies propagate in the
gravitational field.
\bigskip Naturally, the low-latency in the 6G and 7G network will require a detailed
theoretical model of the precise and optimal model of motion of the
satellites on all orbits (near-Earth, medium and high) and an optimal model
of the signal exchange between the satellites, including the ability of the
network system to transmit a signal most effectively by means of a chain of
satellites. Most importantly, such a model will have to incorporate in
itself the effects of the gravitational field in the framework of the
General Relativity Theory.
\bigskip Consequently, the theoretical model should implement in itself the motion of
the system of satellites, the exchange of signals between them and the
characteristics of the gravitational field. Unfortunately, nearly all models
for 5G, 6G and the (future) models of 7G mobile satellite networks do not
account for the action of the gravitational field.
\bigskip In this paper the characteristics of the gravitational field are related to
the notions of space-time distance and geodesic distance and to the
propagation time of the signal, dependent also on the velocities of the
satellites.
\section{Basic features of the theoretical approach of two intersecting null
cones and the multidimensional hyperplane in the framework of the General
Relativity Theory}
\bigskip In this paper we shall try to present a more concrete formulation and
mathematical description of the problem about inter-satellite connections,
which accounts for the physical characteristics of the model:
\bigskip A. The communication and exchange of signals between satellites is performed
in the gravitational field of the near-Earth space. This means that any
signal (laser,light or radio) is propagating on the null cone, obtained after
setting up equal to zero the metric $ds^{2}=0$ in the near-Earth space 
\begin{equation}
ds^{2}=-c^{2}\left( 1+\frac{2V}{c^{2}}\right) (dT)^{2}+\left( 1-\frac{2V_{.}%
}{c^{2}}\right) \left( (dx)^{2}+(dy)^{2}+(dz)^{2}\right) =0\text{,}
\label{DOP255}
\end{equation}%
where $V=\frac{G_{\oplus }M_{\oplus }}{r}$ is the standard gravitational
potential of the Earth. The signal is being emitted by the first satellite
and then received by the second satellite, so it is natural to assign to
each one of these satellites null cones, situated at the origins $%
(T_{1},x_{1},y_{1},z_{1})$ and $(T_{2},x_{2},y_{2},z_{2})$. Thus, the
existence of these two null cones implies that there will be two times of
propagation of the signal - emission time $T_{1}$ and reception time $T_{2}$%
. The time of propagation of the signal is equal to the difference $%
T_{2}-T_{1}$ of the two propagation times. The two null cones have
independent coordinates and do not intersect. However, they can be
intersected by the $6-$dimensional hyperplane of the differential $%
dR_{AB}^{2}$ of the Euclidean distance $R_{AB}^{2}$ between the satellites 
\begin{equation}
dR_{AB}^{2}=2(x_{1}-x_{2})d(x_{1}-x_{2})  \notag \\
\end{equation}%
\begin{equation}
+2(y_{1}-y_{2})d(y_{1}-y_{2})+2(z_{1}-z_{2})d(z_{1}-z_{2})\text{.}
\label{ABC2A22}
\end{equation}%
This is a $6-$dimensional hyperplane in terms of the variables $dx_{1}$, $%
dy_{1}$, $dz_{1}$, $dx_{2}$, $dy_{2}$, $dz_{2}$, which intersects the two
four-dimensional cones. Thus, the system of equations of the two null 
cones ( \ref{DOP255} ) with the corresponding indices "$1$" and "$2$" and the
hyperplane equation ( \ref{ABC2A22} ) in terms of the differentials $%
dT_{1},dT_{2},$ $dx_{1}$, $dy_{1}$, $dz_{1}$, $dx_{2}$, $dy_{2}$, $dz_{2}$
constitute a system of three complicated algebraic equations, which at first
are solved as algebraic ones and subsequently - as differential equations.
These are the basic equations of the s.c. "algebraic geometry approach",
proposed for the first time in the papers \cite{Dimitrov2017} and 
\cite{Dimitrov2019}.
\subsection{Algebraic geometry approach and the new physical notions about
space-time distance, condition for inter-satellite communications and
geodesic distance, travelled by the radio (light or laser) signal}
\bigskip On the base of some solutions of the system of equations and the found
expression for the Euclidean distance, some new physical notions have been
introduced in \cite{Dimitrov2017}, \cite{Dimitrov2019} and also in the
review paper \cite{Dimitrov2023}.
\bigskip The first notion is the space-time distance, which has been found as a
solution of the differential equation for the Euclidean distance.The most
important property of the space-time distance as a General Relativity notion
is that it can be less than zero, equal to zero or greater than zero. In
fact, due to the intersection of the two null cones with the hyperplane, the
Euclidean distance turns out to be different from the usual one and it
becomes the space-time distance.
\bigskip If this space-time distance is required to be comparable to the Euclidean
distance, then the second important notion can be introduced - the
compatibility condition for inter-satellite communications. In \cite%
{Dimitrov2017} (see also the more new corrected and considerably expanded
review paper \cite{Dimitrov2026}) important physical consequences have been
obtained, concerning the eccentricity of the orbit, when satellites on one
plane elliptical orbit communicate - the eccentricity of the orbit for the
case when Inter-Satellite Communications are realized (investigated in the
framework of the Einstein's General Relativity Theory) should be less than a
certain value $e<0.817$.
\bigskip The interesting moment here is that this simple inequality $e<0.817$ turns
out to be consistent with the fact of positivity of the distance, travelled
by the light, radio or laser signal. This distance, which is not only
strictly positive, but also greater than the Euclidean distance is called
the "geodesic distance" and represents the third important physical notion,
introduced in the papers \cite{Dimitrov2017}, \cite{Dimitrov2019}, \cite%
{Dimitrov2023} and \cite{Dimitrov2026}. The peculiar moment here is that
this distance, as well as the previous two notions about space-time distance
and the condition for inter-satellite communications, are defined on the
intersection of the two null cones. In this sense, the physical notion about
the delay of the propagation of the signal is preserved in the new framework
and theoretical approach of intersecting null cones. This is a very
important peculiarity, which has a major importance for the theory of
transmitting signals between satellites on different orbits and also for the
experimental setting of the inter-satellite communications.
\bigskip Since in the mentioned papers the restriction for the eccentricity has
been established for the simplest case of satellites on plane elliptic
orbits, two important questions arise: 1. How are communications between
satellites on different orbits realized, if these orbits are not plane
elliptic orbits with the parametrization ( \ref{Kep} ), but are for example
satellites on space-oriented orbits, parametrized by the expressions 
\ (\ref{K11}), (\ref{K22}) and (\ref{K33}) (written with indices "$1$" and "$2$%
" for the corresponding two satellites ). Here one may assume that from the
two sets of parameters $(f_{1},a_{1},e_{1},\Omega _{1},\omega _{1},I_{1})$
and $(f_{2},a_{2},e_{2},\Omega _{2},\omega _{2},I_{2})$ only the two true
anomaly angles $f_{1}$ and $f_{2}$ are the dynamical ones and change rapidly
and the others do not change and the others do not change.
\subsection{Physical motivation for studying the communications between
satellites on disturbed orbits or on arbitrary space trajectories}
\bigskip Below we shall present some physical reasons to study communications between 
satellites, moving on disturbed orbits, which in fact is a part of the problem 
about inter-satellite communications between satellites, moving along arbitrary space trajectories. 
The theoretical model, which will be developed further will concern the case of communications between 
satellites on arbitrary space trajectories, which is the more general case. In future, 
it will be interesting to demonstrate the transition from the model of communication between
satellites on arbitrary space trajectories to a model about communications
between satellites on disturbed orbits.
\bigskip The physical motivation for studying such more general cases of more general
motions of satellite motions is related to the fact that any satellite in the near-Earth space 
does not move along plane elliptic orbits or space-oriented orbits, because there are various
disturbing forces \cite{Betz}:
\bigskip "- Satellites experience drags;
\bigskip - The Earth is not spherical nor it is even ellipsoidal. Due to this, its
mass distribution is uneven and also time varying due to tidal variations.
Consequently, the Earths gravitational forces on the satellite differ from
those, used to derive the Kepler's laws;
\bigskip - Gravitational forces from other bodies in space, particularly the Moon and
the Sun, perturb the motion that is calculated based on a two-body
(satellite and Earth) model;
\bigskip - Solar radiation pressure exerts forces on the satellite, in addition to
the gravitational forces;
\bigskip - Satellites undergo orbital maneuvers occasionally to keep them in their
desired orbits;
\bigskip As a consequence of the above cited peculiarities, the equations of motion
for the satellites involve additional parameters (with values, changing with
time) beyond Kepler's law. So in general there are more than $7$ parameters".
\subsection{Most general algebraic geometry description of the process of
signal transfer between satellites on different space-oriented orbits with
variable Kepler parameters}
\bigskip If the two sets of orbital coordinates of the two satellites depend on
Kepler parameters (which are assumed to change), then the differentials of
the coordinates $X_{j}^{(1)}=(x_{1},y_{1},z_{1})$ $(j=1,2,3)$ and $%
X_{j}^{(2)}=(x_{2},y_{2},z_{2})$ $(j=1,2,3)$ will depend also on the Kepler
parameters and the equation for the differential of the square of the
Euclidean distance and the two null-cone equations will change \cite%
{Dimitrov2023} \ \ 
\begin{equation}
dR_{AB}^{2}=2\sum\limits_{j=1}^{3}\left( X_{j}^{(1)}-X_{j}^{(2)}\right)
\sum\limits_{i=1}^{6}\left[ \frac{\partial X_{j}^{(1)}}{\partial \Gamma
_{i}^{(1)}}d\Gamma _{i}^{(1)}-\frac{\partial X_{j}^{(2)}}{\partial \Gamma
_{i}^{(2)}}d\Gamma _{i}^{(2)}\right] \text{ .}  \label{15}
\end{equation}%
This equation for the square of the Euclidean distance represents a $12-$%
dimensional hyperplane in terms of the variables $d\Gamma _{i}^{(1)}$ and $%
d\Gamma _{i}^{(2)}$, where $\Gamma _{i}^{(1)}=(f_{1},\Omega _{1},\omega _{1},I_{1},a_{1},e_{1})$ and 
$\Gamma _{i}^{(2)}= (f_{2},a_{2},e_{2},\Omega_{2},\omega _{2},I_{2})$ are the two sets of Kepler orbital parameters. The
equations for the two null cones (when $l=1$ and $l=2$) 
\begin{equation}
-c^{2}(1+\frac{2V_{l}}{c^{2}})\sum\limits_{i\underset{i=k;i\neq k}{,k=1}%
}^{6}\frac{\partial T_{l}}{\partial \Gamma _{i}^{(l)}}\frac{\partial T_{l}}{%
\partial \Gamma _{k}^{(l)}}d\Gamma _{i}^{(l)}d\Gamma _{k}^{(l)}+c^{2}(1-%
\frac{2V_{l}}{c^{2}})\sum\limits_{i\underset{i=k;i\neq k}{,k=1}%
}^{6}\sum\limits_{j=1}^{3}\frac{\partial X_{j}^{(l)}}{\partial \Gamma
_{i}^{(l)}}\frac{\partial X_{j}^{(l)}}{\partial \Gamma _{k}^{(l)}}d\Gamma
_{i}^{(l)}d\Gamma _{k}^{(l)}=0\text{ .}  \label{16}
\end{equation}%
These two $6-$dimensional quadrics in terms of the variables $d\Gamma
_{i}^{(1)}$ $(l=1;i=1,2,3)$ (first quadric) and $d\Gamma _{i}^{(2)}$ $%
(l=2;i=1,2,3)$ (second quadric) will intersect, if the hyperplane equation (%
\ref{15}). However, taking into account the variations $d\Gamma _{i}^{(1)}$
and $d\Gamma _{i}^{(2)}$ of the orbital parameters \ does not solve the
problem, because for these variations $d\Gamma
_{i}^{(1)}=(df_{1},da_{1},de_{1},d\Omega _{1},d\omega _{1},dI_{1})$ and $%
d\Gamma _{i}^{(2)}=$ $(df_{2},da_{2},de_{2},d\Omega _{2},d\omega
_{2},dI_{2}) $ there is a complicated system of nonlinear differential
equations (see for example \cite{Brauer} and \cite{KopEfroimKaplan}),
relating $d\Gamma _{i}^{(1)}$ with $(f_{1},a_{1},e_{1},\Omega _{1},\omega
_{1},I_{1})$ and $d\Gamma _{i}^{(2)}$ with $(f_{2},a_{2},e_{2},\Omega
_{2},\omega _{2},I_{2})$ \ Moreover, the rather complicated from a
mathematical point of view problem about the intersection of the two $6-$%
dimensional quadrics with the $12-$dimensional hyperplane will not be easy
to solve. In principle, in the case of perturbed celestial orbits the method
of variation of constants should be applied. This method is widely discussed
not only in the monographs \cite{Brauer} and \cite{KopEfroimKaplan}, but
also in many other books on celestial mechanics.
\subsection{Inter-satellite links and the new sixth-generation satellite
network system}
\bigskip In this paper the theoretical problem about inter-satellite links and
communications will be discussed in a much broader context, which is
believed to be consistent with the modern views on satellite communications.
According to these views and concepts, sixth-generation mobile communication
(6G) is envisioned as an intelligent interconnection of all devices,
land-based, air-based and space-based, networks \cite{IntegrSatel}. This
means that the inter-satellite links (ISL) connect multiple satellites
together, also all LEO (Low Earth Orbits), MEO (Medium Earth Orbits) and HEO
(High Earth Orbits) satellites. The new generation of satellite systems are
aiming to relocate base stations into space and in this way the wireless
communication reduces the dependence of satellite-to-satellite communication
on ground communication. Thus ISL represent satellites as nodes in a space
communication network, achieving not only data communication and information
exchange, but also position measurements. In this paper, it will be shown
that range measurements will be possible to be performed between moving with
respect to one another satellites, communicating in the near-Earth
gravitational field.
\subsection{Some basic facts about the new theoretical model about
inter-satellite links between satellites on arbitrary space trajectories}
\bigskip If a satellite represents a node in a communication system and if its space
position may be arbitrary (as a result of orbital maneuvers, or subjected to
various disturbing forces, as mentioned in the monograph \cite{Betz}), then
it is reasonable to ask whether the position of a satellite should be
characterized by the Kepler parameters. In this paper a theoretical model
shall be presented, when the motions of two satellites will be described by
their space trajectories $\overrightarrow{r}_{1}=\overrightarrow{r}%
_{1}(x_{1}(F_{1}),y_{1}(F_{1}),z_{1}(F_{1}))$ and $\overrightarrow{r}_{2}=%
\overrightarrow{r}_{2}(x_{2}(F_{2}),y_{2}(F_{2}),z_{2}(F_{2}))$, where $%
\overrightarrow{r}_{1}$ and $\overrightarrow{r}_{2}$ are the radius-vectors
of the two satellites, depending on the generalized dynamical parameters $%
F_{1}$ and $F_{2}$, related to the motion of the satellites. These
parameters can be the eccentric anomaly angle $E$ for the case of plane
elliptical orbit or the true anomaly angle $f$ $\ $for the case of
space-oriented orbits. However, the parameters $F_{1}$ and $F_{2}$ can be
any other. The next development of the theoretical model will be based on
the extension of the number of the dynamical parameters 
\begin{equation}
\overrightarrow{r}_{1}=\overrightarrow{r}%
_{1}(x_{1}(F_{1}^{(1)},....F_{1}^{(n)}),y_{1}(F_{1}^{(1)},....F_{1}^{(n)}),z_{1}(F_{1}^{(1)},....F_{1}^{(n)}))
\label{M11A}
\end{equation}
and 
\begin{equation}
\overrightarrow{r}_{2}=\overrightarrow{r}%
_{2}(x_{2}(F_{2}^{(1)},...F_{2}^{(n)}),y_{2}(F_{2}^{(1)},...F_{2}^{(n)}),z_{2}(F_{2}^{(1)},...F_{2}^{(n)})).
\label{M11B}
\end{equation}
\bigskip In this paper, we shall restrict our attention to the only dynamical
parameters $F_{1}$ and $F_{2}$. The derivative of the radius-vector $%
\overrightarrow{r}_{1}$ with respect to the parameter $F_{1}$ will be $\frac{%
\partial \overrightarrow{\mathbf{r}}_{1}}{\partial F_{1}}=(\frac{\partial
x_{1}}{\partial F_{1}},\frac{\partial y_{1}}{\partial F_{1}},\frac{\partial
z_{1}}{\partial F_{1}})$, respectively for the radius-vector $%
\overrightarrow{r}_{2}$ \ $\frac{\partial \overrightarrow{\mathbf{r}}_{2}}{%
\partial F_{2}}=(\frac{\partial x_{2}}{\partial F_{2}},\frac{\partial y_{2}}{%
\partial F_{2}},\frac{\partial z_{2}}{\partial F_{2}})$. The theoretical
model uses the length (modulus) $\mid \frac{\partial \overrightarrow{\mathbf{%
r}}_{1}}{\partial F_{1}}\mid $of the vector $\frac{\partial \overrightarrow{%
\mathbf{r}}_{1}}{\partial F_{1}}$ 
\begin{equation}
v_{1}=\mid \frac{\partial \overrightarrow{\mathbf{r}}_{1}}{\partial F_{1}}%
\mid =\sqrt[.]{\left( \frac{\partial x_{1}}{\partial F_{1}}\right)
^{2}+\left( \frac{\partial y_{1}}{\partial F_{1}}\right) ^{2}+\left( \frac{%
\partial z_{1}}{\partial F_{1}}\right) ^{2}}\text{,}  \label{M12}
\end{equation}%
which can be considered as a generalized velocity of the first satellite.
Analogous formulae can be written for the modulus (generalized velocity)\ $%
v_{2}=\mid \frac{\partial \overrightarrow{\mathbf{r}}_{2}}{\partial F_{2}}%
\mid $. The implementation of these generalized velocities has its physical
justification in reference to the s.c. "inertial navigation" \cite{Grewal}, 
\cite{Inertial} which uses gyroscopes and accelerometers to maintain an
estimate of the position, velocity, attitude and attitude rates of the
satellites (these notions refer also to aircrafts, ships, submarines, land
vehicles, missiles). Since further the formulae for the accelerations of the
satellites $\frac{\partial ^{2}\left( \overrightarrow{\mathbf{r}}%
_{1}^{.}\right) }{\partial \left( F_{1}\right) ^{2}}$ and $\frac{\partial
^{2}\left( \overrightarrow{\mathbf{r}}_{2}^{.}\right) }{\partial \left(
F_{2}\right) ^{2}}$ will be used, here a clarification is needed. These are
the Newtonian accelerations of the satellites, which do not include the
gravitational accelerations, which cannot be calculated by the navigation
computers. In fact, the position of the satellites is calculated by double
integrating the combined gravitational and Newtonian acceleration, while in
this paper $\frac{\partial ^{2}\left( \overrightarrow{\mathbf{r}}%
_{1}^{.}\right) }{\partial \left( F_{1}\right) ^{2}}$ and $\frac{\partial
^{2}\left( \overrightarrow{\mathbf{r}}_{2}^{.}\right) }{\partial \left(
F_{2}\right) ^{2}}$ will denote only Newtonian accelerations.
\bigskip Recently some new methods appeared in the literature \cite{SINS} about the
SINS (Stripdown Inertial Navigation Systems), which unlike GNSS are
autonomous systems and due to this reason may increasingly operate in
GPS-denied or contested areas. SINS-methods may use again gyroscopes and
accelerometers, but might use as well methods not requiring double
integration of accelerometer output signals.
\bigskip In any case, the velocity and acceleration of the satellite are reliable
characteristics of the model for inter-satellite communications.
\section{Basic physical understanding of the inter-satellite communications
between moving satellites with account also of General Relativity Theory}
\subsection{Independent and dependent variations of the dynamical parameters
of the two satellites}
\bigskip Further we will be dealing with a model, according to which at some initial
value of the dynamical parameter (it can be for example the eccentric
anomaly angle $E$ or the true anomaly angle $f$) the first satellite emits a
signal, which at some subsequent value of the dynamical parameter will be
intercepted by the second satellite. In other words, in this particular case
the "curving" of the trajectory of the signal should reflect also the fact
that signal at some final moment of time should be intercepted by the second
satellite.This means that the first and the second satellite should have
correlated motions. However, for the theoretical model this is not
obligatory. In the framework of our theoretical model, for the case of plane
elliptic orbits a differential equation in full derivatives had been
obtained in \cite{Dimitrov2017} and \cite{Dimitrov2019} for the eccentric
anomaly angles $E_{1}$ and $E_{2}$. Assuming that the variations of these
parameters are independent, the solutions of the corresponding differential
equations result in an expression for the space-time distance, from which
later the geodesic distance will be found. Here it should be stressed that
due to the initial assumptions about the independence of the variations $%
\delta E_{1}$ and $\delta E_{2}$, the found solution for the space-time
distance will not reflect any correlations in the motions of the two
satellites, because it will be fully symmetric with respect to the eccentric
anomaly angles $E_{1}$ and $E_{2}$ and also the eccentricities $e_{1},e_{2}$
of the orbits and the semi-major axis $a_{1},a_{2}$. The symmetry will be
observed also in the expression for the geodesic distance and since this is
the distance, travelled by the light signal, physically this will mean that
the propagation time will be one and the same for signals, travelling from
space point $1$ to space point $2$ and then backwards - from point $2$ to
point $1$.
\bigskip However, the final form of the differential equation might also be written
in the form of a nonlinear differential equation $\frac{dE_{1}}{dE_{2}}%
=f(E_{1},E_{2})$ - for such a case there will be correlations in the motions
of the two satellites. This case has not been investigated for the moment,
but will be performed in future publications.
\subsection{Irreversibility of the process of signal exchange between moving
satellites}
\bigskip Following the general theoretical scheme for calculation of the space-time
distance, the condition for inter-satellite communications and the geodesic
distance, in the paper \cite{Dimitrov2026} it has been proved that the same
physical notions with a similar meaning can be introduced for the more
general case of satellites, moving on space trajectories with given
generalized velocities $v_{1}$ and $v_{2}$. However, the general case will
possess some peculiarities, which shall be related to the principle of
causality, stating that the signal is propagating from point $1$ to point $2$
and thus the propagation time $T_{2}$ should be greater than the emission
propagation time $T_{1}$. Similarly, if a signal is being send from point $2$
to point $1$, then the propagation time $T_{1}$ will be greater in
comparison with the initial (for this second case) time $T_{2}$. At the same
time, each propagation time is a solution of the corresponding null cone
equations which do not change depending on the direction of propagation.
However, in the case of moving with respect to one another satellites the
geodesic distance, travelled by the signal between the satellites in the
direction $1\longrightarrow 2$ will be different from the geodesic distance
in the direction $\ 2\longrightarrow $ $1$, because during the time of
propagation of the signal in the direction $1\longrightarrow 2$ the second
signal-receiving satellite has moved at a certain distance.
\subsection{Pseudo-range definition in the standard theory of GPS
inter-satellite communications without account of General Relativity}
\bigskip In the literature (see, for example \cite{HauschSpriHand}) the s. c.
pseudorange is defined as $p_{r}^{s}(t_{A})=c(\widetilde{t}_{A}-\widetilde{t}%
_{E})$, where $\widetilde{t}_{A}$ and $\widetilde{t}_{E}$ are
correspondingly arrival and emission times, which in principle are affected
by the instrumental delays, receiver and emitter clock offsets and the
relativistic effects. The relativistic effects are in fact caused by the delay
of the signal due to the action of the gravitational field. In a flat
space-time, the delay function $\tau $ will be equal to the geometric range
(the distance between the emitter and the receiver), divided by the velocity
of light $c$. So if the s.c. "initial guess of the receiver position" \cite%
{HauschSpriHand} at the moment of emitting the signal is applied, the
emission time may be approximated by the arrival time and thus the time
delay will be 
\begin{equation}
\tau =\frac{\parallel r^{s}(t_{A})-r_{r}(t_{A})\parallel }{c}\text{ .}
\label{M13}
\end{equation}%
Clearly, in such an approximation of defining only one propagation time $%
t_{A}$, when the delay is calculated on the base of only one \ propagation
time and then updating the receiver position, one can iteratively calculate
the receiver position, but this will not be the actual distance (range)
between emitter and receiver.
\subsection{Standard definition of the range equation and of the world
function in General Relativity}
\bigskip In General Relativity Theory, long time ago in A. Einstein's monograph \cite%
{Einstein} the range equation has been proposed with the following simple
meaning: the velocity of light $c$, multiplied by the difference between the
arrival time $T_{2}$ and the emission time $T_{1}$, gives the distance $R$,
traveled by a light signal $c(T_{2}-T_{1})-R=0$. The range equation,
expressing the constancy of the speed of light in each inertial reference
system, can be written in the form 
\begin{equation}
\mid x_{0}-x_{s}\mid ^{2}-c^{2}(t_{0}-t_{s})^{2}=0\text{ ,}  \label{M14}
\end{equation}%
where $x_{s}^{i}=(t,x_{s}),$ $s=1,2,3,4$ are the coordinates of four
emission events and $x_{0}^{i}=(t,x_{0}^{i})$ are coordinates of the
observer. This equation has been generalized also in the two Synge's
monographs \cite{Synge1} and \cite{Synge2} to the case of a curved
space-time by using the world function \ $\Omega (P_{1},P_{2})$. This
function, defined as 
\begin{equation}
\Omega (P_{1},P_{2})=\frac{1}{2}(u_{2}-u_{1})\int%
\limits_{u_{1}}^{u_{2}}g_{jk}\frac{dx^{j}}{du}\frac{dx^{k}}{du}  \label{M15}
\end{equation}%
represents one half of the square of the space-time distance between the
points $P_{1}$ and $P_{2}$, which lie on the geodesic line $\frac{d^{2}x}{%
du^{2}}+\Gamma _{jk}^{i}\frac{dx^{j}}{du}\frac{dx^{k}}{du}=0$ . Note that
the equality to zero of the world function $\Omega (P_{1},P_{2})=0$ turns
out to be a generalization for the case of a curved space-time of the null
cone equation (\ref{M14}), written in Einstein monograph \cite{Einstein} for
the case of a flat space-time. The equal to zero world function $\Omega
(P_{1},P_{2})$ in fact is a two-point function, defined on the world line,
joining an emission event and an observer (points $P_{1}$ and $P_{2}$,
correspondingly). Note however that the world function (\ref{M15}) presumes
that both ends $u_{1}$ and $u_{2}$ of the geodesic line through the points $%
P_{1}$ and $P_{2}$ are assumed to be fixed.This is the formalism, applied in
the known paper by Bahder \cite{Bahder1}. In order to point out the
difference between the formalism in this paper from the formalism in the
paper \cite{Bahder1}, it is interesting also to note that if two space-time
points $x_{1}^{i}$ and $x_{2}^{i}$ are chosen and \ no distinction is made
between emission points and an observer, the world function $\Omega
(x_{1}^{i},x_{2}^{i})$ in \cite{Bahder1} is calculated to be 
\begin{equation*}
\Omega (x_{1}^{i},x_{2}^{i})=\frac{1}{2}\eta _{ij}\Delta x^{i}\Delta x^{j}+%
\frac{GM}{c^{2}}\left[ \mid x_{2}-x_{1}\mid +\frac{c^{2}\Delta t^{2}}{\mid
x_{2}-x_{1}\mid }\right] \times
\end{equation*}%
\begin{equation}
\times \log \left( \frac{\tan \frac{\Theta _{1}}{2}}{\tan \frac{\Theta _{2}}{%
2}}\right) +\frac{GM}{c^{2}}\mid x_{2}-x_{1}\mid \left( \cos \Theta
_{1}-\cos \Theta _{2}\right) \text{ ,}  \label{M16}
\end{equation}%
where 
\begin{equation}
\cos \Theta _{a}=\frac{x_{a}.\left( x_{2}-x_{1}\right) }{\mid x_{a}\mid
.\mid x_{2}-x_{1}\mid }\text{ , }a=1,2\text{ .}  \label{M17}
\end{equation}%
Since $\eta _{ij}=(-1,+1,+1,+1)$, the world function turns out to be a sum
of a flat space-time part (i.e. an Euclidean distance) plus another part,
related to the investigated in \cite{Bahder1} case of a Schwarzschild metric
for the near-Earth space.
\subsection{A new proposal for a generalization of the notion about geodesic
distance for the case of satellites, moving on arbitrary space trajectories.
Irreversibility of the geodesic distance in the "forward" and in the
"backwards" directions}
\bigskip In the present case of moving signal-receiving and signal-satellites, if a
signal is propagating on the null cone, the end points of the null cone
(attached to the second, signal-receiving satellite) will not be fixed any
more. Consequently, for this case of moving satellites, exchanging signals
between each other and no distinction made between emission point and an
observer (as mentioned, signals can propagate in both directions $%
1\longrightarrow 2$ and $2\longrightarrow 1$), the problem will be how to
generalize the formalism for the case, when both ends of the null geodesics,
joining the satellites will be moving.
\bigskip Further it shall be shown that in the framework of the more general
formalism in this paper, the geodesic distance (the distance, travelled by
the light, radio or laser signal) shall be represented by a flat space-time
part and an additional part 
\begin{equation}
\widehat{R}_{AB.}^{2}=R_{AB.}^{2}+\frac{c}{2}\int S_{1}(r_{1})\frac{D_{1}}{%
\mid \frac{\partial \overrightarrow{\mathbf{r}}_{2}}{\partial F_{2}}\mid }%
\frac{\partial T_{1}}{\partial F_{1}}dF_{1}\text{,}  \label{M40}
\end{equation}%
where the functions $S_{1}(r_{1})$ and $D_{1}$ will be additionally defined
and $\frac{\partial T_{1}}{\partial F_{1}}$ is the derivative of the first
(emission) propagation time with respect to the generalized parameter $F_{1}$%
. The compatibility condition for inter-satellite communications will be
derived in the form \cite{Dimitrov2026} 
\begin{equation}
\int \frac{D_{2}(\frac{\partial T_{2}}{\partial F_{2}})S_{2}(r_{2})}{\mid 
\frac{\partial \overrightarrow{\mathbf{r}}_{1}}{\partial F_{1}}\mid }%
dF_{2}=\int \frac{D_{1}(\frac{\partial T_{1}}{\partial F_{1}})S_{1}(r_{1})}{%
\mid \frac{\partial \overrightarrow{\mathbf{r}}_{2}}{\partial F_{2}}\mid }%
dF_{1}  \label{M36}
\end{equation}%
and will be obtained on the base of the physical assumption that the
space-time distances $\widetilde{R}^{2}(1\longrightarrow 2)$ and $\widetilde{%
R}^{2}(2\longrightarrow 1)$ are equal. In general, however, they will not be
equal, because the emission point $1$ and the reception point $2$ (or vice
versa) belong to two different space trajectories with different
characteristics. Since the compatibility condition will be related to the
geodesic distance, an equivalent form of the geodesic distance (\ref{M40})
will be 
\begin{equation}
\widehat{R}_{AB}^{2}=R_{AB}^{2}+\frac{c}{2}\int S_{2}(r_{2})\frac{D_{2}}{%
\mid \frac{\partial \overrightarrow{\mathbf{r}}_{1}}{\partial F_{1}}\mid }%
\frac{\partial T_{2}}{\partial F_{2}}dF_{2}\text{.}  \label{M41}
\end{equation}%
In view of all these new results, perhaps it will be more appropriate to
call the Bahder's  world function $\Omega (x_{1}^{i},x_{2}^{i})$ \cite{Bahder1}
the geodesic distance.
\subsection{A new proposal for a generalization of the range equation for
the case of geodesic distance between satellites, moving on arbitrary space
trajectories}
\bigskip The next serious problem is how to generalize the causality equation for the
range equation $c(T_{2}-T_{1})-R=0$ to the case of a curved space-time and
moreover, the emission time $T_{1}$ and the reception time $T_{2}$ are
defined for null geodesics with moving boundaries. A closer look at the
second terms in the two representations (\ref{M40}) and (\ref{M41}) shows
that they are symmetric with respect to the interchange of the indices $%
1\Leftrightarrow 2$. Thus, since for the Euclidean distance $%
R_{AB}^{2}=R_{BA}^{2}$, it may seem that the geodesic distance $\widehat{R}%
_{AB}^{2}$ will be equal to the geodesic distance $\widehat{R}_{BA}^{2}$.
Moreover, for the simplified case of plane elliptic orbits, investigated in 
\cite{Dimitrov2017}, \cite{Dimitrov2023} and \cite{Dimitrov2026}, the
difference between the square of the Euclidean and the square of the
geodesic distance is found to be symmetric with respect to the interchange $%
1\Leftrightarrow 2$.
\begin{equation*}
R_{AB}^{2}-\widehat{R}_{AB}^{2}=\frac{1}{2}%
(a_{1}^{2}+a_{2}^{2})-e_{1}e_{2}a_{1}a_{2}
\end{equation*}%
\begin{equation}
+\frac{1}{4}(a_{1}^{2}e_{1}^{2}+a_{2}^{2}e_{2}^{2})-2a_{1}a_{2}\sqrt[.]{%
\left( 1-e_{1}^{2}\right) \left( 1-e_{2}^{2}\right) }\text{.}  \label{ABC47}
\end{equation}
In the present paper the modification of the range equation is proposed to
be in the form $c(T_{2}-T_{1})-\widehat{R}_{AB}^{2}=0$. This is a more
specific definition because it is related with the two intersecting null
cones. More importantly, in this case the geodesic distance $\widehat{R}%
_{AB}^{2}$ is an "intersecting geodesic distance", defined on the
intersection of the two null cones and the hyperplane equation. At the
origins of these null cones are situated the satellites, exchanging signals
between them. Let us clarify that by "exchanging signals" further it shall
be understood the process of sending a signal by the first satellite and its
reception by the second satellite. We shall not investigate the process of
the propagation of the "response" signal, emitted by the second satellite in
the direction of the first satellite, when the motion of the first satellite
during the propagation of the signal has to be taken into account. The
qualitative description of such a process of "forward" and "backward"
(response) propagation of signals is very important for future satellite
networks.
\subsection{Causality definition for the case of communication between
moving satellites on different plane elliptic orbits and the algebraic
geometry formulation of intersecting null cones and a hyperplane}
\bigskip Since the geodesic distance is always positive $\widehat{R}_{AB}^{2}>0$,
then the principle of causality will mean that $T_{2}>T_{1}$, meaning that
the second propagation time is greater than the first one, i.e. the signal
is being send from the first satellite with propagation time $T_{2}$ and is
received by the second one with propagation time $T_{1}$. Thus the causality
condition will be related to the notion of time-orientation, when a future
cone is chosen in opposition of the past cone \cite{Causal}.
\bigskip Now let us briefly consider the simplest case of satellites, defined on
two plane elliptic orbits, which exchange signals \cite{Dimitrov2017}, \cite%
{Dimitrov2026}. The standard simple orbital coordinate parametrization 
Eq.~\eqref{Kep} 
\begin{equation*}
\ x=a(\cos E-e)\text{,}\ y=a\sqrt[.]{1-e^{2}}\sin E\text{ }
\end{equation*}%
will be used (again, written in terms of the first parametrization with the
indice $1$ and then in terms of the second parametrization with the indice $%
2 $). The purpose will be to gain some understanding why the problem about
causality becomes so complicated for such a case of inter-satellite
communications between moving satellites, when the General Relativity
effects are taken into consideration. It might be thought that the
space-time coordinates $(T_{1},x_{1},y_{1},z_{1})$ of the first satellite
and the coordinates $(T_{2},x_{2},y_{2},z_{2})$ of the second satellite are
defined on two Lorentz manifolds. On these two manifolds are defined also
the two null cones (\ref{DOP255}), written with the corresponding indices $%
1 $ and $2$. Additionally, these two null cones are intersected with the
hyperplane (\ref{ABC2A22}) in terms of the differential coordinates $%
(dx_{1},dy_{1},dz_{1},dx_{2},dy_{2},dz_{2})$. The problem from a
mathematical point of view (related to algebraic geometry - see the
monograph \cite{Harris}) is: what is the intersecting variety of the two
four-dimensional null cones with the hyperplane? It turns out that even for
the most simple case of the parametrization (\ref{Kep}), this
intersecting variety will be a complicated multi-dimensional cubic algebraic
curve in terms of the variables $dE_{1},dE_{2},dT_{1},dT_{2}$ \cite%
{Dimitrov2017}, \cite{Dimitrov2026} 
\begin{equation*}
(dR_{AB}^{2})(dE_{1})(dE_{2})+S_{1}(E_{1},E_{2})(dE_{1})^{2}(dE_{2})
\end{equation*}%
\begin{equation*}
+S_{2}(E_{1},E_{2})(dE_{1})(dE_{2})^{2}
\end{equation*}%
\begin{equation}
=P_{1}(E_{1})\left( dT_{1}\right) ^{2}(dE_{2})+P_{2}(E_{2})\text{ }\left(
dT_{2}\right) ^{2}(dE_{1})\text{.}  \label{ABCD17}
\end{equation}%
Consequently, the problem about the relation between the two propagation
times $T_{1}$ and $T_{2}$ and also the definition of the causality problem,
related to this intersecting variety, will be much more complicated.
\section{The notion of simultaneity and measurement of time in known
monographs by specialists in Einstein's Theory of Gravity (Hermann Weyl,
Peter Bergmann, Hermann Bondi)}
\bigskip Here it is useful to remember some basic definitions about the s.c. "length
of time" in the known monograph \cite{Weyl}. The length of time is defined
by H. Weyl as "the distance $AB$, where the length of time includes every
point, which is later than $A$ and is earlier than $B$". In other words, as
Weyl defines the essence of the problem, "if an absolutely isolated physical
system reverts once again to exactly the same state as that in which it was
at some earlier instant, the same succession of states will be repeated in
time and the whole series of events will constitute a cycle." It is
interesting that Hermann Weyl refers to this formulation as a definition of
a clock. In reference to Weyl's definition, it may be noted that by
"reverting to exactly the same state instant" H. Weyl means that the
position of the test body (or particle) has not changed with time.
\bigskip It is interesting that this point of view was the dominant and prevailing
one among the known specialists in General Relativity Theory during the last
century. For example, a similar view has been supported also by Peter
Bergmann in his known monograph \cite{Bergmann}: "There exists an universal
time $t$, which is defined independently of the coordinate system or frame
of reference and that the distance between two simultaneous events is an
invariant quantity the value of which is independent of the coordinate
system." Also P. Bergmann asserts that the notion of simultaneity is defined
independently from the notion of a frame of reference. In fact, Bergmann
expresses Weyl's thought because if simultaneity does not depend on the frame
of reference, then it should be measured at the same space point or
simultaneity may defined with respect to inertial reference systems.
However,in the present case the signals are emitted and received by
satellites, which move in non-inertial systems, because their velocities 
are not constant.
\subsection{Comparison with the invariance of the space-time distance,
introduced in this paper}
\bigskip Further in this paper the following equality will be proved (\ref{M26})%
\begin{equation*}
D_{1}(dT_{1})S_{1}(r_{1})=\mp D_{2}(dT_{2})S_{2}(r_{2})\text{ ,}
\end{equation*}%
which might be considered to be the analogue to the notion of length of time
in Weyl's monograph \cite{Weyl}, but for the more general theory about signal
propagation between moving satellites on arbitrary space trajectories,
developed in this paper. In (\ref{M26}) $D_{1}$ and $D_{2}$ will be
differential operators, related to the space coordinates $%
r_{1}=r_{1}(x_{1},y_{1},z_{1}$ and $r_{2}=r_{2}(x_{2},y_{2},z_{2})$ of the
two satellites and \ $S_{1}(r_{1})$ and $S_{2}(r_{2})$ are expressions,
depending on the gravitational potential of the Earth at the points $r=r_{1}$
and $r=r_{2}$, $dT_{1}$ and $dT_{2}$ are the differentials of the emission
and of the reception propagation times. In some sense, the combination $%
D_{1}(dT_{1})S_{1}(r_{1})$ or $D_{2}(dT_{2})S_{2}(r_{2})$ may be considered
as an invariant quantity, independent from the frame of reference, similarly
also to the definition by Bergmann \cite{Bergmann}.
\bigskip It is interesting that these invariant combinations appeared as a result of
the assumption (\ref{M27}) about constancy of the space-time distance 
\begin{equation*}
\widetilde{R}_{AB}^{2}(1\rightarrow 2)=\widetilde{R}_{BA}^{2}(2\rightarrow
1)=const.\text{,}
\end{equation*}%
which will be proved to lead to the coincidence of the two space vectors $%
\overrightarrow{\mathbf{r}}_{1\text{ }}$ and $\overrightarrow{\mathbf{r}}_{2%
\text{ }}$ and consequently - to the impossibility for the propagation of
any signals. Although such a statement may sound trivial, this is also a
proof that the notion about space-time distance is closely related to the
notion about geodesic distance. And indeed, later on it will be proved that
for non-zero space-time distance the equality (\ref{M26}) will be modified
to the equality (\ref{N2}) 
\begin{equation*}
D_{2}(dT_{2})S_{2}=k_{1,2}(\mid \frac{\partial \overrightarrow{\mathbf{r}}%
_{1}}{\partial F_{1}}\mid ,\mid \frac{\partial \overrightarrow{\mathbf{r}}%
_{2}}{\partial F_{2}}\mid )D_{1}(dT_{1})S_{1}\text{,}
\end{equation*}%
where $k_{1,2}(\mid \frac{\partial \overrightarrow{\mathbf{r}}_{1}}{\partial
F_{1}}\mid ,\mid \frac{\partial \overrightarrow{\mathbf{r}}_{2}}{\partial
F_{2}}\mid )$ is defined as the ratio (\ref{N3}) 
\begin{equation*}
k_{1,2}(\mid \frac{\partial \overrightarrow{\mathbf{r}}_{1}}{\partial F_{1}}%
\mid ,\mid \frac{\partial \overrightarrow{\mathbf{r}}_{2}}{\partial F_{2}}%
\mid ):=\frac{\mid \frac{\partial \overrightarrow{\mathbf{r}}_{2}}{\partial
F_{2}}\mid \mp \mid \frac{\partial \overrightarrow{\mathbf{r}}_{1}}{\partial
F_{1}}\mid }{\mid \frac{\partial \overrightarrow{\mathbf{r}}_{1}}{\partial
F_{1}}\mid \pm \mid \frac{\partial \overrightarrow{\mathbf{r}}_{2}}{\partial
F_{2}}\mid }=\frac{v_{2}\mp v_{1}}{v_{1}\pm v_{2}}\text{.}
\end{equation*}%
\ \ \ \ \ This modification will lead to a non-zero geodesic distance. In
the above formulae, the following new notations are introduced 
\begin{equation}
\mid \frac{\partial \overrightarrow{\mathbf{r}}_{1}}{\partial F_{1}}\mid
=v_{1}\text{ , }\mid \frac{\partial \overrightarrow{\mathbf{r}}_{2}}{%
\partial F_{2}}\mid =v_{2}\text{ \ ,}  \label{ABCD17A}
\end{equation}
which will be used further in deriving the formulaes for the geodesic
distances in the "forward" and in the "backward" directions.
\subsection{\ Comparison with the interpretation of H. Bondi}
\bigskip Bondi in his short monographs \cite{Bondi1}, \cite{Bondi2} describes a
thought experiment in which there are three observers (Alfred, Brian and
David) and also a moving observer (Charles). The observer Charles sets up
his watch by Brian's when Charles shoots past him to be $+T$. In case Brian
(the middle observer) is moving away from Alfred (the left observer), his
watch will show time $0+T+T=2T$. Further a constant $k$ is defined which is
the ratio of the interval of reception to the interval of transmission, when
Brian and Alfred exchange flashes of light. The left observer Alfred will
receive the two flashes of light separated by an interval $kT$. All the
observers in Bondi's thought experiment are inertial and when $%
k=1 $, the two observers are at relative rest, when $k>1$- they are receding
from each other and when $k<1$ they are approaching each other. For example,
when a fourth moving observer Charles is approaching the left observer
Alfred, the light pulses which Charles receives from Alfred and Brian (the
middle observer) will be separated by an interval $\frac{T}{k}$.
\bigskip Further, the distance of Brian (middle observer) from Alfred (left observer)
at the moment of responding is $\frac{1}{2}(k^{2}-1)T$. If the moment of
response is counted by Alfred (left observer) as occurring half-way between
transmission and reception, then this will be at the time $\frac{1}{2}%
(k^{2}+1)T$. The ratio of the two time intervals 
\begin{equation}
v=\frac{k^{2}-1}{k^{2}+1}  \label{NN3}
\end{equation}%
will represent Brian's (middle observer) velocity $v$. Formulae (%
\ref{NN3}) is very similar to the derived in this paper formulae (\ref{N3}).
\subsubsection{The ratio (\protect\ref{NN3}) of the two time intervals in
Bondi's monographs and its analogue in the present paper (case of
non-constant space-time distance)}
\bigskip The similarity becomes evident if one sets up $k^{2}=\frac{v_{2}}{v_{1}}$
and thus the derived in this paper new formulae (\ref{N3}) can be written as
\begin{equation}
k_{1,2}=\frac{k^{2}\mp 1}{1\pm k^{2}}\text{ .}  \label{NN4}
\end{equation}%
However, the new formulae (\ref{NN4}) possesses also some differences.
Firstly, signals are transmitted between moving observers (satellites) and
since the satellite velocities can easily be calculated, it is clear that
the light (radio or laser) signal is emitted and received in\ non-inertial
systems, for which the velocity depends on the some dynamical parameter.
This can be seen for example for the most simple case of plane elliptic
orbit, where the velocity is given by the formulae (\ref{M106A2})%
\begin{equation*}
v_{E}=\frac{na}{\sqrt[.]{1-e\cos E}}\sqrt[.]{1-e^{2}\cos ^{2}E}=na\sqrt[.]{%
1+e\cos E}\text{. }
\end{equation*}%
For the general case of motion on arbitrary orbits, further the formulae (%
\ref{M106}) $K_{F_{1}}:=\frac{\partial \overrightarrow{\mathbf{r}_{1}}}{%
\partial F_{1}}.\frac{\partial ^{2}\overrightarrow{\mathbf{r}_{1}}\text{ }}{%
\partial F_{1}^{2}}$ for the function $K_{F_{1}}$ will be used, representing
the scalar product of the velocity vector components with the acceleration
vector components. It is important that further this expression will play an
important role in the formulae for the geodesic distance.
\bigskip Secondly, for this more complicated case of two-way exchange of signals, one
may also have $\ k>0$, $k<0$ and $k=0$. Both satellites are moving, so $k=%
\sqrt[.]{\frac{v_{2}}{v_{1}}}$ is a real function. This issue will be
clarified in the next paragraph, when the formulaes for $k_{1}$ and $k_{2}$
will be obtained.
\bigskip Thirdly, because of the signs $\mp $ in the nominator and $\ \pm $ $\ $in
the denominator of (\ref{NN4}), the two options for $k_{1,2}$ are $k_{1}=%
\frac{k^{2}-1}{1+k^{2}}$ (the upper signs in (\ref{NN4})) and $k_{2}=\frac{%
k^{2}+1}{1-k^{2}}$ (the lower signs in (\ref{NN4}). It is interesting to
check that $k_{2}=-\frac{1}{k_{1}}$. This reminds of the transformation $%
k\longrightarrow \frac{1}{k}$ in Bondi's formulae (\ref{NN3}),
but yet the minus sign in $k_{2}=-\frac{1}{k_{1}}$ is essential because it
might be related to the non-equivalence of the propagation of signals in the
"forward" and in the "backward" direction. Now, after the definition of $%
k_{1}$ and $k_{2}$, one may clarify the issue with the different signs of $%
k_{1}$ and $k_{2}$ depending on $k=\sqrt[.]{\frac{v_{2}}{v_{1}}}$.
\bigskip Fourthly, if one defines $\overline{k}^{2}=\frac{v_{1}}{v2}$, then instead
of formulae (\ref{NN4}) $k_{1,2}=\frac{k^{2}\mp 1}{1\pm k^{2}}$ one will
derive the formulae 
\begin{equation}
k_{1,2}=\frac{1\mp \overline{k}^{2}}{\overline{k}^{2}\pm 1}\text{ .}
\label{NNN4}
\end{equation}%
Then the analysis, which will be performed in the next section will be similar
to the analysis for the case of formulae (\ref{NN4}).
\subsubsection{The different signs of the coefficient functions $k_{1}$
and $k_{2}$ depending on the \ cases $v_{2}>v_{1}$ and $v_{2}<v_{1}$}
\bigskip If $v_{2}>v_{1}$, then $k^{2}>1$ and consequently $k_{1}=\frac{k^{2}-1}{%
1+k^{2}}>0$. But then $k_{2}=\frac{k^{2}+1}{1-k^{2}}<0$. The opposite
case when $v_{2}<v_{1}$ is 
\begin{equation}
k^{2}<1\Rightarrow k_{1}=\frac{k^{2}-1}{1+k^{2}}<0\text{ }\Longrightarrow
k_{2}=\frac{k^{2}+1}{1-k^{2}}>0\text{ .}  \label{N2A0}
\end{equation}%
So in both cases $v_{2}>v_{1}$ and $v_{2}<v_{1}$ the signs of $k_{1}$ and $%
k_{2}$ are opposite. Also interesting, when $v_{1}=v_{2}$, $k$ will be $k=1$%
, correspondingly $k_{1}=0$, but $k_{2}$ cannot be determined because $k_{2}=%
\frac{k^{2}+1}{1-k^{2}}\longrightarrow \infty $. There could be a natural
physical interpretation for this situation: $k_{1}$ is a coefficient
function related to the "forward" propagation of the signal from the first
moving satellite $1$ with a velocity $v_{1}$ towards the second satellite $2$
with a velocity $v_{2}=v_{1}$. The coefficient function $k_{2}$ cannot be
determined because if the satellites $1$ and $2$ move with equal velocities $%
v_{1}$ and $v_{2}$, the second satellite $2$ cannot send a "backward" signal
in the direction of satellite $1$, since this satellite will be unable to
receive this signal. Consequently, a two-way exchange of signals will be
impossible for the case $v_{1}=v_{2}$.
\bigskip Fourthly, $k_{1,2}$ in (\ref{NN4}) does not have the meaning of any velocity
as $v$ in (\ref{NN3}), because $k_{1,2}$ is a function of proportionality
between $D_{1}(dT_{1})S_{1}(r_{1})$ and $D_{2}(dT_{2})S_{2}(r_{2})$ in the
equality (\ref{N2}). Moreover, $k=\sqrt[.]{\frac{v_{2}}{v_{1}}}$ as the
square root of the ratio of two velocities is a function of the two
dynamical parameters $F_{1}$ and $F_{2}$. Note also that the equality (\ref%
{NN4}) for the two different values $k_{1}$ and $k_{2}=-\frac{1}{k_{1}}$ can
be written as 
\begin{equation}
D_{2}(dT_{2})S_{2}=k_{1}D_{1}(dT_{1})S_{1}\text{ ,}  \label{N2A1}
\end{equation}%
\begin{equation}
D_{1}(dT_{1})S_{1}=-k_{1}D_{2}(dT_{2})S_{2}\text{ .\ }  \label{N2A2}
\end{equation}
Now an important conclusion can be made - if one assumes that both these
equalities are fulfilled at one at the same time, then by combining them it
will follow that $k_{1}^{2}=-1$ or $k_{1}=\pm i$. Let us take the plus sign.
Then from the defining equality $k_{1}=\frac{k^{2}-1}{1+k^{2}}$ it will
follow that $i=\frac{k^{2}-1}{1+k^{2}}$ or 
\begin{equation}
k^{2}=\frac{v_{2}}{v_{1}}=\frac{1+i}{1-i}=\frac{1}{2}(2i)=i\text{ .}
\label{N2A3}
\end{equation}%
However, according to the definition for the velocity (\ref{M12}) $%
v_{1}=\mid \frac{\partial \overrightarrow{\mathbf{r}}_{1}}{\partial F_{1}}%
\mid $, velocities can only be real-valued functions and not imaginary ones.
Consequently, the assumption that both relations (\ref{N2A1}) and (\ref{N2A2}%
) can be simultaneously fulfilled is not true. The correct understanding
should be that the first relation (\ref{N2A1}) is fulfilled when the signal
travels from satellite $1$ to satellite $2$ ("forward" direction) and the
second relation (\ref{N2A2}) is fulfilled when the signal travels from
satellite $2$ to satellite $1$ ("backward" direction).
\subsection{Some new physical settings in the present paper}
\bigskip In both mentioned monographs \cite{Weyl}, \cite{Bergmann} the definition of
simultaneity is based on physical intuition, not on technical realization or
on any detailed calculations. Therefore, one of the purposes of this paper
will be to give a more different interpretation, based on a new
mathematical approach, which incorporates three basic equations, treated at
first as algebraic ones and subsequently solved as differential equations.
The two of these equations are of two null cones $ds_{1}^{2}=0$ and $%
ds_{2}^{2}=0$, which define four-dimensional null surfaces (cones). If they
are intersected by the hyperplane equation for the differential of the
Euclidean distance (this is equation (\ref{ABC2A22}) and its multi-parameter
version (\ref{15})), then the two null cones will define (null) reference
systems, which will be no longer independent one from another. The reason is
that each space point on the trajectory $\overrightarrow{r}_{1}=%
\overrightarrow{r}_{1}(x_{1}(F_{1}),y_{1}(F_{1}),z_{1}(F_{1}))$ of the first
satellite will be related to a space point on the space trajectory $%
\overrightarrow{r}_{2}=\overrightarrow{r}%
_{2}(x_{2}(F_{2}),y_{2}(F_{2}),z_{2}(F_{2}))$ of the second satellite.
\bigskip Let us point out also that such a theoretical setting can be considered as a
generalization of a basic approach, developed long time in the known
monograph by Wolfgang Pauli \cite{Pauli}.
\subsection{The initial idea about the two null cones as a basic approach in
Special Theory of Relativity (Wolfgang Pauli). Comparison with the approach
in the present paper}
\bigskip The proposal of W. Pauli in \cite{Pauli} is to relate the coordinates $%
x,y,z,t$ and $x^{%
{\acute{}}%
},y^{%
{\acute{}}%
},z^{%
{\acute{}}%
.},t^{%
{\acute{}}%
}$ of two reference systems $K$ and $K^{%
{\acute{}}%
}$ by means of the two null cone equations 
\begin{equation}
x^{2}+y^{2}+z^{2}-c^{2}t^{2}=0\text{ }  \label{Pauli1}
\end{equation}%
\begin{equation}
x^{%
{\acute{}}%
2}+y^{%
{\acute{}}%
2}+z^{%
{\acute{}}%
2}-c^{2}t^{%
{\acute{}}%
2}=0\text{ ,}  \label{Pauli2}
\end{equation}%
which move uniformly rectilinearly with respect to one another. In this
paper and in the approach, developed in previous papers \cite{Dimitrov2017}, 
\cite{Dimitrov2019}, \cite{Dimitrov2023}, \cite{Dimitrov2026}, the two null
cone equations are of more general kind $ds_{1}^{2}=0=g_{\alpha \beta
}^{(1)}dx^{\alpha (1)}dx^{\beta (1)}$ and $ds_{2}^{2}=0=g_{\alpha \beta
}^{(2)}dx^{\alpha (2)}dx^{\beta (2)}$ and the metric functions $g_{\alpha
\beta }^{(1)}$, $g_{\alpha \beta }^{(2)}$ are related to the metric in the
near-Earth space-time (\ref{DOP255}). In this sense, the use of these two
null-cone equations in the present paper (further these will be the
equations (\ref{ABC1}) and (\ref{ABC2})) may be considered as a
generalization of the Special Relativity approach of W. Pauli for the case of
General Relativity. \ 
\bigskip Further, Pauli proposes to construct the relation between the two null cones
in the form 
\begin{equation}
\left( x^{%
{\acute{}}%
2}+y^{%
{\acute{}}%
2}+z^{%
{\acute{}}%
2}-c^{2}t^{%
{\acute{}}%
2}\right) =\chi \left( x^{2}+y^{2}+z^{2}-c^{2}t^{2}\right) \text{ ,}
\label{Pauli3}
\end{equation}%
which is physically justifiable because of the fundamental assumption about
the homogeneity of space-time. This means also that $\chi $, assumed to
depend on the velocity, should be independent of the direction of velocity
and therefore is a constant. Additional assumptions in Pauli$^{%
{\acute{}}%
}$s approach are:
\bigskip 1A. The idea about the homogeneity is related also to the proposal by Pauli
that "the determination of the simultaneity of spatially separated events
can be carried out entirely with the help of measuring rods without the use
of clocks." Due to this reason, the contraction of lengths at rest in the
system $K^{%
{\acute{}}%
}$ and observed in $K$ is equal to that of lengths at rest in $K$ and
observed in $K^{%
{\acute{}}%
}$.
\bigskip 2A. Any motion, parallel to the $x-$axis must remain so after the
transformation;
\bigskip 3A. Displacement of the coordinate axis should contain the Galilean
transforms 
\begin{equation}
x^{%
{\acute{}}%
}=x-vt\text{ , }t^{%
{\acute{}}%
}=t\text{ .}  \label{Pauli4}
\end{equation}%
\bigskip 4A. As a result of the transformation of the variables in the two null
cones in (\ref{Pauli3}), the obtained Lorentz transformations 
\begin{equation}
x^{%
{\acute{}}%
}=\frac{x-vt}{\sqrt[.]{1-\beta ^{2}}}\text{ , \ }y^{%
{\acute{}}%
}=y\text{ , }z^{%
{\acute{}}%
}=z\text{ , }t^{%
{\acute{}}%
}=\frac{t-\frac{v}{c^{2}}x}{\sqrt[.]{1-\beta ^{2}}}\text{ , \ }\beta =\frac{v%
}{c}  \label{Pauli5}
\end{equation}%
should be invertible with respect to the interchange of the variables $x^{%
{\acute{}}%
}\Longleftrightarrow x$, $y^{%
{\acute{}}%
}\Longleftrightarrow y$, $z^{%
{\acute{}}%
}\Longleftrightarrow z$ , $t^{%
{\acute{}}%
}\Longleftrightarrow t$.
\subsection{The different assumptions in this paper \ }
\bigskip In this paper (also in \cite{Dimitrov2026}) none of the above mentioned
assumptions will be fulfilled:
\bigskip 1B. The two null cones are defined \ in the framework of the General
Relativity Theory as 
\begin{equation}
ds_{1}^{2}=0=g_{\alpha \beta }^{(1)}dx^{\alpha (1)}dx^{\beta (1)\text{ }}%
\text{, \ }ds_{2}^{2}=0=g_{\alpha \beta }^{(2)}dx^{\alpha (2)}dx^{\beta (2)}%
\text{ }  \label{Pauli6}
\end{equation}%
and they are not proportional as the null cones in (\ref{Pauli3}), but they
are related through the hyperplane equation (\ref{ABC2A22}) for the
differential of the square of the Euclidean distance. Assumptions can be
made about the two sets $(x_{1},y_{1},z_{1})$ and $(x_{2},y_{2},z_{2})$ of 
the space coordinates to be the orbital coordinates (for plane elliptic or 
for space-oriented orbits) of the two satellites, but the transformations 
(\ref{K11}), (\ref{K22}), (\ref{K33}) (written for the first satellite with indice $1$ and for the
second satellite with indice $2$) for the case of space-oriented orbits do
not have the property to be invertible. These transformations have some
interesting topological properties, which are summarized in the review paper 
\cite{Dimitrov2024}. The two null cones (\ref{Pauli6}) are defined at two
different space-time points of one and the same metric-in the case this is
the null cone metric (\ref{DOP255}) in the near-Earth space. However, the
necessity to develop communications between Earth-orbiting and Moon-orbiting
satellites might require the development of theoretical  models for
inter-satellite communications between these two types of satellites, when
the two null cones are defined by means of different metrics. In particular,
one of the reasons for the implementation of such two null cones formalism
is related to the investigated problem about the "lunar time" \cite{KopKaplan}, \cite{AshbyPatla}, \cite{TuryshAJ}.
\bigskip 2B. Now a problem appears whether such a simplified definition can be
applied in order to define the important physical notion about a "clock" for
the case considered, when the time variable may be even multi-valued. The
complexity of the problem is illustrated \ by the calculation of the
dependence of $dT_{2}$ on the differential $dT_{1}$ for the general case of
arbitrary parametrizations of the two sets $x_{1},y_{1},z_{1}$ and $%
x_{2},y_{2},z_{2}$ of space coordinates. These may be the orbital
coordinates, but may be also arbitrary ones, accounting for example the
action of various disturbing forces \cite{Dimitrov2017}, \cite{Dimitrov2026}.
\bigskip One of the basic results in this paper (see also \cite{Dimitrov2017} and 
\cite{Dimitrov2026}) will be the derivation of the formulae for dependence
of the differential of the second (arrival) propagation time on the first
(emission) propagation time (\ref{DOPC7}) 
\begin{equation*}
dT_{2}=\frac{1}{c}\sqrt[.]{\widetilde{A}_{1}(dT_{1})^{2}+\widetilde{A}%
_{2}(dT_{1})+\widetilde{A}_{3}}\text{,}
\end{equation*}%
where $\widetilde{A}_{1}$, $\widetilde{A}_{2}$ and $\widetilde{A}_{3}$ are
complicated expressions, determined in Appendix A and depending on the
coordinates $x_{1},y_{1},z_{1},x_{2},y_{2},z_{2}$, on the differentials of
the space coordinates and also on the differential $dR_{AB}^{2}$ \ of the
square of the Euclidean distance.
\bigskip The above expression, although rather complicated, gives insight to some
physical consequences:
\bigskip 1. If the first propagation time $T_{1}$ is dependent on the coordinates $%
x_{1},y_{1},z_{1}$ and on the first generalized parameter $F_{1}$, then in
the general case the second propagation time $T_{2}$ depends on the second
set of coordinates $x_{2},y_{2},z_{2}$, $F_{2}$ and also on the first set of
coordinates $x_{1},y_{1},z_{1}$ and on the generalized parameter $F_{1}$. It
should be noted that the model about inter-satellite communications between
satellites on arbitrary space trajectories is build essentially on the
assumption about the independence of $T_{2}$ on $x_{1},y_{1},z_{1},F_{1}$.
However, if a chain of $N$ satellites has to transmit a message (signal)
from the first satellite to the $N-$th, then it will be proved in a future
publication that the coordinates of the $(k+1)-$st satellite will be
dependent on the coordinates of the preceding $k-$th satellite. Similarly,
such a dependence of the coordinates of the first satellite on the
coordinates of the second satellite can appear when the integration in the
expressions for the two propagation times is performed not along each of the
radius-vectors in the two orbital planes of movement of the satellites, but
along the line, joining the two satellites.
\bigskip 2. Expression (\ref{DOPC7}) is not symmetric with respect to the
interchange of the indices $1\Leftrightarrow 2$. This means that after
solving (\ref{DOPC7}) with respect to $T_{2}$ and finding the difference $%
T_{2}-T_{1}$, it will not be the same as the difference $T_{1}-T_{2}$
because from (\ref{DOPC7}) after the interchange $1\Leftrightarrow 2$ it
follows that \cite{Dimitrov2026} 
\begin{equation}
dT_{1}=\frac{1}{c}\sqrt[.]{\overline{A}_{1}(dT_{2})^{2}+\overline{A}%
_{2}(dT_{2})+\overline{A}_{3}}\text{,}  \label{DOPC8}
\end{equation}%
where $\overline{A}_{1},\overline{A}_{2},\overline{A}_{3}$ are the same
expressions as $\widetilde{A}_{1}$, $\widetilde{A}_{2}$ and $\widetilde{A}%
_{3}$ in (\ref{DOPC7}), but with interchanged indices $1\Leftrightarrow 2$.
In this sense, the transformation (\ref{DOPC8}) is not the inverted one to
the transformation (\ref{DOPC7}), so there is no similarity to the case of
the time transformations in the Lorentz transformations (\ref{Pauli5}) 
\begin{equation}
t^{%
{\acute{}}%
}=\frac{t-\frac{v}{c^{2}}x}{\sqrt[.]{1-\beta ^{2}}}\text{ , }t^{.}=\frac{t^{%
{\acute{}}%
}+\frac{v}{c^{2}}x^{%
{\acute{}}%
}}{\sqrt[.]{1-\beta ^{2}}}\text{ , \ }\beta =\frac{v}{c}\text{,}  \label{Pauli7}
\end{equation}%
which are invertible with respect to one another. It is important to note
that $\overline{A}_{1},\overline{A}_{2},\overline{A}_{3}$ and $\widetilde{A}%
_{1}$, $\widetilde{A}_{2},$ $\widetilde{A}_{3}$ depend on the differential 
 $dR_{AB}$ of the Euclidean distance.
 The dependence on the Euclidean distance has the following physical
consequence: if the initial propagation time $T_{1}$ is a solution of the
first null cone equation and the signal is being send to a satellite on a
second orbit, then the second propagation time $T_{2}$ will be found from
the equation (\ref{DOPC7}) and thus it will no longer be a solution of the
second null cone equation. Since the frequency of the signal is inversely
proportional to time, this means also that the frequency of the received by
the second satellite signal will depend on the distance between the two
satellites. Thus from a theoretical point of view it may be verified whether
and at what distances communications between satellites by means of infrared
light will be most efficient.
\subsection{Propagation time, distance between the satellites and frequency
of the signal}
\bigskip The process of determination of the final value of the dynamical parameter
for the case of transmission of signals between satellites will be based on
the solution of the full system of the three algebraic equations for the two
null cones and the hyperplane equation for the differential of the Euclidean
distance between the satellites. Due to this reason, the second reception
propagation time will depend also on the Euclidean distance. Thus the time
of propagation $T_{2}-T_{1}$ will depend on the Euclidean distance and since
the propagation time is inversely proportional to the frequency of the
signal, not so large propagation times will correspond to higher frequencies
of the signal and correspondingly - to smaller wavelengths. Emitting a
signal at higher frequencies, as noted in the literature (see \cite%
{PrattAllnut}) is essential, because if sufficiently high frequencies are
used, then the range can be measured to millimeter accuracy. The active
determination of the range is achieved by emitting a pulse or a sequence of
pulses and then observing the time delay of the signal. This is important
for the multiple access systems, using time division. Multiple
Access System allows several users to use one communication channel, dividing
the shared channel into separate time slots. There is one more important reason to investigate from a theoretical point
of view how the propagation time depends on the distance
between the signal-emitting and signal-receiving satellite. This is related
to the high-rate data transmission by means of infrared light \cite%
{LaserCommun}. Infrared light has a wide wavelength range, spanning from $%
700 $ $nm$ $(nanometers)$ to $1$ $mm$ (millimeter). These wavelengths are
higher that those for the visible red light and are subdivided into three main
categories: NIR (Near-Infrared) - from $700$ $nm$ to $2500$ $nm$; MIR
(Mid-Infrared) - from $2500$ $nm$ to $12,000$ $nm$; FIR (Far-Infrared)- from 
$12,000$ $nm$ to $1,000,000$ $nm$ (corresponds to $1$ $mm$). If infrared
light from the NIR-spectrum is used, it will correspond to higher
frequencies, by means of which, as mentioned in \cite{PrattAllnut}, ranges
can be measured with high accuracy. Since frequency is inversely
proportional to the propagation time, for high frequencies it will
correspond to smaller propagation time. On the contrary, if lower-frequency
infrared-light is used, then the propagation time will be greater and the
signal will travel between more distant satellites in a framework of a
large network. There is also another important aspect of the propagation time, related to
the problem about the higher-rate transfer rate. If this is achieved at
shorter wavelengths (i.e. higher frequencies of the laser signal ) and since
the propagation time is inversely proportional to the frequency, then the
time, which depends on the two sets of Kepler parameters, should be
minimized in order to enhance the fast-rate transmission of the signal.
\subsection{Propagation time differences in the forward and in the
backward directions, the range equations and the different forward and
backward geodesic distances}
  \bigskip In both cases of propagation of the signal in the directions 
$1\longrightarrow 2$ (forward direction) and $2\longrightarrow 1$ (backward
direction), the propagation time differences $T_{2}-T_{1}$ and $T_{1}-T_{2}$
can be calculated from the corresponding expressions (\ref{DOPC7}) and 
(\ref{DOPC8}) 
\begin{equation}
c(T_{2}-T_{1})-\widehat{R}_{AB}^{2}=0\text{ , }c(T_{1}-T_{2})-\widehat{R}%
_{BA}^{2}=0\text{ \ .}  \label{DOPC9}
\end{equation}%
Since these time differences will be different, from the above range
equations it will follow that the geodesic distance $\widehat{R}_{AB}^{2}$
in the "forward" direction will be different from the geodesic distance $%
\widehat{R}_{BA}^{2}$ in the "backward" direction. But then from the above
two ranging equations it becomes evident that the two geodesic distances in
the "forward" and "backward" directions should be different, i.e. . $%
\widehat{R}_{AB}^{2}\neq \widehat{R}_{BA}^{2}$. However, in all the previous
publications \cite{Dimitrov2017}, \cite{Dimitrov2019}, \cite{Dimitrov2023}
the notions about "space-time" distance and "geodesic" distance had been
introduced on the base of the symmetry of the corresponding intervals (a
very short summary of these notions has been presented also in the report 
\cite{Dimitrov2025b}). Only the two cases of plane elliptic orbits and
space-oriented orbits have been considered. Thus the important question remains: does the geodesic distance $\widehat{R}%
_{AB}^{2}$ remains symmetric in the general case? It should be kept in mind
also that all the notions about space-time-distance and geodesic distance
have been deduced on the base of solving the corresponding differential
equation in full derivatives - a case, which excludes the dependence of the
dynamical parameters $E_{1},E_{2}$ or $f_{1},f_{2}$ on each other. So
further the cases when $E_{1}=E_{1}(E_{2})$ and $f_{1}=f_{1}(f_{2})$ should
be investigated, because such cases will introduce a definite "asymmetry"
between these dynamical parameters.
\subsection{The asymmetry for the geodesic distance $\protect\widehat{R}%
_{AB}(1\longrightarrow 2)$ in the "forward" direction compared to the
geodesic distance $\protect\widehat{R}_{BA}(2\longrightarrow 1)$ in the
"backward" direction}
In the paper \cite{Dimitrov2026} it has been shown that the definition of
the notion about space-time distance is more peculiar for the case of
exchange of signals between satellites on arbitrary space orbits. The reason
is that the space-time distance $\widetilde{R}_{AB}(1\longrightarrow 2)$ in
the "forward" direction is different from the space-time distance $%
\widetilde{R}_{AB}(2\longrightarrow 1)$ in the backward direction. When
defining the geodesic distance (and especially, the geodesic distance
between arbitrary moving points) this would lead to imposing an additional
condition about the equality between the differentials $d\widetilde{R}%
_{AB}(1\longrightarrow 2)=d$ $\widetilde{R}_{AB}(2\longrightarrow 1)$. After
solving this differential equation in full derivatives with respect to the
generalized dynamic parameters $F_{1}$ and $F_{2}$, this will lead to
in-equivalency in the definition of the geodesic distances $\widehat{R}%
_{AB}(1\longrightarrow 2)$ and $\widehat{R}_{BA}(2\longrightarrow 1)$.
Finally the geodesic distance is calculated to be  
\begin{equation*}
\widehat{R}_{AB}^{2}=\int \widetilde{K}(v_{1},v_{2}).\frac{v_{1}}{K_{F_{1}}}%
.\exp \left[ -K_{F_{1}}.\frac{v_{2}}{v_{1}^{2}}.\frac{%
(v_{1}-v_{2}-2v_{1}^{2})}{(v_{1}v_{2}-v_{1}^{2}+v_{2}^{2})}\right] dv_{1}%
\text{ ,}
\end{equation*}%
where $\ v_{1}=$ $\mid \frac{\partial \overrightarrow{\mathbf{r}}_{1}}{%
\partial F_{1}}\mid $ and $v_{2}=$ $\mid \frac{\partial \overrightarrow{%
\mathbf{r}}_{2}}{\partial F_{2}}\mid $ are the generalized velocities
according to formulae (\ref{M12}) and $\widetilde{K}(v_{1},v_{2})$ and $K$ $%
_{F_{1}}$ are the following functions \cite{Dimitrov2026} (\ref{M107A}) 
\begin{equation*}
\widetilde{K}(v_{1},v_{2}):=\frac{\left[ 2\mid \frac{\partial 
\overrightarrow{\mathbf{r}}_{1}}{\partial F_{1}}\mid \mid \frac{\partial 
\overrightarrow{\mathbf{r}}_{2}}{\partial F_{2}}\mid +\mid \frac{\partial 
\overrightarrow{\mathbf{r}}_{1}}{\partial F_{1}}\mid ^{2}-\mid \frac{%
\partial \overrightarrow{\mathbf{r}}_{2}}{\partial F_{2}}\mid ^{2}\right] }{%
\left( \mid \frac{\partial \overrightarrow{\mathbf{r}}_{1}}{\partial F_{1}}%
\mid +\mid \frac{\partial \overrightarrow{\mathbf{r}}_{2}}{\partial F_{2}}%
\mid \right) }
\end{equation*}%
\begin{equation*}
=\frac{2v_{1}v_{2}+v_{1}^{2}-v_{2}^{2}}{v_{1}+v_{2}}\text{ ,}
\end{equation*}%
the function $K_{F_{1}}$ is given by the formulaes (\ref{M106AA3}), (\ref%
{M106AA4})%
\begin{equation*}
K_{F_{1}}:=\frac{\partial \overrightarrow{\mathbf{r}_{1}}}{\partial F_{1}}.%
\frac{\partial ^{2}\overrightarrow{\mathbf{r}_{1}}\text{ }}{\partial
F_{1}^{2}}=\frac{1}{2}\frac{\partial }{\partial F_{1}}\left( \frac{\partial 
\overrightarrow{\mathbf{r}_{1}}}{\partial F_{1}}\right) ^{2}
\end{equation*}%
\begin{equation*}
=\frac{1}{2}\frac{\partial }{\partial F_{1}}\left[ \left( \frac{\partial
x_{1}}{\partial F_{1}}\right) ^{2}+\left( \frac{\partial y_{1}}{\partial
F_{1}}\right) ^{2}+\left( \frac{\partial z_{1}}{\partial F_{1}}\right) ^{2}%
\right] =\frac{1}{2}\frac{\partial v_{F_{1}}^{2}}{\partial F_{1}}\text{ .}
\end{equation*}%
The important conclusion which can be made is that since the functions $%
\widetilde{K}(v_{1},v_{2})$ (\ref{M107A}), $K_{F_{1}}$ (\ref{M106AA3}) and the
exponential function in the formulae for the geodesic distance are all not symmetric with respect to
the velocities $v_{1}$ and $v_{2}$, the expression for the geodesic distance
 is not
symmetric under the interchange $v_{1}\Leftrightarrow v_{2}$ and therefore $%
\widehat{R}_{AB}^{2}\neq \widehat{R}_{BA}^{2}$. Therefore, the direction of
propagation of the signal (light, radio or laser) is essential for this case
of moving ends of geodesics, when each end point is situated on a different
space trajectory.
\section{Space-time distance, compatibility condition and geodesic distance
for the simplest case of plane elliptic orbits}
\subsection{The case of independent variations of the two dynamical
parameters - the two eccentric anomaly angles $E_{1}$ and $E_{2}$}
\bigskip In this section the formalism about the mathematical theory of exchanging
signals between two satellites, moving on two plane elliptic orbits shall be
briefly reminded, and we shall introduce the notions about space-time
distance, compatibility condition and geodesic distance with the purpose to
remind how these notions were introduced in previous papers \cite%
{Dimitrov2017}, \cite{Dimitrov2019} and the review paper \cite{Dimitrov2023}%
. The detailed proofs can be seen in these papers, so they will not be given
here. 
\bigskip The main purpose of this paper (and also \cite{Dimitrov2026}) will be to
prove that in the same manner these notions can be introduced also for the
general case of satellites, moving along arbitrary space trajectories.
\bigskip The two null cones, situated at the corresponding space-time points $%
(x_{1},y_{1},T_{1})$\bigskip and $(x_{2},y_{2},T_{2})$ are parametrized by
the simple Kepler two-dimensional coordinate transformations 
\begin{equation}
x_{1}=a_{1}(\cos E_{1}-e_{1})\text{, }y_{1}=a_{1}\sqrt[.]{1-e_{1}^{2}}\sin
E_{1}\text{,}  \label{ABC3}
\end{equation}%
\begin{equation}
x_{2}=a_{2}(\cos E_{2}-e_{2})\text{, }y_{2}=a_{2}\sqrt[.]{1-e_{2}^{2}}\sin
E_{2}\text{.}  \label{ABC4}
\end{equation}%
The null cone at the signal-emitting satellite is 
\begin{equation}
ds_{1}^{2}=0=-(c^{2}+2V_{1})(dT_{1})^{2}+  \notag \\
\end{equation}%
\begin{equation}
\\
+(1-\frac{2V_{1}}{c^{2}})\left( (dx_{1})^{2}+(dy_{1})^{2}+(dz_{1})^{2}\right)
\label{ABC1}
\end{equation}%
and the null cone metric for the second signal - receiving satellite is 
\begin{equation}
ds_{2}^{2}=0=-(c^{2}+2V_{2})(dT_{2})^{2}+  \notag \\
\end{equation}%
\begin{equation}
\\
+(1-\frac{2V_{2}}{c^{2}})\left(
(dx_{2})^{2}+(dy_{2})^{2}+(dz_{2})^{2}\right) \text{ .}  \label{ABC2}
\end{equation}%
Both cones are intersected by the $6$-dimensional hyperplane equation in
terms of the differentials $dx_{1}$, $dy_{1}$, $dz_{1}$, $dx_{2}$, $dy_{2}$, 
$dz_{2}$ 
\begin{equation}
dR_{AB}^{2}=2(x_{1}-x_{2})d(x_{1}-x_{2})+  \notag \\
\end{equation}%
\begin{equation}
+2(y_{1}-y_{2})d(y_{1}-y_{2})+2(z_{1}-z_{2})d(z_{1}-z_{2})\text{ .}
\end{equation}%
\begin{equation}
dR_{AB}^{2}=2(x_{1}-x_{2})d(x_{1}-x_{2})+  \notag \\
\end{equation}%
\begin{equation}
+2(y_{1}-y_{2})d(y_{1}-y_{2})+2(z_{1}-z_{2})d(z_{1}-z_{2})\text{ .}
\label{ABC2A22}
\end{equation}%
For the case of plane elliptic orbits $z_{1}=z_{2}=0$. The following
relation can be obtained between the differentials $d(x_{1}^{2}+y_{1}^{2})$
and $\left[ (dx_{1})^{2}+(dy_{1})^{2}\right] $ 
\begin{equation}
d(x_{1}^{2}+y_{1}^{2})=\frac{2e_{1}\sin E_{1}}{(1+e_{1}\cos E_{1})}\frac{%
\left[ (dx_{1})^{2}+(dy_{1})^{2}\right] }{dE_{1}}\text{ .}  \label{ABC12}
\end{equation}%
The differentials $\left[ (dx_{1})^{2}+(dy_{1})^{2}\right] $ can be
expressed from the first null cone. Repeating the same calculations for the
differentials with the indice $"2"$, finally it can be derived 
\begin{equation*}
dR_{AB}^{2}+S_{1}(E_{1},E_{2})dE_{1}+S_{2}(E_{1},E_{2})dE_{2}=
\end{equation*}%
\begin{equation}
=P_{1}(E_{1})\frac{\left( dT_{1}\right) ^{2}}{dE_{1}}+P_{2}(E_{2})\frac{%
\left( dT_{2}\right) ^{2}}{dE_{2}}\text{ ,}  \label{ABC17}
\end{equation}%
which is in fact equation (\ref{ABCD17}). In (\ref{ABC17}) $P_{1}(E_{1})$ is
the expression 
\begin{equation}
P_{1}(E_{1}):=\frac{2e_{1}\sin E_{1}}{(1+e_{1}\cos E_{1})}\frac{%
c^{2}(c^{2}+2V_{1})}{(c^{2}-2V_{1})}  \label{ABC15}
\end{equation}%
and $P_{2}(E_{2})$ is the same expression, but with changed indice $%
1\longrightarrow 2$. Now it can be noted that from the first null cone
equation $P_{1}(E_{1})\left( dT_{1}\right) ^{2}$ can be expressed as 
\begin{equation}
\left( dT_{1}\right) ^{2}P_{1}(E_{1})=a_{1}^{2}\left( 1-e_{1}\cos
E_{1}\right) 2e_{1}\sin E_{1}\left( dE_{1}\right) ^{2}\text{ }  \label{ABC36}
\end{equation}%
and then substituted into equation (\ref{ABC17}). The same expression with
the indices $2$ can be written from the second null cone and again it can be
substituted into (\ref{ABCD17}). The final equation obtained is 
\cite{Dimitrov2017}, \cite{Dimitrov2019}, \cite{Dimitrov2026} \ 
\begin{equation}
dR_{AB}^{2}=F_{1}(E_{1},E_{2})dE_{1}+F_{2}(E_{1},E_{2})dE_{2}\text{ ,}
\label{ABC36A}
\end{equation}%
where $F_{1}(E_{1},E_{2})$ and $F_{2}(E_{1},E_{2})$ are the expressions 
\begin{equation}
F_{1}(E_{1},E_{2}):=2e_{1}a_{1}^{2}\left( 1-e_{1}\cos E_{1}\right) \sin
E_{1}-S_{1}(E_{1},E_{2})\text{,}  \label{ABC37}
\end{equation}%
\begin{equation}
F_{2}(E_{1},E_{2})=2e_{2}a_{2}^{2}\left( 1-e_{2}\cos E_{2}\right) \sin
E_{1}-S_{2}(E_{1},E_{2})\text{.\ \ }  \label{ABC38}
\end{equation}%
Expressions $S_{1}(E_{1},E_{2})$ and $S_{2}(E_{1},E_{2})$ in (\ref{ABC37})
and (\ref{ABC38}) are given by the formulaes 
\begin{equation*}
S_{1}(E_{1},E_{2}):=-2[a_{1}a_{2}\sqrt[.]{(1-e_{1}^{2})(1-e_{2}^{2})}\sin
E_{2}\cos E_{1}+
\end{equation*}%
\begin{equation}
+a_{1}a_{2}\sin E_{1}\cos E_{2}-e_{2}a_{1}a_{2}\sin E_{1}]\text{ .}
\label{ABC8}
\end{equation}%
The expression for $S_{2}(E_{1},E_{2})$ is the same as the previous one, but
with interchanged $E_{1}\Longleftrightarrow E_{2}$, i.e. $%
S_{2}(E_{1},E_{2})=S_{1}(E_{2},E_{1})$.
\bigskip Equation (\ref{ABC36A}) is the final differential equation to be solved in
terms of the dynamical variables $E_{1}$,$E_{2}$, related to the motions of
the two satellites. It is important to stress that in such a mathematical
setting the variations of the eccentric anomaly angles are assumed to be
independent. In the next sections it will be demonstrated that this is not
the only theoretical option.
\subsection{Determination of the space-time distance, the compatibility
condition and the geodesic distance under independent variations of the two
eccentric anomaly angles $E_{1}$ and $E_{2}$}
\subsubsection{Derivation of the formulae for the space-time distance as a
General Relativity notion}
\bigskip The solutions of the differential equation in full derivatives has been
found in \cite{Dimitrov2017}, \cite{Dimitrov2019}, \cite{Dimitrov2023} and 
\cite{Dimitrov2026} and this solution is no longer related to the Euclidean
distance $R_{AB}$, but to the space-time distance $\widetilde{R}_{AB}^{2}$
(further it will be denoted only with the "tilde" sign) $\ $%
\begin{equation*}
\widetilde{R}_{AB}^{2}=\left( -2e_{1}a_{1}^{2}\cos E_{1}-2e_{2}a_{2}^{2}\cos
E_{2}\right)
\end{equation*}%
\begin{equation*}
+\left( 2e_{2}a_{1}a_{2}\cos E_{1}+2e_{1}a_{1}a_{2}^{.}\cos E_{2}\right)
\end{equation*}%
\begin{equation*}
+\frac{1}{2}\left( e_{1}^{2}a_{1}^{2}\cos \left( 2E_{1}\right)
+e_{2}^{2}a_{2}^{2.}\cos \left( 2E_{2}\right) \right) -2a_{1}a_{2}\cos
E_{1}\cos E_{2}
\end{equation*}%
\begin{equation}
+2a_{1}a_{2}\sqrt[.]{\left( 1-e_{1}^{2}\right) \left( 1-e_{2}^{2}\right) }%
\sin E_{1}\sin E_{2}\text{.}  \label{ABC43}
\end{equation}%
The most important property of the space-time distance (symmetric in the
indices $E_{1}\Leftrightarrow E_{2}$) is that it can be negative, equal to
zero or positive. Thus, the space-time distance is a General Relativity
notion and it is a result of the intersection of the two null cone equations
with the hyperplane equation for the differential of the square of the
Euclidean distance.
\subsubsection{Derivation of the compatibility condition and the geodesic
distance}
Further, if the space-time distance is equal to the Euclidean distance $%
\widetilde{R}_{AB}^{2}=R_{AB}^{2}$, the compatibility condition is obtained
in the form \cite{Dimitrov2017}, \cite{Dimitrov2019}, \cite{Dimitrov2026} 
\begin{equation*}
4a_{1}a_{2}\sqrt[.]{\left( 1-e_{1}^{2}\right) \left( 1-e_{2}^{2}\right) }%
\sin E_{1}\sin E_{2}
\end{equation*}%
\begin{equation}
=a_{1}^{2}+a_{2}^{2}+(a_{2}e_{2}-a_{1}e_{1})^{2}-\frac{1}{2}\left(
e_{1}^{2}a_{1}^{2}+e_{2}^{2}a_{2}^{2}\right) \text{.}  \label{ABC45}
\end{equation}%
If this compatibility condition is substituted into the expression (\ref%
{ABC43}) for the space-time distance $\widetilde{R}_{AB}^{2}$, then the
geodesic distance is obtained, which is the actual distance, travelled by
the signal. The geodesic distance, which further shall be denoted with a
hat-sign $\widehat{R}_{AB}^{2}$, is given by the formulae \cite{Dimitrov2017}%
, \cite{Dimitrov2019}, \cite{Dimitrov2026} 
\begin{equation*}
\widehat{R}_{AB}^{2}=\frac{1}{2}(a_{1}^{2}+a_{2}^{2})+\frac{1}{2}\left(
a_{2}e_{2}-a_{1}e_{1}\right) ^{2}+\frac{1}{4}\left(
a_{1}^{2}e_{1}^{2}+a_{2}^{2}e_{2}^{2}\right)
\end{equation*}%
\begin{equation*}
-\left( 2e_{1}a_{1}^{2}\cos E_{1}+2e_{2}a_{2}^{2}\cos E_{2}\right)
\end{equation*}%
\begin{equation*}
-\left( e_{1}^{2}a_{1}^{2}\sin ^{2}E_{1}+e_{2}^{2}a_{2}^{2}\sin
^{2}E_{2}\right) -2a_{1}a_{2}\cos E_{1}\cos E_{2}
\end{equation*}%
\begin{equation}
+2a_{1}a_{2}\left( e_{2}\cos E_{1}+e_{1}\cos E_{2}\right) \text{.}
\label{ABC46}
\end{equation}%
For all general and partial cases (for example $e_{1}=e_{2}$, $E_{1}=E_{2}$, 
$a_{1}=a_{2}$), it has been proved in the papers \cite{Dimitrov2017} and 
\cite{Dimitrov2026} that the geodesic distance $\widehat{R}_{AB}^{2}$ is
always positive and greater than the Euclidean distance.
\subsection{Asymmetric definition of the space-time distance for
space-oriented orbits with eccentric anomaly angles $E_{1}$ and $E_{2}$ as
dynamical parameters, which depend on one another. Derivation of the cubic
algebraic equation for $\frac{dE_{2}}{dE_{1}}$}
\bigskip If the expression for the square of the differential $(dT_{2})^{2}$
in its standard form (assuming that both propagation times $%
T_{1}(E_{1},E_{2})~$and $T_{2}(E_{1},E_{2})$ depend on the eccentric anomaly
angles $E_{1}$ and $E_{2}$) 
\begin{equation*}
\left( dT_{2}\right) ^{2}=\left( \frac{\partial T_{2}(E_{1},E_{2})}{\partial
E_{1}}\right) ^{2}(dE_{1})^{2}
\end{equation*}%
\begin{equation}
+2\frac{\partial T_{2}(E_{1},E_{2})}{\partial E_{1}}\frac{\partial
T_{2}(E_{1},E_{2})}{\partial E_{2}}dE_{1}dE_{2}  \notag \\
\end{equation}%
\begin{equation}
\\
+\left( \frac{\partial T_{2}(E_{1},E_{2})}{\partial E_{2}}\right)
^{2}(dE_{2})^{2}  \label{ABC18}
\end{equation}
\bigskip is substituted into (\ref{ABC17}) \ and both sides are divided by $%
(dE_{1})^{2}$, then the following cubic polynomial with respect to $\frac{%
dE_{2}}{dE_{1}}$ is obtained 
\begin{equation}
Q_{1}\left( \frac{dE_{2}}{dE_{1}}\right) ^{3}+Q_{2}\left( \frac{dE_{2}}{%
dE_{1}}\right) ^{2}+Q_{3}\left( \frac{dE_{2}}{dE_{1}}\right) +Q_{4}=0\text{.}
\label{ABC19}
\end{equation}%
In (\ref{ABC19}) the functions $Q_{1}(E_{1},E_{2})$, $Q_{2}(E_{1},E_{2})$, $%
Q_{3}(E_{1},E_{2})$ and $Q_{4}(E_{1},E_{2})$ are given by formulaes (\ref%
{BBB1}), (\ref{BBB2}), (\ref{BBB3}) and (\ref{BBB4}) in Appendix $A$. It
should be noted that the assumption about the dependence of $T_{2}$ on both
eccentric anomaly angles $E_{1}$ and $E_{2}$ is more unusual. In previous
papers \cite{Dimitrov2022} and in the review paper \cite{Dimitrov2025}, the
propagation time $T$ has been calculated only on the base of the assumption
that $T_{1}=T_{1}(E_{1})$ and $T_{2}=T_{2}(E_{2})$. However, in the
Introduction on the base of formulae (\ref{DOPC7}) it was shown that the
second propagation time $T_{2}$ should depend not only on the coordinates $%
x_{2},y_{2},z_{2}$, but also on the coordinates $x_{1},y_{1},z_{1}$. This
concerns the case when a signal is emitted by a satellite and is received by
another satellite.
\bigskip Equation (\ref{ABC19}) can be transformed into an equation, depending on the
derivatives $\frac{\partial T_{1}}{\partial E_{1}}$ and $\frac{\partial T_{2}%
}{\partial E_{1}}$ 
\begin{equation}
\left[ \frac{\partial T_{2}}{\partial E_{1}}+\frac{P_{1}(E_{1})}{%
2P_{2}(E_{2})}\frac{\partial T_{1}}{\partial E_{1}}\left( \frac{dE_{2}}{%
dE_{1}}\right) \right] ^{2}=K\text{,}  \label{ABC21}
\end{equation}%
where $K$ can be represented as 
\begin{equation}
K:=G_{1}(E_{1},E_{2})\left( \frac{dE_{2}}{dE_{1}}\right)
^{2}+G_{2}(E_{1},E_{2})\left( \frac{dE_{2}}{dE_{1}}\right) \text{.}
\label{ABC22}
\end{equation}%
The functions $G_{1}(E_{1},E_{2})$ and $G_{2}(E_{1},E_{2})$ are given by
expressions (\ref{BBB5}) and (\ref{BBB6}) in Appendix $A$. Equation (\ref%
{ABC21}) enables to express the unknown derivative $\frac{\partial T_{2}}{%
\partial E_{1}}$ as 
\begin{equation}
\frac{\partial T_{2}}{\partial E_{1}}=-\frac{P_{1}}{2P_{2}}\frac{\partial
T_{1}}{\partial E_{1}}\frac{dE_{2}}{dE_{1}}+\epsilon \sqrt[.]{K}\text{,}
\label{ABC23}
\end{equation}%
where $\epsilon =\pm 1$.
\bigskip The interchange of the indices $1\Leftrightarrow 2$ allows to find 
\begin{equation*}
\frac{\partial T_{1}}{\partial E_{2}}=-\frac{P_{2}}{2P_{1}}\frac{\partial
T_{2}}{\partial E_{2}}\frac{dE_{1}}{dE_{2}}
\end{equation*}%
\begin{equation}
+\epsilon \sqrt[.]{\overline{G}_{1}(E_{2},E_{1})\left( \frac{dE_{1}}{dE_{2}}%
\right) ^{2}+\overline{G}_{2}(E_{2},E_{1})\left( \frac{dE_{1}}{dE_{2}}%
\right) }\text{,}  \label{ABC24}
\end{equation}%
where $\overline{G}_{1}(E_{2},E_{1})$ (expression (\ref{BBB7}) in Appendix $%
A $) and $\overline{G}_{2}(E_{2},E_{1})$ (expression (\ref{BBB8}) in
Appendix $A$) are the functions $G_{1}(E_{1},E_{2})$ and $G_{2}(E_{1},E_{2})$%
, but with interchanged indices $1$ and $2$. If the derivative $\frac{%
\partial T_{1}}{\partial E_{2}}$ is known, then the dependence of the second
propagation time on $E_{2}$ can be found as 
\begin{equation}
\frac{\partial T_{2}}{\partial E_{2}}=-\frac{1}{2}\left( \frac{dE_{1}}{dE_{2}%
}\right) \left( 1+4\frac{\partial T_{1}}{\partial E_{2}}\right) +\epsilon 
\sqrt[.]{N}\text{,}  \label{ABC25}
\end{equation}%
where the function $N$ is formulae (\ref{BBB8}) in Appendix $A$. Note a very
important property of formulae (\ref{ABC25}) when it is assumed that $\frac{%
\partial T_{1}}{\partial E_{2}}=0$ - since the function $N$ acquires the
form 
\begin{equation*}
N:=\left[ \frac{1}{4}+\frac{S_{1}(E_{1},E_{2})}{P_{2}(E_{2})}\right] \left( 
\frac{dE_{1}}{dE_{2}}\right) ^{2}
\end{equation*}%
\begin{equation*}
+\frac{S_{2}(E_{1},E_{2})}{P_{2}(E_{2})}+\frac{1}{P_{2}(E_{2})}\frac{%
\partial R_{AB}^{2}}{\partial E_{2}}
\end{equation*}%
\begin{equation}
-\frac{2P_{1}(E_{1})}{P_{2}(E_{2})}\frac{dE_{1}}{dE_{2}}\left( \frac{%
\partial T_{1}(E_{1})}{\partial E_{1}}\right) ^{2}\text{,}  \label{ABC25D1}
\end{equation}%
the expression (\ref{ABC25}) becomes 
\begin{equation}
\frac{\partial T_{2}}{\partial E_{2}}=-\frac{1}{2}\frac{dE_{1}}{dE_{2}}%
+\epsilon \sqrt[.]{N}\text{ .}  \label{ABC25D2}
\end{equation}%
Thus, the derivative of the propagation time $\frac{\partial T_{2}}{\partial
E_{2}}$ depends not only on  $E_{1}=E_{1}(E_{2})$, but also on
the derivative of the first propagation time $\frac{\partial T_{1}(E_{1})}{%
\partial E_{1}}$ and also on the change of the Euclidean distance $\frac{%
\partial R_{AB}^{2}}{\partial E_{2}}$. Even if the standard model of
independence between the two eccentric anomaly angles $E_{1}$ and $E_{2}$ is
considered (i.e. $\frac{dE_{1}}{dE_{2}}=0$), yet the derivative $\frac{%
\partial T_{2}}{\partial E_{2}}$ will depend in a complicated way on $\frac{%
\partial T_{1}(E_{1})}{\partial E_{1}}$ and on $\frac{\partial R_{AB}^{2}}{%
\partial E_{2}}$.
\bigskip  The above model can easily be generalized to the case, when instead of the
eccentric anomaly angles $E_{1}$ and $E_{2}$, the generalized parameters $%
F_{1}$ and $F_{2}$ can be used. For such a case, the functions $P_{2}$ and $%
P_{1}$ in (\ref{ABC24}) should be again calculated.
\subsection{The case of equal eccentric anomaly angles $E_{1}=E_{2}$, equal
semi-major axis $a_{1}=a_{2}$ and equal eccentricities of the orbits $e_{1}=e 
_{\protect\substack{ 2  \\ .}}$ and the effect of General Relativity Theory}
\bigskip This simple case, considered only in the papers \cite{Dimitrov2017} and \cite%
{Dimitrov2026}, will demonstrate the effects of the General Relativity
Theory, concerning the propagation of signals in the gravitational field.
The Euclidean distance between two points $(x_{1},y_{1})$ and $(x_{2},y_{2})$
for the plane orbit parametrization (\ref{Kep}) is given by the formulae 
\begin{equation*}
R_{AB}^{2}=\left[ \left( a_{1}\cos E_{1}-a_{2}\cos E_{2}\right) +\left(
a_{2}e_{2}-a_{1}e_{1}\right) \right] ^{2}
\end{equation*}%
\begin{equation}
+\left[ a_{1}\sqrt[.]{1-e_{1}^{2}}\sin E_{1}-a_{2}\sqrt[.]{1-e_{2}^{2}}\sin
E_{2}\right] ^{2}\text{.}  \label{ABC27}
\end{equation}%
Obviously, for $E_{1}=E_{2}$, $a_{1}=a_{2}$ and $e_{1}=e_{\substack{ 2  \\ .
}} $ this distance is equal to zero. Formulae (\ref{ABC15}) for $P(E)$ can
be written as 
\begin{equation}
P(E)=P_{1}(E_{.}):=\frac{2e_{.}\sin E_{.}}{(1+e_{.}\cos E_{.})}\frac{%
c^{2}(c^{2}+2V_{.})}{(c^{2}-2V_{.})}\text{,}  \label{ABCM27}
\end{equation}%
expression (\ref{ABC8}) for $E_{1}=E_{2}$ acquires the form 
\begin{equation}
S(E)=S_{1}(E)=S_{2}(E)=-a^{2}(2-e^{2})\sin (2E)+2ea^{2}\sin E\text{.\ }
\label{ABCM8A}
\end{equation}%
The expressions for $F_{1}(E_{1},E_{2})$ and $F_{2}(E_{1},E_{2})$ in (\ref%
{ABC36A}) 
\begin{equation*}
dR_{AB}^{2}=F_{1}(E_{1},E_{2})dE_{1}+F_{2}(E_{1},E_{2})dE_{2}
\end{equation*}%
for $E_{1}=E_{2}$ become 
\begin{equation}
F(E)=F_{1}(E_{1})=F_{2}(E_{2}):=a^{2}(2-e^{2})\sin (2E)\text{ }
\label{ABC37A}
\end{equation}%
and thus 
\begin{equation}
dR_{AB}^{2}=2a^{2}(2-e^{2})\sin (2E)dE\text{. \ }  \label{ABC36AB}
\end{equation}%
From the last expression the space-time and geodesic distance can be
obtained and it is clear that they will be different from zero. This can be
confirmed by a direct calculation of the geodesic distance $\widehat{R}%
_{AB}^{2}$ according to formulae (\ref{ABC46}) for the case $E_{1}=E_{2}=E$, 
$e_{1}=e_{2}=e$, $a_{1}=a_{2}=a$. The result is different from zero and is
equal to 
\begin{equation}
\widehat{R}_{AB}^{2}=a^{2}-\frac{3}{2}e^{2}a^{2}-(1-e^{2})2a^{2}\cos ^{2}E%
\text{.}  \label{ABC46A}
\end{equation}%
From here, the geodesic distance $\widehat{R}_{AB}^{2}$ is equal to zero
only when 
\begin{equation}
\cos ^{2}E=\frac{(1-\frac{3}{2}e^{2})}{(1-e^{2})}\text{.}  \label{ABC46B}
\end{equation}%
It can be proved that for $\sin ^{2}E\geq \sin ^{2}E_{\lim }$ (corresponding
to $\cos ^{2}E\leq \cos ^{2}E_{\lim }$), where $\sin ^{2}E_{\lim }=\frac{%
2-e^{2}}{4(1-e^{2})}$ and $\cos ^{2}E_{\lim }=\frac{(1-\frac{3}{2}e^{2})}{%
(1-e^{2})}$, the geodesic distance is positive, i.e. $\widehat{R}%
_{AB}^{2}\geq 0$ . The value for $\sin ^{2}E_{\lim }$ exactly corresponds to
the value (\ref{ABC46B}), so for this value $\cos ^{2}E_{\lim }$ the
compatibility condition (\ref{ABC45}) for inter-satellite communications is
consistent with a zero geodesic distance. The positivity of $\cos ^{2}E$ in
fact follows from expression $(1-\frac{3}{2}e^{2})$ in the nominator of (\ref%
{ABC46B}) or in other words - by the restriction $e^{2}<\frac{2}{3}$ on the
ellipticity of the orbit, which ensures the positivity of the geodesic
distance and thus makes possible the inter-satellite communications.
\subsection{Interpretation of the possibility for inter-satellite
communications for the case $E_{1}=E_{2}=E$, $e_{1}=e_{2}=e$, $a_{1}=a_{2}=a$%
}
\bigskip An important question arises whether there can be a communication between
two satellites or the propagation time, which is to be found is related to
one satellite \cite{Dimitrov2026}. The answer is that this case can also be
considered as a communication between two satellites, moving along one
elliptical orbit. The first satellite can be considered to be on the point
of intersection of the large semi-major axis and the orbit ellipse. The
second orbit (with $e_{1}=e_{2}$, $a_{1}=a_{2}$) can be considered to be
rotated at a true anomaly angle\bigskip $f$ anticlockwise, where $f$ is
given from the known formulae from celestial mechanics \cite{KopEfroimKaplan}
\begin{equation}
\tan \frac{f}{2}=\sqrt{\frac{1-\cos f}{1+\cos f}}=\sqrt{\frac{1+e}{1-e}}\tan 
\frac{E}{2}\text{.}  \label{K133}
\end{equation}%
Then the second large semi-major axis will cross the first orbit at some
point, which will have also a true anomaly angle $f$ and consequently
according to the above formulae - the same eccentric anomaly angle $E$. So
the second satellite may be considered to be situated at this point on the
first orbit.
\bigskip In such a setting, the restriction on the ellipticity $e^{2}<\frac{2}{3}$
and the condition for positivity of the geodesic distance $\widehat{R}%
_{AB}^{2}\geq 0$ have the meaning that such a configuration of the
disposition of the satellites represents the minimal angular distance $%
f_{\lim }$ between the satellites (calculated from (\ref{K133}) $\tan \frac{%
f_{\lim }}{2}=\sqrt{\frac{1+e}{1-e}}\tan \frac{E_{\lim }}{2}$), at which the
communication between the two satellites is possible (with account of the
General Relativity effects).
\subsection{Emission and reception propagation times for the case of
satellites on one plane orbit}
\bigskip Expression (\ref{ABC23}) 
\begin{equation*}
\frac{\partial T_{2}}{\partial E_{1}}=-\frac{P_{1}}{2P_{2}}\frac{\partial
T_{1}}{\partial E_{1}}\frac{dE_{2}}{dE_{1}}+\epsilon \sqrt[.]{K}
\end{equation*}%
for $\frac{\partial T_{2}}{\partial E}$ (i.e. when $E_{1}=E_{2}=E$) can be
rewritten as 
\begin{equation}
\frac{\partial T_{2}}{\partial E}=-\frac{1}{2}\frac{\partial T_{1}}{\partial
E}+\varepsilon \sqrt[.]{G_{1}(E,E)+G_{2}(E,E)}\text{.}  \label{ABC29}
\end{equation}%
In (\ref{ABC23}) the function $K$ \ is the previously defined function (\ref%
{ABC22}). For $E_{1}=E_{2}$ the function $K$ is equal to the expression
under the square root in (\ref{ABC29}). Expressions $G_{1}(E,E)$ and $%
G_{2}(E,E)$ can be found from formulaes (\ref{BBB5}) and (\ref{BBB6}) in
Appendix $A$ 
\begin{equation*}
G_{1}(E,E)+G_{2}(E,E)
\end{equation*}%
\begin{equation}
=-\frac{1}{4}\left( \frac{\partial T_{1}}{\partial E}\right) ^{2}-\frac{1}{2}%
\frac{\partial T_{1}}{\partial E}+\left[ \frac{S}{P}+\frac{1}{P}%
2a^{2}(2-e^{2})\sin (2E)\right] \text{,}  \label{ABC30}
\end{equation}%
where the last term in the bracket under the square root is denoted as $%
\widehat{T}(E)$ and since $S$ and $P$ are given by expressions (\ref{ABC15})
and (\ref{ABCM8A}) respectively, $\widehat{T}(E)$ will depend on the
parameters $(a,e,E)$ and on the gravitational potential $V$.
\begin{equation}
\widehat{T}(E):=\frac{S}{P}+\frac{1}{P}2a^{2}(2-e^{2})\sin (2E)\text{.}
\label{ABCD30}
\end{equation}%
Note that $\widehat{T}(E)$ is not related to any propagation time, due to
which it is denoted with a "hat" sign, in order to distinguish it from the
propagation time $T(E)$. After performing the integration in (\ref{ABC29}),
one can obtain ($\epsilon =\pm 1$) 
\begin{equation}
T_{2}=-\frac{1}{2}T_{1}+\epsilon \int dE\sqrt[.]{-\frac{1}{4}\left( \frac{%
\partial T_{1}}{\partial E}\right) ^{2}-\frac{1}{2}\frac{\partial T_{1}}{%
\partial E}+\widehat{T}(E)}\text{.}  \label{ABC31}
\end{equation}%
From (\ref{ABC25}) 
\begin{equation*}
\frac{\partial T_{2}}{\partial E_{2}}=-\frac{1}{2}\left( \frac{dE_{1}}{dE_{2}%
}\right) \left( 1+4\frac{\partial T_{1}}{\partial E_{2}}\right) +\epsilon 
\sqrt[.]{N}
\end{equation*}%
for $E_{1}=E_{2}=E$ one can also derive 
\begin{equation*}
\frac{\partial T_{2}}{\partial E}=-\frac{1}{2}\left( 1+4\frac{\partial T_{1}%
}{\partial E}\right)
\end{equation*}%
\begin{equation}
+\epsilon \sqrt[.]{-\frac{7}{4}\left( \frac{\partial T_{1}}{\partial E}%
\right) ^{2}+\frac{1}{2}\left( \frac{\partial T_{1}}{\partial E}\right) +%
\frac{2S}{P}+\frac{1}{4}+\frac{2a^{2}(2-e^{2})\sin (2E)}{P}}\text{.}
\label{ABC32}
\end{equation}
\subsubsection{The resulting asymmetry in the propagation times $T_{2}-T_{1}$
and $T_{1}-T_{2}$ and the quartic algebraic equation with respect to $\frac{%
\partial T_{1}}{\partial E}$}
\bigskip Note that after performing the integration, it is interesting to note that
the propagation time $T_{2}-T_{1}$ for a signal in the "forward" direction
(i.e. from satellite  $1$ to the second satellite  $2$%
) is different from the propagation time $T_{1}-T_{2}$ in the "backward"
direction (i.e. from satellite  $2$ to the second satellite $1$). 
The reason is that expression (\ref{ABC32}) is not 
symmetric under the interchange $T_{2}\longleftrightarrow T_{1}$.  
\bigskip The asymmetry in $T_{1}$ and $T_{2}$ is clearly seen if after introducing
the notation 
\begin{equation}
Q(E):=\frac{2S}{P}+\frac{1}{4}+\frac{2a^{2}(2-e^{2})\sin (2E)}{P}\text{ ,}
\label{ABC32A1}
\end{equation}%
the expression for $\frac{\partial T_{2}}{\partial E}$ in (\ref{ABC32}) can
be represented as 
\begin{subequations}
\begin{equation}
\frac{\partial T_{2}}{\partial E}=-\frac{1}{2}\left( 1+4\frac{\partial T_{1}%
}{\partial E}\right) +\epsilon \sqrt[.]{-\frac{7}{4}\left( \frac{\partial
T_{1}}{\partial E}\right) ^{2}+\frac{1}{2}\left( \frac{\partial T_{1}}{%
\partial E}\right) +Q(E)}\text{.}  \label{ABC32A2}
\end{equation}%
From the equality of expressions (\ref{ABC29}) and (\ref{ABC32}) for \ $%
\frac{\partial T_{2}}{\partial E}$ and with account also of the expression (%
\ref{ABC30}) for $G_{1}+G_{2}$, one can obtain the following quartic
algebraic equation with respect to $\frac{\partial T_{1}}{\partial E}$ 
\end{subequations}
\begin{equation*}
\frac{1}{2}N\left( 1+4\frac{\partial T_{1}}{\partial E}\right) ^{2}=N^{2}+%
\frac{1}{16}\left( 1+4\frac{\partial T_{1}}{\partial E}\right)
^{4}+(G_{1}+G_{2})^{2}
\end{equation*}%
\begin{equation}
-2N(G_{1}+G_{2})-\frac{1}{2}\left( 1+4\frac{\partial T_{1}}{\partial E}%
\right) ^{2}(G_{1}+G_{2})=0\text{.}  \label{ABC33}
\end{equation}%
This equation can be solved as an algebraic equation with respect to $\frac{%
\partial T_{1}}{\partial E}$ and then, after integration, the function $%
T_{1}=T_{1}(E)$ can be found. Since a quartic equation (i.e. equation of the 
fourth order) with respect to $\frac{\partial T_{1}}{\partial E}$ has four
roots, the initial propagation time $T_{1}$ can be determined in different
ways.
\bigskip It is interesting to note that further four different types of geodesic
distances will be obtained. It may be thought that this may have relation to
the four different solutions of the algebraic equation (\ref{ABC33}) with
respect to $\frac{\partial T_{1}}{\partial E}$. This is a problem, which needs 
further serious investigation. 
\section{Finding the dependence of the second propagation time on the first
propagation time and on the variable Euclidean distance in the general case
of space-oriented orbits and arbitrary space trajectories(not necessarily
orbits)}
\label{sec:propag space}
\bigskip In this section we return again to the model when each set of the space
coordinates $(x_{1},y_{1},z_{1})$ and $(x_{2},y_{2},z_{2})$ depends on the
corresponding generalized parameters $F_{1}$ and $F_{2}$. However, instead
of $F_{1}$ and $F_{2}$, the coordinates may depend on some other dynamical
parameter, related to the motion of the two satellites. Such a model will not be 
considered in this paper.  Further it shall be assumed
that the space coordinates $(x_{1},y_{1},z_{1})$ depend only on the
dynamical parameter $F_{1}$ and the other set of space coordinates $%
(x_{2},y_{2},z_{2})$ depends only on the second dynamical parameter $F_{2}$.
In fact, a more realistic model should account for the dependence $%
(x_{1}(F_{1},F_{2}),y_{1}(F_{1},F_{2}),z_{1}(F_{1},F_{2}))$ and $%
(x_{2}(F_{1},F_{2}),y_{2}(F_{1},F_{2}),z_{2}(F_{1},F_{2}))$. Such a model
will not be considered in this paper. Also, the model can easily be
generalized to the case, when the coordinates depend on many dynamical
parameters $F_{1},F_{2}....,F_{n}$.
\bigskip Further a more general model shall be presented, based on the dependence of
the two radius-vectors on the space coordinates. Let us first represent the
solution of the first equation (\ref{ABC1}) for the first propagation time $%
T_{1}$ in the form 
\begin{equation}
T_{1}=\frac{1}{c}\int_{E_{per}^{(1)}}^{\widetilde{E}_{1}}\sqrt{\widetilde{Q}%
_{1}(V_{1},x_{1},y_{1},z_{1},E_{1})}dE_{1}+T_{1}^{init}\text{,}
\label{DOPC11}
\end{equation}%
where $\widetilde{Q}_{1}(V_{1},x_{1},y_{1},z_{1},E_{1})$ is the expression 
\begin{equation}
\widetilde{Q}_{1}:=\frac{(c^{2}-2V_{1})}{(c^{2}+2V_{1})}\left(
(dx_{1})^{2}+(dy_{1})^{2}+(dz_{1})^{2}\right) \text{.}  \label{DOPC22}
\end{equation}
\bigskip Let us write the hyperplane equation (\ref{ABC2A22}) for the differential of
the square of the Euclidean distance as 
\begin{equation}
2(z_{1}-z_{2})\widetilde{S}_{1}dF_{1}-2(z_{1}-z_{2})\widetilde{S}%
_{2}dF_{2}=dR_{AB}^{2}\text{,}  \label{DOPC4}
\end{equation}%
where $\widetilde{S}_{i}$ $(i=1,2)$ is the introduced notation for 
\begin{equation*}
\widetilde{S}_{i}:=\left( \frac{\partial z_{i}}{\partial F_{i}}\right) +%
\frac{(x_{1}-x_{2})}{(z_{1}-z_{2})}\frac{\partial x_{i}}{\partial F_{i}}
\end{equation*}%
\begin{equation}
+\frac{(y_{1}-y_{2})}{(z_{1}-z_{2})}\frac{\partial y_{i}}{\partial F_{i}}%
\text{.}  \label{DOPC5}
\end{equation}%
Expressing the differential $dF_{2}$ from (\ref{DOPC4}), substituting it
into expression (\ref{ABC2}) for the second propagation time, and
introducing also the notation ($i=1,2$) 
\begin{equation}
\widetilde{R}_{i}:=\frac{1}{c}\left( \frac{c^{2}-2V_{i}}{c^{2}+2V_{i}}%
\right) \left[ \left( \frac{\partial x_{i}}{\partial F_{i}}\right)
^{2}+\left( \frac{\partial y_{i}}{\partial F_{i}}\right) ^{2}+\left( \frac{%
\partial z_{i}}{\partial F_{i}}\right) ^{2}\right] \text{,}  \label{DOPC6}
\end{equation}%
one can represent expression (\ref{ABC2}) for the second null cone in the
form 
\begin{equation}
dT_{2}=\frac{1}{c}\sqrt[.]{\widetilde{A}_{1}(dT_{1})^{2}+\widetilde{A}%
_{2}(dT_{1})+\widetilde{A}_{3}}=\frac{1}{c}\sqrt[.]{\widetilde{Q}_{2}(....)}%
\text{.}  \label{DOPC7}
\end{equation}%
In (\ref{DOPC7}) $\widetilde{A}_{1}$, $\widetilde{A}_{2}$ and $\widetilde{A}%
_{3}$ are complicated expressions, depending on the derivatives of the space
coordinates, on the expression (\ref{DOPC5}) and (\ref{DOPC6}) for $%
\widetilde{S}_{i}$ and $\widetilde{R}_{i}$ and also on the differential $%
dR_{AB}^{2}$ \ of the square of the Euclidean distance. The expressions for $%
\widetilde{A}_{1}$, $\widetilde{A}_{2}$ and $\widetilde{A}_{3}$ will be
given in Appendix A. It can be noted also  from (\ref{DOPC11}) and
(\ref{DOPC4}) that the two differentials $dF_{1}$ and $dF_{2}$ can be expressed
as 
\begin{equation}
dF_{1}=\frac{dT_{1}}{\widetilde{R}_{1}}\text{, }dF_{2}=\frac{\widetilde{S}%
_{1}}{\widetilde{S}_{2}}\frac{dT_{1}}{\widetilde{R}_{1}}-\frac{dR_{AB}^{2}}{%
2(z_{1}-z_{2})\widetilde{S}_{2}}\text{ \ \ }  \label{DOPC8}
\end{equation}%
and the differential $dF_{2}$ depends on the differential of the square of
the Euclidean distance $dR_{AB}^{2}$. For arbitrary space trajectories of
the two satellites, the Euclidean distance $R_{AB}^{2}$ in (\ref{DOPC8})
should be substituted by the expression 
\begin{equation}
R_{AB}^{2}=(x_{1}-x_{2})^{2}+(y_{1}-y_{2})^{2}+(z_{1}-z_{2})^{2}\text{.}
\label{DOPC88}
\end{equation}%
From the system of equations (\ref{DOPC8}) the following nonlinear equation
is obtained for the dependence of the generalized dynamical parameter $F_{2}$
on the generalized dynamical parameter $F_{1}$ 
\begin{equation}
\frac{dF_{2}}{dF_{1}}=\frac{\widetilde{S}_{1}}{\widetilde{S}_{2}}-\frac{%
dR_{AB}^{2}}{2(z_{1}-z_{2})}\frac{\widetilde{R}_{1}\widetilde{S}_{2}}{dT_{1}}%
\text{ .}  \label{DOPC99}
\end{equation}%
An interesting case for application of this complicated equation will be
exchange of signals between satellites orbiting on two space-oriented
orbits, for which the orbital coordinates are defined by the transformations
(\ref{K11}), (\ref{K22}) and (\ref{K33}) (written with the indices $1$ and $%
2 $) and the true anomaly angles $f_{1}$and $f_{2}$ are assumed to play the
role of the dynamical parameters $F_{1}$ and $F_{2}$. It should be noted
that the space-time distance $\widetilde{R}_{AB}^{2}$ and the geodesic
distance $\widehat{R}_{AB}^{2}$ can be determined, following the formalism
in the previous sections \ for the two-dimensional plane elliptic orbits.
However, further a general formalism for finding the space-time distance and
the geodesic distance will be presented, when the two satellites move along
arbitrary space trajectories, not necessarily representing orbits,
characterized by the Kepler parameters.
\bigskip In any case, after finding the solutions $F_{2}=F_{2}(F_{1})$ of the
nonlinear differential equation (\ref{DOPC99}) and substituting them in the
corresponding expressions for the space-time distance and the geodesic
distance, these expressions will not be symmetric with respect to the
dynamical parameters $F_{1}\Longleftrightarrow F_{2}$.
\bigskip An interesting problem to be investigated further is how the causality
condition $T_{2}>T_{1}$ or $T_{1}>T_{2}$ can be defined for the range
equation (\ref{DOPC99}) 
\begin{equation*}
c(T_{2}-T_{1})-\widehat{R}_{AB}^{2}=0\text{ , }c(T_{1}-T_{2})-\widehat{R}%
_{BA}^{2}=0\text{ .}
\end{equation*}%
In both cases the fulfillment of the causality condition in the "forward"
direction ($1\mapsto 2$) and in the "backward" direction ($2\mapsto 1$) will
be related to the inequality of the corresponding geodesic distances $%
\widehat{R}_{AB}^{2}$ and $\widehat{R}_{BA}^{2}$, which will not be a result
of any assumptions, but will be confirmed by the strict mathematical
calculations and also is a physical consequence of the propagation of signals in the 
"forward" and "backwards" directions. As noted, the physical reason is hidden in the 
non-inertiality of the reference systems, associated with the two satellites, emitting and 
receiving the signals.  
\subsection{Two satellites orbiting on plane elliptic orbits and the
nonlinear equation for the dependence of the second eccentric anomaly angle $%
E_{2}$ on the first anomaly angle $E_{1}$}
\bigskip For satellites on plane elliptic orbits the expression (\ref{DOPC88}) for $%
R_{AB}^{2}$ acquires the form (\ref{ABC27}) and the differential $%
dR_{AB}^{2} $ should be taken in the form (\ref{ABC36A}) 
\begin{equation*}
dR_{AB}^{2}=F_{1}(E_{1},E_{2})dE_{1}+F_{2}(E_{1},E_{2})dE_{2}\text{ .}
\end{equation*}%
For this case, equation (\ref{DOPC4}) should be replaced by 
\begin{equation}
2\widetilde{S}_{1}^{(1)}dE_{1}-2\widetilde{S}_{2}^{(2)}dE_{2}=dR_{AB}^{2}%
\text{,}  \label{DOPC100}
\end{equation}%
where 
\begin{equation}
S_{i}^{(i)}\equiv (x_{1}-x_{2})\frac{\partial x_{i}}{\partial E_{i}}%
+(y_{1}-y_{2})\frac{\partial x_{i}}{\partial E_{i}}\text{ \ \ }i=1,2\text{ .}
\label{DOPC101}
\end{equation}%
The analogue of the nonlinear differential equation (\ref{DOPC99}) for this
case will be obtained from (\ref{DOPC100}) and will have the following form 
\begin{equation}
\frac{dE_{2}}{dE_{1}}=\frac{2\widetilde{S}_{1}^{(1)}-F_{1}(E_{1},E_{2})}{2%
\widetilde{S}_{2}^{(2)}+F_{2}(E_{1},E_{2})}\text{ \ ,}  \label{DOPC102}
\end{equation}%
where $F_{1}(E_{1},E_{2})$ and $F_{2}(E_{1},E_{2})$ are the previously
defined expressions (\ref{ABC37}) 
\begin{equation*}
F_{1}(E_{1},E_{2}):=2e_{1}a_{1}^{2}\left( 1-e_{1}\cos E_{1}\right) \sin
E_{1}-S_{1}(E_{1},E_{2})
\end{equation*}%
and (\ref{ABC38})%
\begin{equation*}
F_{2}(E_{1},E_{2})=2e_{2}a_{2}^{2}\left( 1-e_{2}\cos E_{2}\right) \sin
E_{1}-S_{2}(E_{1},E_{2})\text{.\ \ }
\end{equation*}%
In (\ref{ABC37}) and (\ref{ABC38}) the expressions for $S_{1}(E_{1},E_{2})$
and $S_{2}(E_{1},E_{2})$ are defined according to formulae (\ref{ABC8}).
Note that if the nonlinear differential equation (\ref{DOPC102}) is solved
and the dependence $E_{2}=E_{2}(E_{1})$ is found, then if substituted in
formulaes (\ref{ABC43}) for the space-time distance, (\ref{ABC45}) for the
compatibility condition and (\ref{ABC46}) for the geodesic distance, then
there will no longer be a symmetry with respect to the interchange $%
E_{1}\Longleftrightarrow E_{2}$.
\subsection{Second propagation time for the case of changing Euclidean
distance between the satellites - some physical considerations}
\label{sec:SecChang}
\bigskip The found formulae (\ref{DOPC7}) for the differential of the second
propagation time 
\begin{equation*}
dT_{2}=\frac{1}{c}\sqrt[.]{\widetilde{A}_{1}(dT_{1})^{2}+\widetilde{A}%
_{2}(dT_{1})+\widetilde{A}_{3}}=\frac{1}{c}\sqrt[.]{\widetilde{Q}_{2}(....)}
\end{equation*}%
clearly illustrates the difference when the propagation time is found only
as a solution of the null cone equation in comparison with the present
case, when the differential of the second propagation time $T_{2}$ depends
also on the coefficient functions $\widetilde{A}_{1}$, $\widetilde{A}_{2}$,$%
\widetilde{A}_{3}$, given in Appendix A. For example, the function $%
\widetilde{A}_{1}$ in (\ref{GGGG1}) depends on the expressions for $\left( 
\frac{\partial z_{2}}{\partial E_{1}}\right) ^{2}$ and $\left( \frac{%
\partial z_{2}}{\partial E_{2}}\right) ^{2}$, which depend in a complicated
way on $dR_{AB}^{2}$ through the formulaes (\ref{GGGG2}) and (\ref{GGGG3}).
The coefficient function $\widetilde{A}_{2}$ in (\ref{GGGG4}) depends on $%
dR_{AB}^{2}$, but also on the expressions for $\left( \frac{\partial z_{2}}{%
\partial E_{1}}\right) ^{2}$ and $\left( \frac{\partial z_{2}}{\partial E_{2}%
}\right) ^{2}$and the function (\ref{GGGG5}) for $\widetilde{A}_{3}$ depends
on $\left( dR_{AB}^{2}\right) ^{2}$. Consequently, if the first propagation
time $T_{1}$ is known and also the changing Euclidean distance $R_{AB}^{2}$
between the two satellites is known, then the second propagation time $T_{2}$
can also be found after complicated integration of the differential equation
(\ref{DOPC7}). In (\ref{DOPC7}) the parametrizations of the two sets $%
x_{1},y_{1},z_{1}$ and $x_{2},y_{2},z_{2}$ might not necessarily depend on
the parameters of their plane elliptic or space-oriented orbits, but may
depend on some parameter, related to the motion of the satellites.
\bigskip From a physical point of view, it is interesting to note that if both
propagation times $T_{1}$ and $T_{2}$ are known, then the differential $%
dR_{AB}^{2}$ of the Euclidean distance can be found after treating formulae (%
\ref{DOPC7}) as a complicated algebraic relation. This method shall not be
presented in the present paper. This example clearly illustrates the
advantage of investigating the relativistic version of the ranging equation,
which will be $c(T_{2}-T_{1})-\widehat{R}_{AB}^{2}=0$, where $\widehat{R}%
_{AB}^{2}$ will be the geodesic distance, found according to the
prescription and the methods, developed in the previous sections.
\section{Derivation of the quadratic equation with respect to the space-time
distance for the general case of moving satellites on arbitrary space
trajectories}
\label{sec:GenCaseTwoSatel}
\subsection{Quadratic equation from the two null cone equations and the
hyperplane equation for the differential of the square of the Euclidean
distance}
\bigskip Let us consider the general case of two moving satellites along space
trajectories \ $r_{1}=r_{1}(t_{1},F_{1})$ and $r_{2}=r_{2}(t_{2},F_{2})$,
where it has been assumed that $F_{1}$ and $F_{2}$ are dynamical variables,
related to the motion of the satellites.Therefore, they can be the two
eccentric anomaly angles $E_{1}$ and $E_{2}$ for the case of plane orbital
motion or they can be the true anomaly angles $f_{1}$ and $f_{2}$ for the
case of space-oriented orbits, characterized by the six Kepler parameters $%
(f,a,e,\Omega ,I,\omega )$. The two null cone equations are given by the
expressions (\ref{ABC1}), (\ref{ABC2}) with origins
at the two satellites and also the hyperplane equation (\ref{ABC2A22}).
\bigskip Our aim will be to prove that the solution of the system of equations (\ref%
{ABC1}), (\ref{ABC2}), (\ref{ABC2A22}) with respect to $R_{AB}^{2}$ will
enable to define the space-time distance $\widetilde{R}_{AB}^{2}$, which
will be different from the Euclidean distance \cite{Dimitrov2026}. In other
words, for this general case the algorithm for finding the space-time
distance, the compatibility condition and the geodesic distance will turn
out to be the same as for the case of plane elliptic orbits. Further it will
be assumed also that the dependence of the propagation times $T_{1}$ and $%
T_{2}$ on the space coordinates $x_{1},y_{1},z_{1}$ and $x_{2},y_{2},z_{2}$
will be found after \ finding the solutions of the two null cone equations (\ref%
{ABC1}) and (\ref{ABC2}).
\bigskip Since the two sets of space coordinates depend on the dynamical parameters $%
F_{1}$ and $F_{2}$, determining the motion of the satellites, the
differentials of the space coordinates can be written as 
\begin{equation}
d\overrightarrow{\mathbf{r}}_{1}(F_{1})=\frac{\partial \overrightarrow{%
\mathbf{r}}_{1}}{\partial F_{1}}dF_{1}\text{, }d\overrightarrow{\mathbf{r}}%
_{2}(F_{2})=\frac{\partial \overrightarrow{\mathbf{r}}_{2}}{\partial F_{2}}%
dF_{2}\text{,}  \label{M4}
\end{equation}%
where $\overrightarrow{\mathbf{r}}_{1}=(x_{1},y_{1},z_{1})$ and $%
\overrightarrow{\mathbf{r}}_{2}=(x_{2},y_{2},z_{2})$. It can easily be
checked that the differential of the square of the Euclidean distance can be
rewritten in the form%
\begin{equation}
dR_{AB}^{2}=2(\overrightarrow{\mathbf{r}}_{1}.d\overrightarrow{\mathbf{r}}%
_{1})+2(\overrightarrow{\mathbf{r}}_{2}.d\overrightarrow{\mathbf{r}}_{2})-2(%
\overrightarrow{\mathbf{r}}_{1}.d\overrightarrow{\mathbf{r}}_{2})-2(%
\overrightarrow{\mathbf{r}}_{2}.d\overrightarrow{\mathbf{r}}_{1})  \label{M5}
\end{equation}%
\begin{equation}
=D_{1}(F_{1},F_{2},\overrightarrow{\mathbf{r}}_{1},\overrightarrow{\mathbf{r}%
}_{2})dF_{1}+D_{2}(F_{1},F_{2},\overrightarrow{\mathbf{r}}_{1},%
\overrightarrow{\mathbf{r}}_{2})dF_{2}\text{,}  \label{M6}
\end{equation}%
where $D_{1}(F_{1},F_{2},\overrightarrow{\mathbf{r}}_{1},\overrightarrow{%
\mathbf{r}}_{2})$ and $D_{2}(F_{1},F_{2},\overrightarrow{\mathbf{r}}_{1},%
\overrightarrow{\mathbf{r}}_{2})$ denote the expressions 
\begin{equation}
D_{1}(F_{1},F_{2},\overrightarrow{\mathbf{r}}_{1},\overrightarrow{\mathbf{r}}%
_{2}):=\overrightarrow{\mathbf{r}}_{1}.\frac{\partial \overrightarrow{%
\mathbf{r}}_{1}}{\partial F_{1}}-\overrightarrow{\mathbf{r}}_{2}.\frac{%
\partial \overrightarrow{\mathbf{r}}_{1}}{\partial F_{1}}\text{, }
\label{M7A}
\end{equation}%
\begin{equation}
D_{2}(F_{1},F_{2},\overrightarrow{\mathbf{r}}_{1},\overrightarrow{\mathbf{r}}%
_{2}):=\overrightarrow{\mathbf{r}}_{2}.\frac{\partial \overrightarrow{%
\mathbf{r}}_{2}}{\partial F_{2}}-\overrightarrow{\mathbf{r}}_{1}.\frac{%
\partial \overrightarrow{\mathbf{r}}_{2}}{\partial F_{2}}\text{.}
\label{M7B}
\end{equation}%
Expressing $(dF_{2})^{2}$ from (\ref{M5}) and (\ref{M6}) as a function of $%
dR_{AB}^{2}$ and substituting in the second null cone equation (\ref{ABC2}),
one can obtain the following equation in full derivatives with respect to $%
dT_{2}$ and $dF_{1}$ 
\begin{equation}
ds_{2}^{2}=0=-c^{2}(1+2\frac{V_{2}}{c^{2}})(dT_{2})^{2}+(1-\frac{2V_{2}}{%
c^{2}})\mid \frac{\partial \overrightarrow{\mathbf{r}}_{1}}{\partial F_{1}}%
\mid ^{2}  \notag \\
\end{equation}%
\begin{eqnarray}
&& \\
&&\times \left[ \frac{\left( dR_{AB}^{2}\right) ^{2}}{(D_{2})^{2}}+\frac{%
\left( D_{1}\right) ^{2}}{\left( D_{2}\right) ^{2}}(dF_{1})^{2}-2\frac{%
dR_{AB}^{2}}{D_{2}^{2}}D_{1}dF_{1}\right] \text{.}  \label{M8}
\end{eqnarray}
\bigskip The first null cone equation (\ref{ABC1}) can be represented in the form 
\begin{equation}
ds_{1}^{2}=0=-c^{2}(1+2\frac{V_{1}}{c^{2}})(dT_{1})^{2}+(1-2\frac{V_{1}}{%
c^{2}})\mid \frac{\partial \overrightarrow{\mathbf{r}}_{1}}{\partial F_{1}}%
\mid ^{2}(dF_{1})^{2}\text{,}  \label{M9}
\end{equation}%
from where an expression can be found for $dF_{1}$ (respectively also for $%
dF_{2}$ from the second null cone equation (\ref{ABC2})) 
\begin{equation}
dF_{1}=\frac{c(dT_{1})}{\mid \frac{\partial \overrightarrow{\mathbf{r}}_{1}}{%
\partial F_{1}}\mid }S_{1}(r_{1})\text{,\ }dF_{2}=\frac{c(dT_{2})}{\mid 
\frac{\partial \overrightarrow{\mathbf{r}}_{2}}{\partial F_{2}}\mid }%
S_{2}(r_{2})\text{.}  \label{M10}
\end{equation}%
In (\ref{M10}) $S_{1}(r_{1})$, $S_{2}(r_{2})$) ,$\mid \frac{\partial 
\overrightarrow{\mathbf{r}}_{1}}{\partial F_{1}}\mid $ denote the
expressions 
\begin{equation}
S_{1}(r_{1})=\sqrt[.]{\frac{c^{2}+2V_{1}}{c^{2}-2V_{1}}}\text{, }%
S_{2}(r_{2})=\sqrt[.]{\frac{c^{2}+2V_{2}}{c^{2}-2V_{2}}}\text{\ }
\label{M11}
\end{equation}%
and 
\begin{equation}
\mid \frac{\partial \overrightarrow{\mathbf{r}}_{1}}{\partial F_{1}}\mid =%
\sqrt[.]{\left( \frac{\partial x_{1}}{\partial F_{1}}\right) ^{2}+\left( 
\frac{\partial y_{1}}{\partial F_{1}}\right) ^{2}+\left( \frac{\partial z_{1}%
}{\partial F_{1}}\right) ^{2}}\text{.}  \label{M12}
\end{equation}%
In (\ref{M12}) $\mid \frac{\partial \overrightarrow{\mathbf{r}}_{1}}{%
\partial F_{1}}\mid $ represents the length of the vector $\frac{\partial 
\overrightarrow{\mathbf{r}}_{1}}{\partial F_{1}}=(\frac{\partial x_{1}}{%
\partial F_{1}},\frac{\partial y_{1}}{\partial F_{1}},\frac{\partial z_{1}}{%
\partial F_{1}})$, while the notation $\frac{\partial r_{1}}{\partial F_{1}}$
for the derivative of the scalar quantity $r_{1}=r_{1}(F_{1})=\sqrt[.]{%
x_{1}^{2}+y_{1}^{2}+z_{1}^{2}}$ will mean that 
\begin{equation}
\frac{\partial r_{1}}{\partial F_{1}}=\frac{1}{r_{1}}\left( x_{1}\frac{%
\partial x_{1}}{\partial F_{1}}+y_{1}\frac{\partial y_{1}}{\partial F_{1}}%
+z_{1}\frac{\partial z_{1}}{\partial F_{1}}\right) =\frac{1}{r_{1}}\left( 
\overrightarrow{\mathbf{r}}_{1}.\frac{\partial \overrightarrow{\mathbf{r}}%
_{1}}{\partial F_{1}}\right) \text{.}  \label{M13}
\end{equation}%
The gravitational potentials $V_{1}$ and $V_{2}$ in (\ref{M11}) are defined
as 
\begin{equation}
V_{1}=\frac{GM}{r_{1}}\text{ , }V_{2}=\frac{GM}{r_{2}}\text{ ,}  \label{M15A}
\end{equation}%
so both satellites are moving in the gravitational field, created by one
celestial body (the Earth). Of course, much more realistic will be the
two-body problem (Earth-Moon) or even the three-body problem
(Earth-Moon-Sun), but the calculations will be much more complicated.
\bigskip Substituting expression (\ref{M10}) for the differential $dF_{1}$ in
equation (\ref{M8}) for the square of the Euclidean distance, one can obtain
a quadratic equation with respect to the differential of the square of the
Euclidean distance $dR_{AB}^{2}$ \cite{Dimitrov2026}%
\begin{equation}
\widehat{P}_{1}(F_{1},F_{2})\left( dR_{AB}^{2}\right) ^{2}+2\widehat{P}%
_{2}(F_{1},F_{2})dR_{AB}^{2}+\widehat{P}_{3}(F_{1},F_{2})=0\text{,}
\label{M16}
\end{equation}%
where $\widehat{P}_{1}(F_{1},F_{2})$, $\widehat{P}_{2}(F_{1},F_{2})$, $%
\widehat{P}_{3}(F_{1},F_{2})$ denote the following expressions 
\begin{equation}
\widehat{P}_{1}(F_{1},F_{2}):=\mid \frac{\partial \overrightarrow{\mathbf{r}}%
_{1}}{\partial F_{1}}\mid ^{2}\frac{(1-\frac{2V_{2}}{c^{2}})}{D_{2}^{2}}%
\text{,}  \label{M17}
\end{equation}%
\begin{equation}
\widehat{P}_{2}(F_{1},F_{2}):=-(1-\frac{2V_{2}}{c^{2}})\frac{D_{1}}{D_{2}^{2}%
}.\mid \frac{\partial \overrightarrow{\mathbf{r}}_{1}}{\partial F_{1}}\mid
.c(dT_{1})S_{1}\text{,}  \label{M18}
\end{equation}
\begin{equation}
\widehat{P}_{3}(F_{1},F_{2}):=-c^{2}(1+\frac{2V_{2}}{c^{2}})(dT_{2})^{2} 
\notag \\
\end{equation}%
\begin{eqnarray}
&& \\
&&+(1-\frac{2V_{2}}{c^{2}})\frac{D_{1}^{2}c^{2}(dT_{1})^{2}S_{1}^{2}}{%
D_{2}^{2}}\text{ .}  \label{M19A}
\end{eqnarray}
\subsection{Space-time distance as a solution of the quadratic equation for
the differential of the square of the Euclidean distance}
\bigskip The quadratic equation (\ref{M16}) is important for the proof that the
distance $dR_{AB}^{2}$ can be also negative, which is characteristic feature
for the space-time distance. Because of this, further we shall denote the
solution of the quadratic equation (\ref{M16}) for $dR_{AB}^{2}$ by $d%
\widetilde{R}_{AB}^{2}$ \cite{Dimitrov2026}:
.%
\begin{equation}
d\widetilde{R}_{AB}^{2}=-\frac{\widehat{P}_{2}}{2\widehat{P}_{1}}\mp \sqrt[.]%
{\left( \frac{\widehat{P}_{2}}{2\widehat{P}_{1}}\right) ^{2}-\frac{\widehat{P%
}_{3}}{\widehat{P}_{1}}}  \label{M20}
\end{equation}%
\begin{equation}
=d\widetilde{R}_{AB}^{2}(1\rightarrow 2)=\frac{cD_{1}(dT_{1})S_{1}(r_{1})}{%
\mid \frac{\partial \overrightarrow{\mathbf{r}}_{1}}{\partial F_{1}}\mid }%
\pm \frac{cD_{2}(dT_{2})S_{2}(r_{2})}{\mid \frac{\partial \overrightarrow{%
\mathbf{r}}_{1}}{\partial F_{1}}\mid })\text{.}  \label{M21}
\end{equation}%
The first term in (\ref{M21}) corresponds to the term $-\frac{\widehat{P}_{2}%
}{2\widehat{P}_{1}}$ in (\ref{M20}) and the second term in (\ref{M21}) - to
the term $\sqrt[.]{\left( \frac{\widehat{P}_{2}}{2\widehat{P}_{1}}\right)
^{2}-\frac{\widehat{P}_{3}}{\widehat{P}_{1}}}$ in (\ref{M20}). Since $dT_{2}$
will be always positive (the positivity of the propagation time as a
solution of the null cone equation was proved in the papers \cite%
{Dimitrov2022} and \cite{Dimitrov2025}) and $S_{2}(r_{2})$ and $\mid \frac{%
\partial \overrightarrow{\mathbf{r}}_{1}}{\partial F_{1}}\mid $ are also
positive, the expression under the square root in (\ref{M20}) can be
negative only when $D_{2}=$ $(\overrightarrow{\mathbf{r}}_{2}-%
\overrightarrow{\mathbf{r}}_{1}).\frac{\partial \overrightarrow{\mathbf{r}}%
_{2}}{\partial F_{2}}<0$. Then the expression for $d\widetilde{R}_{AB}^{2}$
will be imaginary and $\widetilde{R}_{AB}^{2}$ will be also imaginary.
Consequently, the found expression has the properties, characteristic for
the space-time distance, because if the differential $d\widetilde{R}%
_{AB}^{2} $ is imaginary, the space-time distance will be also imaginary.
\subsection{Symmetry properties of the two expressions for the space-time
distance}
\label{sec:SymPropSpaceDist}
\bigskip Unlike the previously considered case for a signal, exchanged between two
satellites moving on plane elliptic orbits, when there was a symmetry with
respect to the interchange of the indices $1\longleftrightarrow 2$, the
expression (\ref{M21}) will not have this symmetry. This means also that
under the interchange $\overrightarrow{\mathbf{r}}_{1}\longleftrightarrow 
\overrightarrow{\mathbf{r}}_{2}$, when the expressions in (\ref{M21}) change
as 
\begin{equation}
D_{1}\longleftrightarrow D_{2}\text{, }S_{1}(r_{1})\longleftrightarrow
S_{2}(r_{2})\text{ , }T_{1}\longleftrightarrow T_{2}\text{ ,}  \label{M24}
\end{equation}%
the minus sign (if we take the case of the lower "minus" sign) in front of
the second term in the expression (\ref{M21}) for $d\widetilde{R}%
_{AB}^{2}(1\rightarrow 2)$ will reverse its sign. Consequently, the two
expressions for $d\widetilde{R}_{AB}^{2}(1\rightarrow 2)$ and $d\widetilde{R}%
_{BA}^{2}(2\rightarrow 1)$ will not be equal, because also of the term $\mid 
\frac{\partial \overrightarrow{\mathbf{r}}_{1}}{\partial F_{1}}\mid $ in the
denominator of (\ref{M21}) 
\begin{equation*}
d\widetilde{R}_{AB}^{2}(1\rightarrow 2)=\frac{cD_{1}(dT_{1})S_{1}(r_{1})}{%
\mid \frac{\partial \overrightarrow{\mathbf{r}}_{1}}{\partial F_{1}}\mid }%
\pm \frac{cD_{2}(dT_{2})S_{2}(r_{2})}{\mid \frac{\partial \overrightarrow{%
\mathbf{r}}_{1}}{\partial F_{1}}\mid })\text{,}
\end{equation*}%
which will change to$\mid \frac{\partial \overrightarrow{\mathbf{r}}_{2}}{%
\partial F_{2}}\mid $ in the expression for $dR_{BA}^{2}(2\rightarrow 1)$ 
\begin{equation}
d\widetilde{R}_{BA}^{2}(2\rightarrow 1)=\frac{cD_{2}(dT_{2})S_{2}(r_{2})}{%
\mid \frac{\partial \overrightarrow{\mathbf{r}}_{2}}{\partial F_{2}}\mid }%
\pm \frac{cD_{1}(dT_{1})S_{1}(r_{1})}{\mid \frac{\partial \overrightarrow{%
\mathbf{r}}_{2}}{\partial F_{2}}\mid })\neq d\widetilde{R}%
_{AB}^{2}(1\rightarrow 2)\text{.}  \label{M25}
\end{equation}%
The non-equivalence of the two expressions for $d\widetilde{R}%
_{AB}^{2}(1\rightarrow 2)$ and $d\widetilde{R}_{BA}^{2}(2\rightarrow 1)$ is
related to the fact that concerning the differential of the space-time
distance, there is no equivalency in the process of sending a signal from a
satellite-emitting satellite from point $A$ to the signal-receiving
satellite at point $B$, and then "backwards" - from the signal-receiving
satellite at point $B$ back to the signal-emitting satellite at point $A$.
This non-equivalence of the two differentials \ $d\widetilde{R}%
_{AB}^{2}(1\rightarrow 2)$ and $d\widetilde{R}_{BA}^{2}(2\rightarrow 1)$
displays itself later on in the determination of the geodesic distance in
the two directions \ $\widehat{R}_{AB}^{2}(1\rightarrow 2)$ and $\widehat{R}%
_{BA}^{2}(2\rightarrow 1)$. From a physical point of view, this
non-equivalency is natural because according to Einstein's 
Principle of Relativity, it does not have any importance if a light signal
is being sent from one inertial observer to another and then in the
backwards direction. The more complicated problem in the investigated case is that both
satellites move non-uniformly and with different orbital velocities, so both signals in
the forward ($1\mapsto 2$) direction and in the backward one ($%
2\longrightarrow 1$) can be considered to be send in non-inertial reference
systems.
\subsubsection{Consequences for the definition of the geodesic distance}
\bigskip Related to the problem of defining the physical notion of geodesic distance,
which shall be discussed in the next sections, this non-equivalency would mean
that it will be natural to expect that the two geodesic distances from point 
$A$ to point $B$ and then from point $B$ to point $A~$will be different.
\bigskip We shall find an expression for the geodesic distance in the next sections
and from physical considerations, we shall try to specify the properties of
the geodesic distance. Further, we shall keep the notations $\widetilde{R}%
_{AB}^{2}(1\rightarrow 2)$ and $\widetilde{R}_{BA}^{2}(2\rightarrow 1)$ for
the space-time distances between space points $A$ and $B$ and between points 
$B$ and $A$, respectively, which are different from the usual Euclidean
distance.
\subsection{The case of equal to zero differentials of the space-time
distances $d\protect\widetilde{R}_{AB}^{2}(1\rightarrow 2)$ and $d\protect%
\widetilde{R}_{BA}^{2}(2\rightarrow 1)$}
\label{sec:ZeroDifferSpTimeDistances}
\bigskip Note also that both differentials of the space-time distances $d\widetilde{R}%
_{AB}^{2}(1\rightarrow 2)$ and $d\widetilde{R}_{BA}^{2}(2\rightarrow 1)$
from (\ref{M21}) and (\ref{M25}) are equal to zero for 
\begin{equation}
D_{1}(dT_{1})S_{1}(r_{1})=\mp D_{2}(dT_{2})S_{2}(r_{2})  \label{M26}
\end{equation}%
and thus do not depend on the length of the vectors $\mid \frac{\partial 
\overrightarrow{\mathbf{r}}_{1}}{\partial F_{1}}\mid $ and $\mid \frac{%
\partial \overrightarrow{\mathbf{r}}_{2}}{\partial F_{2}}\mid $. In the
introduction a physical interpretation was given that this equality is the
analogue of the time intervals of emission and reception of the signal in
the thought experiment, proposed in the monographs by Bondi \cite{Bondi1}
and \cite{Bondi2}.
\bigskip Further, after finding the geodesic distance, it shall be proved that the
equality and constancy of the two space-time distances 
\begin{equation}
\widetilde{R}_{AB}^{2}(1\rightarrow 2)=\widetilde{R}_{BA}^{2}(2\rightarrow
1)=const.  \label{M27}
\end{equation}%
may be a natural definition for the space-time distance, but only for a
certain specific case, when the square of the radius-vector remains
constant, i.e. $\left( \overrightarrow{\mathbf{r}}_{1}\right) ^{2}=\left( 
\overrightarrow{\mathbf{r}}_{2}\right) ^{2}$. Consequently, it shall be
proved that for variable space-time distances $\widetilde{R}%
_{AB}^{2}(1\rightarrow 2)$ and $\widetilde{R}_{BA}^{2}(2\rightarrow 1)$ the
equality (\ref{M26}) will be modified.
\section{Solutions for the space-time distances for the case of
inter-satellite communications between satellites on different space
trajectories (not necessarily orbits)}
\label{sec:SolutSpaceTime}
\bigskip Now the solution for the two space-time distances shall be found from the
two equations in full derivatives (\ref{M21}) and (\ref{M25}).
\bigskip Let us write the differential equality (\ref{M21}) in the form 
\begin{equation*}
d\widetilde{R}_{AB}^{2}(1\rightarrow 2)=\frac{\partial \widetilde{R}_{AB}^{2}%
}{\partial F_{1}}dF_{1}+\frac{\partial \widetilde{R}_{AB}^{2}}{\partial F_{2}%
}dF_{2}
\end{equation*}%
\begin{equation}
=\frac{cD_{1}(\frac{\partial T_{1}}{\partial F_{1}})S_{1}(r_{1})}{\mid \frac{%
\partial \overrightarrow{\mathbf{r}}_{1}}{\partial F_{1}}\mid }dF_{1}\pm 
\frac{cD_{2}(\frac{\partial T_{2}}{\partial F_{2}})S_{2}(r_{2})}{\mid \frac{%
\partial \overrightarrow{\mathbf{r}}_{1}}{\partial F_{1}}\mid }dF_{2}\text{,}
\label{M28}
\end{equation}%
provided that propagation time $T_{1}$ depends only on the dynamical
parameter $F_{1}$ and the propagation time $T_{2}$ depends on the dynamical
parameter $F_{2}$. 
\begin{equation}
T_{1}=T_{1}(F_{1})\text{,}T_{2}=T_{2}(F_{2})\text{.}  \label{M29}
\end{equation}
\bigskip We shall solve the differential equation (\ref{M28}) in full derivatives
first by means of the standard method without multipliers \cite{DiffEq}.
This means that the following differential equations are to be fulfilled 
\begin{equation}
\frac{\partial \widetilde{R}_{AB}^{2}(1\rightarrow 2)}{\partial F_{1}}=\frac{%
cD_{1}(\frac{\partial T_{1}}{\partial F_{1}})S_{1}(r_{1})}{\mid \frac{%
\partial \overrightarrow{\mathbf{r}}_{1}}{\partial F_{1}}\mid }\text{ , }%
\frac{\partial \widetilde{R}_{AB}^{2}(1\rightarrow 2)}{\partial F_{2}}=\text{%
\ }\frac{cD_{2}(\frac{\partial T_{2}}{\partial F_{2}})S_{2}(r_{2})}{\mid 
\frac{\partial \overrightarrow{\mathbf{r}}_{1}}{\partial F_{1}}\mid }\text{.}
\label{M30}
\end{equation}%
After integrating the first equation with respect to the dynamical parameter 
$F_{1}$ and then differentiating the obtained equation with respect to $%
F_{2} $, the following equation is obtained 
\begin{equation}
\frac{\partial \widetilde{R}_{AB}^{2}(1\rightarrow 2)}{\partial F_{2}}%
=c\int \frac{\frac{\partial D_{1}}{\partial F_{2}}(\frac{\partial T_{1}}{%
\partial F_{1}})S_{1}(r_{1})}{\mid \frac{\partial \overrightarrow{\mathbf{r}}%
_{1}}{\partial F_{1}}\mid }dF_{1}+\frac{\partial \Psi (F_{2})}{\partial F_{2}%
}\text{.\ \ }  \label{M31}
\end{equation}%
In the formulae for the integration with respect to $F_{1}$ (the first
formulae in (\ref{M30}), it has been taken into account that a new function $%
\Psi (F_{2})$ has to be added. It should be taken into consideration in the
derivation of the formulae (\ref{M31}) for $\frac{\partial \widetilde{R}%
_{AB}^{2}(1\rightarrow 2)}{\partial F_{2}}$. Setting up equal eq. (\ref{M31}%
) to the second equation in (\ref{M30}) and performing integration by parts,
the following final equation is obtained for the space-time distance $%
\widetilde{R}_{AB}^{2}$%
\begin{equation}
\widetilde{R}_{AB}^{2}(1\rightarrow 2)=\pm c\int \frac{D_{2}(\frac{\partial
T_{2}}{\partial F_{2}})S_{2}(r_{2})}{\mid \frac{\partial \overrightarrow{%
\mathbf{r}}_{1}}{\partial F_{1}}\mid }dF_{2}\text{.}  \label{M32}
\end{equation}%
Details about the calculation are given in Appendix B. Since the
differentials of the two space-time distances $d\widetilde{R}%
_{AB}^{2}(1\rightarrow 2)$ and $d\widetilde{R}_{BA}^{2}(2\rightarrow 1)$ in (%
\ref{M24}) were different due to the absence of any symmetry under the
replacement of indices $1$ and $2$ (equivalent to the displacement of space
points $A$ and $B$), there is no reason to expect that there will be such
symmetry in the formulae (\ref{M32}) for $\widetilde{R}_{AB}^{2}(1%
\rightarrow 2)$.
\subsection{Compatibility conditions for inter-satellite communications and
the two integral equations for determination of the propagation times $T_{1}$
and $T_{2}$}
\label{sec:CompatCondInteg}
\bigskip As previously mentioned, the two space-time distances $\widetilde{R}%
_{AB}^{2}(1\rightarrow 2)$ and $\widetilde{R}_{BA}^{2}(2$ $\rightarrow 1)$
are different. This means that the space-time distance $\widetilde{R}%
_{BA}^{2}(2\rightarrow 1)$ in analogy with formulae (\ref{M32}) should be
written as 
\begin{equation}
\widetilde{R}_{BA}^{2}(2\rightarrow 1)=\pm c\int \frac{D_{1}(\frac{\partial
T_{1}}{\partial F_{1}})S_{1}(r_{1})}{\mid \frac{\partial \overrightarrow{%
\mathbf{r}}_{2}}{\partial F_{2}}\mid }dF_{1}\text{.}  \label{M33}
\end{equation}%
However, according to the prescription in the preceding sections, the
compatibility condition for inter-satellite communications is determined by
assuming that the space-time distance is comparable to the Euclidean
distance, which is symmetric with respect to the exchange of space points $A$
and $B$. This means that both space-time distances are equivalent to the
Euclidean distance, i.e. 
\begin{equation}
\widetilde{R}_{AB}^{2}(1\rightarrow 2)=R^{2}\text{, }\widetilde{R}%
_{BA}^{2}(2\rightarrow 1)=R^{2}\text{.\ }  \label{M34}
\end{equation}%
Therefore, the space-time distance $\widetilde{R}_{.}^{2}$ can be defined as
the mean value of the space-time distances 
\begin{equation}
\widetilde{R}_{.}^{2}=\frac{1}{2}(\widetilde{R}_{AB}^{2}(1\rightarrow 2)+%
\widetilde{R}_{BA}^{2}(2\rightarrow 1))\text{.}  \label{M35}
\end{equation}%
Taking into account equalities (\ref{M32}) - (\ref{M35}), the condition for
inter-satellite communications can be written in the symmetric form 
\begin{equation}
\int \frac{D_{2}(\frac{\partial T_{2}}{\partial F_{2}})S_{2}(r_{2})}{\mid 
\frac{\partial \overrightarrow{\mathbf{r}}_{1}}{\partial F_{1}}\mid }%
dF_{2}=\int \frac{D_{1}(\frac{\partial T_{1}}{\partial F_{1}})S_{1}(r_{1})}{%
\mid \frac{\partial \overrightarrow{\mathbf{r}}_{2}}{\partial F_{2}}\mid }%
dF_{1}\text{.}  \label{M36}
\end{equation}%
After performing the integration by parts in (\ref{M32}) and taking into
account (\ref{M34}) and (\ref{M35}), the condition for inter-satellite
communications can be written in the form of an integral equation with
respect to the propagation time $T_{2}$ \cite{IntegEq}. Here only the plus
sign has been taken, but from (\ref{M32}) evidently the minus sign can also
be taken. \ 
\begin{equation}
\mp c\int T_{2}\frac{\partial }{\partial F_{2}}\left[ S_{2}(r_{2})D_{2}%
\right] dF_{2}=\mid \frac{\partial \overrightarrow{\mathbf{r}}_{1}}{\partial
F_{1}}\mid .R^{2}\mp cS_{2}(r_{2})D_{2}T_{2}\text{.}  \label{M37}
\end{equation}%
Similarly, from the integral equality (\ref{M33}) after integration by parts
another integral equation with respect to the propagation time $T_{1}$ can
be obtained 
\begin{equation}
\mp c\int T_{1}\frac{\partial }{\partial F_{1}}\left[ S_{1}(r_{1})D_{1}%
\right] dF_{1}=\mid \frac{\partial \overrightarrow{\mathbf{r}}_{2}}{\partial
F_{2}}\mid .R^{2}\mp cS_{1}(r_{1})D_{1}T_{1}\text{.}  \label{M38}
\end{equation}
\section{Defining the geodesic distance for the general case of two moving
satellites on arbitrary space trajectories}
\label{sec:DefGeodDistGeneral}
\bigskip According to the prescription in the preceding sections, the geodesic
distance is defined when the condition for inter-satellite communications is
substituted into the equation, defining the space-time distance (\ref{M35}) $%
\widetilde{R}_{.}^{2}=\frac{1}{2}(\widetilde{R}_{AB}^{2}(1\rightarrow 2)+%
\widetilde{R}_{BA}^{2}(2\rightarrow 1))$. However, when the space-time
distance is comparable to the Euclidean distance, i.e. $\widetilde{R}%
_{AB}^{2}(1\rightarrow 2)=R_{AB}^{2}$ and $\widetilde{R}_{BA}^{2}(2%
\rightarrow 1)=R_{BA}^{2}$ and equalities (\ref{MM9}) and (\ref{MM10}) are
taken into account, the geodesic distance (further it will be denoted by $%
\widehat{R}^{2}$) can be expressed as 
\begin{equation}
\widehat{R}_{AB}^{2}=\pm \frac{c}{2}\int S_{2}(r_{2})\frac{D_{2}}{\mid 
\frac{\partial \overrightarrow{\mathbf{r}}_{1}}{\partial F_{1}}\mid }\frac{%
\partial T_{2}}{\partial F_{2}}dF_{2}\mp \frac{c}{2}\int S_{1}(r_{1})\frac{%
D_{1}}{\mid \frac{\partial \overrightarrow{\mathbf{r}}_{2}}{\partial F_{2}}%
\mid }\frac{\partial T_{1}}{\partial F_{1}}dF_{1}\text{.}  \label{M39}
\end{equation}%
In view of the equality (\ref{M37}) and also equality from (\ref{MM9}) from
Appendix B 
\begin{equation*}
\widetilde{R}_{AB}^{2}(1\rightarrow 2)=\pm c\int S_{2}(r_{2})\frac{D_{2}}{%
\mid \frac{\partial \overrightarrow{\mathbf{r}}_{1}}{\partial F_{1}}\mid }%
\frac{\partial T_{2}}{\partial F_{2}}dF_{2}\text{\ \ }
\end{equation*}%
and taking the positive sign in (\ref{M39}), the above equality for the
geodesic distance can be rewritten as 
\begin{equation}
\widehat{R}_{AB.}^{2}=R_{AB.}^{2}+\frac{c}{2}\int S_{1}(r_{1})\frac{D_{1}}{%
\mid \frac{\partial \overrightarrow{\mathbf{r}}_{2}}{\partial F_{2}}\mid }%
\frac{\partial T_{1}}{\partial F_{1}}dF_{1}\text{.}  \label{M40}
\end{equation}%
The sign in front of the second term is plus, because of the equality (\ref%
{M36}). If the right-hand side of (\ref{M36}) is used, then the geodesic
distance can be written also as 
\begin{equation}
\widehat{R}_{AB}^{2}=R_{AB}^{2}+\frac{c}{2}\int S_{2}(r_{2})\frac{D_{2}}{%
\mid \frac{\partial \overrightarrow{\mathbf{r}}_{1}}{\partial F_{1}}\mid }%
\frac{\partial T_{2}}{\partial F_{2}}dF_{2}\text{.}  \label{M41}
\end{equation}%
If the second terms in the integral parts in (\ref{M40}) and (\ref{M41}) are
positive, then evidently the geodesic distance $\widehat{R}^{2}$ will be
greater than the Euclidean distance $R$. However, the minus sign in (\ref%
{M39}) can also be taken, which is a consequence also of the equalities (\ref%
{M28}) and (\ref{M32}). Then it may happen that the geodesic distance $%
\widehat{R}^{2}$ can be written either as 
\begin{equation}
\widehat{R}^{2}=\frac{c}{2}\int \frac{D_{1}S_{1}(r_{1})}{\mid \frac{%
\partial \overrightarrow{\mathbf{r}}_{2}}{\partial F_{2}}\mid }\frac{%
\partial T_{1}}{\partial F_{1}}dF_{1}\text{ }-\frac{c}{2}\int \frac{%
D_{2}S_{2}(r_{2})}{\mid \frac{\partial \overrightarrow{\mathbf{r}}_{1}}{%
\partial F_{1}}\mid }\frac{\partial T_{2}}{\partial F_{2}}dF_{2}\text{ }
\label{M42}
\end{equation}%
or in the form 
\begin{equation}
\widehat{R}^{2}=\frac{c}{2}\int \frac{D_{2}S_{2}(r_{1})}{\mid \frac{%
\partial \overrightarrow{\mathbf{r}}_{1}}{\partial F_{1}}\mid }\frac{%
\partial T_{2}}{\partial F_{2}}dF_{2}\text{ }-\frac{c}{2}\int \frac{%
D_{1}S_{1}(r_{1})}{\mid \frac{\partial \overrightarrow{\mathbf{r}}_{2}}{%
\partial F_{2}}\mid }\frac{\partial T_{1}}{\partial F_{1}}dF_{1}\text{. }
\label{M43}
\end{equation}%
Thus, depending on the motion of the two satellites, one of the cases will
be realized and the corresponding positive geodesic distance will be ensured.
\subsection{Geodesic distance for the case of constant space-time distances $%
\protect\widetilde{R}_{AB}^{2}(1\rightarrow 2)$ and}
\subsection{$\protect\widetilde{R}_{BA}^{2}(2$ $\rightarrow 1)$}
\label{sec:GeodConstSpTime}
\bigskip The aim of this section will be to prove that the assumption about constant
space-time distances $\widetilde{R}_{AB}^{2}(1\rightarrow 2)$ and 
$\widetilde{R}_{BA}^{2}(2$ $\rightarrow 1)$ will lead to an impossibility to
determine the geodesic distance because it will tend to infinity (since the
two radius vectors $\overrightarrow{\mathbf{r}}_{1}$ and $\overrightarrow{%
\mathbf{r}}_{2}$ will coincide) and thus, no inter-satellite communications
will be possible.
\bigskip For the purpose, the second term in the geodesic equation (\ref{M40}) will
be transformed by means of the first relation in the equations (\ref{M10})%
\begin{equation*}
dF_{1}=\frac{c(dT_{1})}{\mid \frac{\partial \overrightarrow{\mathbf{r}}_{1}}{%
\partial F_{1}}\mid }S_{1}(r_{1})\text{,\ }dF_{2}=\frac{c(dT_{2})}{\mid 
\frac{\partial \overrightarrow{\mathbf{r}}_{2}}{\partial F_{2}}\mid }%
S_{2}(r_{2})\text{.}
\end{equation*}%
Thus the term will be written as 
\begin{equation}
\frac{c}{2}\int S_{1}\frac{D_{1}}{\mid \frac{\partial \overrightarrow{%
\mathbf{r}}_{2}}{\partial F_{2}}\mid }\frac{\partial T_{1}}{\partial F_{1}}%
dF_{1}=\frac{c}{2}\int S_{1}\frac{\partial T_{1}}{\partial F_{1}}\left[ 
\frac{S_{1}D_{1}}{\mid \frac{\partial \overrightarrow{\mathbf{r}}_{2}}{%
\partial F_{2}}\mid }\frac{\partial T_{1}}{\partial F_{1}}dF_{1}\right]
\label{M44}
\end{equation}%
\begin{equation}
=\frac{c}{2}S_{1}\frac{\partial T_{1}}{\partial F_{1}}\int \frac{S_{2}D_{2}%
}{\mid \frac{\partial \overrightarrow{\mathbf{r}}_{2}}{\partial F_{2}}\mid }%
\frac{\partial T_{2}}{\partial F_{2}}dF_{2}\text{ ,}  \label{M45}
\end{equation}%
where the condition (\ref{M26}) \ 
\begin{equation*}
D_{1}(dT_{1})S_{1}(r_{1})=\mp D_{2}(dT_{2})S_{2}(r_{2})
\end{equation*}%
for the equality to zero of the differentials $d\widetilde{R}%
_{AB}^{2}(1\rightarrow 2)$ and $d\widetilde{R}_{BA}^{2}(2$ $\rightarrow 1)$
has been applied.
\bigskip Thus, in an analogous way and making use of the second relation in (\ref{M10}%
), the second term in the second representation (\ref{M41}) of the geodesic
equation can be transformed as 
\begin{equation}
\frac{c}{2}\int S_{2}(r_{1})\frac{D_{2}}{\mid \frac{\partial 
\overrightarrow{\mathbf{r}}_{1}}{\partial F_{1}}\mid }\frac{\partial T_{2}}{%
\partial F_{2}}dF_{2}=\frac{c}{2}\int S_{2}\frac{\partial T_{2}}{\partial
F_{2}}\left[ \frac{S_{2}D_{2}}{\mid \frac{\partial \overrightarrow{\mathbf{r}%
}_{1}}{\partial F_{1}}\mid }\frac{\partial T_{2}}{\partial F_{2}}dF_{2}%
\right]  \label{M46}
\end{equation}%
\begin{equation}
=\frac{c}{2}S_{2}\frac{\partial T_{2}}{\partial F_{2}}\int \frac{S_{1}D_{1}%
}{\mid \frac{\partial \overrightarrow{\mathbf{r}}_{1}}{\partial F_{1}}\mid }%
\frac{\partial T_{1}}{\partial F_{1}}dF_{1}\text{.}  \label{M47}
\end{equation}%
But from the equality of the second terms in the representations (\ref{M40})
and (\ref{M41}) of the geodesic equation another equality can also be
obtained 
\begin{equation}
\frac{1}{\mid \frac{\partial \overrightarrow{\mathbf{r}}_{2}}{\partial F_{2}}%
\mid }\int S_{1}D_{1}\frac{\partial T_{1}}{\partial F_{1}}dF_{1}=\frac{1}{%
\mid \frac{\partial \overrightarrow{\mathbf{r}}_{1}}{\partial F_{1}}\mid }%
\int S_{2}D_{2}\frac{\partial T_{2}}{\partial F_{2}}dF_{2}\text{. \ }
\label{M48}
\end{equation}%
Let us remind again that in accord with the initial assumptions, the
generalized velocity $\mid \frac{\partial \overrightarrow{\mathbf{r}}_{2}}{%
\partial F_{2}}\mid $ does not depend on the generalized parameter $F_{1}$
and the generalized velocity $\mid \frac{\partial \overrightarrow{\mathbf{r}}%
_{1}}{\partial F_{1}}\mid $ does not depend on the parameter $F_{2}$.
Consequently, the two generalized velocities remain outside the left-hand
side integral in (\ref{M48}) and the right-hand side of (\ref{M48})
respectively. Again, taking into account the integral relation (we take the
positive sign), obtained from (\ref{M26}), it can be derived 
\begin{equation}
\int S_{1}D_{1}\frac{\partial T_{1}}{\partial F_{1}}dF_{1}=\int S_{2}D_{2}%
\frac{\partial T_{2}}{\partial F_{2}}dF_{2}\text{. \ }  \label{M49}
\end{equation}%
From the last two equalities (\ref{M48}) and (\ref{M49}) it follows that 
\begin{equation}
\mid \frac{\partial \overrightarrow{\mathbf{r}}_{1}}{\partial F_{1}}\mid
=\mid \frac{\partial \overrightarrow{\mathbf{r}}_{2}}{\partial F_{2}}\mid 
\text{. }  \label{M50}
\end{equation}%
Another equality can be obtained after comparing equalities (\ref{M45}) and (%
\ref{M47}) 
\begin{equation}
S_{1}\frac{\partial T_{1}}{\partial F_{1}}=S_{2}\frac{\partial T_{2}}{%
\partial F_{2}}\text{.}  \label{M51}
\end{equation}%
Taking into account again (\ref{M26}), it follows that 
\begin{equation}
D_{1}dF_{1}=D_{2}dF_{2}\text{, }  \label{M52}
\end{equation}%
where the defining relations (\ref{M7A}) for $D_{1}$ and (\ref{M7B}) for $%
D_{2}$ were previously defined as 
\begin{equation*}
D_{1}(F_{1},F_{2},\overrightarrow{\mathbf{r}}_{1},\overrightarrow{\mathbf{r}}%
_{2}):=\overrightarrow{\mathbf{r}}_{1}.\frac{\partial \overrightarrow{%
\mathbf{r}}_{1}}{\partial F_{1}}-\overrightarrow{\mathbf{r}}_{2}.\frac{%
\partial \overrightarrow{\mathbf{r}}_{1}}{\partial F_{1}}\text{, }
\end{equation*}%
\begin{equation*}
D_{2}(F_{1},F_{2},\overrightarrow{\mathbf{r}}_{1},\overrightarrow{\mathbf{r}}%
_{2}):=\overrightarrow{\mathbf{r}}_{2}.\frac{\partial \overrightarrow{%
\mathbf{r}}_{2}}{\partial F_{2}}-\overrightarrow{\mathbf{r}}_{1}.\frac{%
\partial \overrightarrow{\mathbf{r}}_{2}}{\partial F_{2}}\text{.}
\end{equation*}%
By means of the above equalities, formulae (\ref{M52}) can be represented as 
\begin{equation*}
\frac{\partial }{\partial F_{1}}\left[ \frac{1}{2}\left( \overrightarrow{%
\mathbf{r}}_{1}\right) ^{2}-\left( \overrightarrow{\mathbf{r}}_{1}.%
\overrightarrow{\mathbf{r}}_{2}\right) \right] dF_{1}
\end{equation*}%
\begin{equation}
=\frac{\partial }{\partial F_{2}}\left[ \frac{1}{2}\left( \overrightarrow{%
\mathbf{r}}_{2}\right) ^{2}-\left( \overrightarrow{\mathbf{r}}_{1}.%
\overrightarrow{\mathbf{r}}_{2}\right) \right] dF_{2}\text{ ,}  \label{M53}
\end{equation}%
from where $\left( \overrightarrow{\mathbf{r}}_{1}\right) ^{2}=\left( 
\overrightarrow{\mathbf{r}}_{2}\right) ^{2}$. Now if expression (\ref{M52})
is substituted in the geodesic distance formulae (\ref{M40}) 
and also the simple formulae for the Euclidean distance 
\begin{equation}
R_{AB}^{2}=\left( \overrightarrow{\mathbf{r}}_{1}\right) ^{2}+\left( 
\overrightarrow{\mathbf{r}}_{2}\right) ^{2}-2\left( \overrightarrow{\mathbf{r%
}}_{1}.\overrightarrow{\mathbf{r}}_{2}\right)  \label{M54}
\end{equation}%
is taken into account, then the geodesic distance can be expressed as 
\begin{equation*}
\widehat{R}_{AB}^{2}=\left( \overrightarrow{\mathbf{r}}_{1}\right)
^{2}+\left( \overrightarrow{\mathbf{r}}_{2}\right) ^{2}-2\left( 
\overrightarrow{\mathbf{r}}_{1}.\overrightarrow{\mathbf{r}}_{2}\right)
\end{equation*}%
\begin{equation}
+\frac{c}{2}\int S_{1}\frac{\partial T_{1}}{\partial F_{1}}\frac{\partial }{%
\partial F_{1}}\left[ \frac{1}{2}\left( \overrightarrow{\mathbf{r}}%
_{1}\right) ^{2}-\left( \overrightarrow{\mathbf{r}}_{1}.\overrightarrow{%
\mathbf{r}}_{2}\right) \right] dF_{1}\text{.}  \label{M55}
\end{equation}%
Clearly, when $\overrightarrow{\mathbf{r}}_{1}=\overrightarrow{\mathbf{r}}%
_{2}$, both the terms on the first and on the second line in (\ref{M55})
equal to zero. Consequently, the assumption for a constant space-time
distance (although defined uniquely in both directions, i.e. $\widetilde{R}%
_{AB}^{2}(1\rightarrow 2)=$ $\widetilde{R}_{BA}^{2}(2$ $\rightarrow 1)$) is
equivalent to coinciding points of disposition of the satellites and to zero
geodesic distance. Thus, there is impossibility for any inter-satellite
communication.
\section{Variable space-time distance for the case of equality of the
space-time distance in both directions}
\label{sec:VarSpTimeDist}
\subsection{The case of equal differentials of the two space-time distances $%
d\protect\widetilde{R}_{AB}^{2}(1\rightarrow 2)$ and $d\protect\widetilde{R}%
_{BA}^{2}(2$ $\rightarrow 1)$}
\label{sec:VarSpTimeDistEqDiffer}
\bigskip Now the space-time distance will be defined by assuming the equality of the
differential $d\widetilde{R}_{AB}^{2}(1\rightarrow 2)$ to the differential $d%
\widetilde{R}_{BA}^{2}(2$ $\rightarrow 1)$. The two differentials are given
respectively by the formulae (\ref{M21}) 
\begin{equation*}
d\widetilde{R}_{AB}^{2}(1\rightarrow 2)=\frac{cD_{1}(dT_{1})S_{1}(r_{1})}{%
\mid \frac{\partial \overrightarrow{\mathbf{r}}_{1}}{\partial F_{1}}\mid }%
\pm \frac{cD_{2}(dT_{2})S_{2}(r_{2})}{\mid \frac{\partial \overrightarrow{%
\mathbf{r}}_{1}}{\partial F_{1}}\mid })
\end{equation*}%
and formulae (\ref{M25}) 
\begin{equation*}
d\widetilde{R}_{BA}^{2}(2\rightarrow 1)=\frac{cD_{2}(dT_{2})S_{2}(r_{2})}{%
\mid \frac{\partial \overrightarrow{\mathbf{r}}_{2}}{\partial F_{2}}\mid }%
\pm \frac{cD_{1}(dT_{1})S_{1}(r_{1})}{\mid \frac{\partial \overrightarrow{%
\mathbf{r}}_{2}}{\partial F_{2}}\mid })\neq d\widetilde{R}%
_{AB}^{2}(1\rightarrow 2)\text{.}
\end{equation*}%
From the equality of the two space-time differentials (\ref{M21}) and (\ref%
{M25}), the following equality can be obtained 
\begin{equation*}
\mid \frac{\partial \overrightarrow{\mathbf{r}}_{1}}{\partial F_{1}}\mid .%
\left[ cD_{2}(dT_{2})S_{2}\pm D_{1}(dT_{1})S_{1}\right]
\end{equation*}%
\begin{equation}
=\mid \frac{\partial \overrightarrow{\mathbf{r}}_{2}}{\partial F_{2}}\mid %
\left[ -cD_{1}(dT_{1})S_{1}\pm D_{2}(dT_{2})S_{2}\right] \text{.}  \label{N1}
\end{equation}%
Rearranging the terms, the a simple relation can be obtained 
\begin{equation}
D_{2}(dT_{2})S_{2}=k_{1,2}(\mid \frac{\partial \overrightarrow{\mathbf{r}}%
_{1}}{\partial F_{1}}\mid ,\mid \frac{\partial \overrightarrow{\mathbf{r}}%
_{2}}{\partial F_{2}}\mid )D_{1}(dT_{1})S_{1}\text{,}  \label{N2}
\end{equation}%
where $k_{1,2}(\mid \frac{\partial \overrightarrow{\mathbf{r}}_{1}}{\partial
F_{1}}\mid ,\mid \frac{\partial \overrightarrow{\mathbf{r}}_{2}}{\partial
F_{2}}\mid )$ is defined as the ratio 
\begin{equation}
k_{1,2}(\mid \frac{\partial \overrightarrow{\mathbf{r}}_{1}}{\partial F_{1}}%
\mid ,\mid \frac{\partial \overrightarrow{\mathbf{r}}_{2}}{\partial F_{2}}%
\mid ):=\frac{\mid \frac{\partial \overrightarrow{\mathbf{r}}_{2}}{\partial
F_{2}}\mid \mp \mid \frac{\partial \overrightarrow{\mathbf{r}}_{1}}{\partial
F_{1}}\mid }{\mid \frac{\partial \overrightarrow{\mathbf{r}}_{1}}{\partial
F_{1}}\mid \pm \mid \frac{\partial \overrightarrow{\mathbf{r}}_{2}}{\partial
F_{2}}\mid }\text{.}  \label{N3}
\end{equation}%
Further $k_{1}=k$ will denote the above ratio with the lower signs in the
nominator and in the denominator (respectively $+$ and $-$) and $k_{2}$ -
the ratio with the upper signs $-$ in the nominator and $+$ in the
denominator. Thus 
\begin{equation}
k_{1}=k=\frac{\mid \frac{\partial \overrightarrow{\mathbf{r}}_{2}}{\partial
F_{2}}\mid +\mid \frac{\partial \overrightarrow{\mathbf{r}}_{1}}{\partial
F_{1}}\mid }{\mid \frac{\partial \overrightarrow{\mathbf{r}}_{1}}{\partial
F_{1}}\mid -\mid \frac{\partial \overrightarrow{\mathbf{r}}_{2}}{\partial
F_{2}}\mid }\text{ , }k_{2}=\frac{\mid \frac{\partial \overrightarrow{%
\mathbf{r}}_{2}}{\partial F_{2}}\mid -\mid \frac{\partial \overrightarrow{%
\mathbf{r}}_{1}}{\partial F_{1}}\mid }{\mid \frac{\partial \overrightarrow{%
\mathbf{r}}_{1}}{\partial F_{1}}\mid +\mid \frac{\partial \overrightarrow{%
\mathbf{r}}_{2}}{\partial F_{2}}\mid }=-\frac{1}{k_{1}}\text{ }=-\frac{1}{k}%
\text{.}  \label{N4}
\end{equation}%
Evidently, relation (\ref{N2}) is the analogue of relation (\ref{M26}) $%
D_{1}(dT_{1})S_{1}=\mp D_{2}(dT_{2})S_{2}$, which was derived for the case
constant space-time distances $\widetilde{R}_{AB}^{2}(1\rightarrow 2)$ and $d%
\widetilde{R}_{BA}^{2}(2$ $\rightarrow 1)$. For the second case with upper
indices, the second relation instead of (\ref{N2}) is 
\begin{equation}
D_{2}(dT_{2})S_{2}=-\frac{1}{k_{.}}D_{1}(dT_{1})S_{1}\text{.}  \label{N5}
\end{equation}
\subsection{Modified expression for the geodesic distance and cases, when it
is positive}
\label{sec:ModifGeodDist}
\bigskip By means of expressions (\ref{N2}) and (\ref{N5}), formulaes (\ref{M42}) and
(\ref{M43}) for the geodesic distance can be written in the following way 
\begin{equation}
\widehat{R}_{AB}^{2}=\frac{c}{2\mid \frac{\partial \overrightarrow{\mathbf{r}%
}_{2}}{\partial F_{2}}\mid }\int \frac{D_{1}S_{1}\frac{\partial T_{1}}{%
\partial F_{1}}\left[ \mid \frac{\partial \overrightarrow{\mathbf{r}}_{1}}{%
\partial F_{1}}\mid -k\mid \frac{\partial \overrightarrow{\mathbf{r}}_{2}}{%
\partial F_{2}}\mid \right] }{\mid \frac{\partial \overrightarrow{\mathbf{r}}%
_{1}}{\partial F_{1}}\mid }dF_{1}\text{.}  \label{N6}
\end{equation}%
It can easily be checked that the formulae for (\ref{M43}) is the same as
the one for (\ref{M42}), but with a minus sign for the expression in the
square brackets, i.e. 
\begin{equation}
\widehat{R}_{AB}^{2}=\frac{c}{2\mid \frac{\partial \overrightarrow{\mathbf{r}%
}_{2}}{\partial F_{2}}\mid }\int \frac{D_{1}S_{1}\frac{\partial T_{1}}{%
\partial F_{1}}\left[ k\mid \frac{\partial \overrightarrow{\mathbf{r}}_{2}}{%
\partial F_{2}}\mid -\mid \frac{\partial \overrightarrow{\mathbf{r}}_{1}}{%
\partial F_{1}}\mid \right] }{\mid \frac{\partial \overrightarrow{\mathbf{r}}%
_{1}}{\partial F_{1}}\mid }dF_{1}\text{.}  \label{N7}
\end{equation}%
Now the following question naturally arises: in which cases formulaes (\ref%
{N6}) and (\ref{N7}) are positive? Note also that this expression for the
geodesic distance does not depend on the second propagation time $T_{2}$,
but only on the generalized velocities $\mid \frac{\partial \overrightarrow{%
\mathbf{r}}_{1}}{\partial F_{1}}\mid $ and $\mid \frac{\partial 
\overrightarrow{\mathbf{r}}_{2}}{\partial F_{2}}\mid $. In this sense, the
only indication that this is a geodesic distance between a space point $1$
on the trajectory of the first satellite to the space point $2$ on the
trajectory of the second satellite is the dependence on the second
generalized velocity $\mid \frac{\partial \overrightarrow{\mathbf{r}}_{2}}{%
\partial F_{2}}\mid $.
\subsection{ Positivity of the first expression (\protect\ref{M42}) for the
geodesic distance}
\label{sec:PositFirstGeodDist}
\bigskip Let us remind that previously expression (\ref{M42}) has been defined as 
\begin{equation*}
\widehat{R}^{2}=\frac{c}{2}\int \frac{D_{1}S_{1}(r_{1})}{\mid \frac{%
\partial \overrightarrow{\mathbf{r}}_{2}}{\partial F_{2}}\mid }\frac{%
\partial T_{1}}{\partial F_{1}}dF_{1}\text{ }-\frac{c}{2}\int \frac{%
D_{2}S_{2}(r_{2})}{\mid \frac{\partial \overrightarrow{\mathbf{r}}_{1}}{%
\partial F_{1}}\mid }\frac{\partial T_{2}}{\partial F_{2}}dF_{2}\text{ }
\end{equation*}
\bigskip Let us write the expression $\mid \frac{\partial \overrightarrow{\mathbf{r}}%
_{1}}{\partial F_{1}}\mid -k\mid \frac{\partial \overrightarrow{\mathbf{r}}%
_{2}}{\partial F_{2}}\mid $ in the nominator of (\ref{N6}) as 
\begin{equation*}
\mid \frac{\partial \overrightarrow{\mathbf{r}}_{1}}{\partial F_{1}}\mid -%
\text{ }k\mid \frac{\partial \overrightarrow{\mathbf{r}}_{2}}{\partial F_{2}}%
\mid =\left( \mid \frac{\partial \overrightarrow{\mathbf{r}}_{1}}{\partial
F_{1}}\mid +\mid \frac{\partial \overrightarrow{\mathbf{r}}_{2}}{\partial
F_{2}}\mid \right)
\end{equation*}%
\begin{equation}
+\frac{2\mid \frac{\partial \overrightarrow{\mathbf{r}}_{1}}{\partial F_{1}}%
\mid \mid \frac{\partial \overrightarrow{\mathbf{r}}_{2}}{\partial F_{2}}%
\mid }{\mid \frac{\partial \overrightarrow{\mathbf{r}}_{2}}{\partial F_{2}}%
\mid -\mid \frac{\partial \overrightarrow{\mathbf{r}}_{1}}{\partial F_{1}}%
\mid }\text{ .}  \label{N8}
\end{equation}%
Clearly, for $\mid \frac{\partial \overrightarrow{\mathbf{r}}_{2}}{\partial
F_{2}}\mid >\mid \frac{\partial \overrightarrow{\mathbf{r}}_{1}}{\partial
F_{1}}\mid $ expression (\ref{N8}) and also the geodesic distance (\ref{N6})
are positive when the generalized velocity $\mid \frac{\partial 
\overrightarrow{\mathbf{r}}_{2}}{\partial F_{2}}\mid $ of the second
satellite-receiving signal is greater than the generalized velocity $\mid 
\frac{\partial \overrightarrow{\mathbf{r}}_{1}}{\partial F_{1}}\mid $ of the
first, signal-emitting satellite. Note also that when $\mid \frac{\partial 
\overrightarrow{\mathbf{r}}_{2}}{\partial F_{2}}\mid =\mid \frac{\partial 
\overrightarrow{\mathbf{r}}_{1}}{\partial F_{1}}\mid $ expressions (\ref{N8}%
) and (\ref{N6}) tend to infinity and thus the geodesic distance is not
properly determined.
\bigskip It is possible also that for certain relations the expression (\ref{N6}) and
(\ref{N8}) will be positive also for $\mid \frac{\partial \overrightarrow{%
\mathbf{r}}_{2}}{\partial F_{2}}\mid <\mid \frac{\partial \overrightarrow{%
\mathbf{r}}_{1}}{\partial F_{1}}\mid $. However, further the guiding
principle will be that the positivity of the geodesic distance should be
ensured for all possible cases. That is why, we shall not investigate when
expression (\ref{N8}) will be positive for $\mid \frac{\partial 
\overrightarrow{\mathbf{r}}_{2}}{\partial F_{2}}\mid <\mid \frac{\partial 
\overrightarrow{\mathbf{r}}_{1}}{\partial F_{1}}\mid $ . For such a case,
the first term in (\ref{N8}) will be again positive, but the second term will
be negative due to the negative denominator.
\bigskip Instead, we shall prove that for the second possible inequality $\mid \frac{%
\partial \overrightarrow{\mathbf{r}}_{2}}{\partial F_{2}}\mid <\mid \frac{%
\partial \overrightarrow{\mathbf{r}}_{1}}{\partial F_{1}}\mid $ it will be
more appropriate to choose the second formulae (\ref{N7}) as adequate for
expressing the geodesic distance.
\subsection{Positivity of the second expression (\protect\ref{M43}) for the
geodesic distance}
\label{sec:PositSecondGeodDist}
\bigskip For investigating the second geodesic equation (\ref{N7}) for the case $\mid 
\frac{\partial \overrightarrow{\mathbf{r}}_{2}}{\partial F_{2}}\mid <\mid 
\frac{\partial \overrightarrow{\mathbf{r}}_{1}}{\partial F_{1}}\mid $, let
us represent the expression in the square brackets in the nominator as 
\begin{equation}
k\mid \frac{\partial \overrightarrow{\mathbf{r}}_{2}}{\partial F_{2}}\mid
-\mid \frac{\partial \overrightarrow{\mathbf{r}}_{1}}{\partial F_{1}}\mid =%
\frac{Z_{1}}{Z_{2}}\text{, }  \label{N9}
\end{equation}%
where $Z_{1}$ and $Z_{2}$ denote the expressions 
\begin{equation}
Z_{1}:=2\mid \frac{\partial \overrightarrow{\mathbf{r}}_{1}}{\partial F_{1}}%
\mid .\mid \frac{\partial \overrightarrow{\mathbf{r}}_{2}}{\partial F_{2}}%
\mid -\left[ \mid \frac{\partial \overrightarrow{\mathbf{r}}_{1}}{\partial
F_{1}}\mid ^{2}-\mid \frac{\partial \overrightarrow{\mathbf{r}}_{2}}{%
\partial F_{2}}\mid ^{2}\right] \text{ ,}  \label{N10}
\end{equation}%
\begin{equation}
Z_{2}:=\mid \frac{\partial \overrightarrow{\mathbf{r}}_{1}}{\partial F_{1}}%
\mid ^{.}-\mid \frac{\partial \overrightarrow{\mathbf{r}}_{2}}{\partial F_{2}%
}\mid \text{.}^{.}  \label{N11}
\end{equation}%
Introducing the notations 
\begin{equation}
v_{1}=\mid \frac{\partial \overrightarrow{\mathbf{r}}_{1}}{\partial F_{1}}%
\mid \text{, }v_{2}=\mid \frac{\partial \overrightarrow{\mathbf{r}}_{2}}{%
\partial F_{2}}\mid \text{ }  \label{N12}
\end{equation}%
and noting that the denominator $Z_{2}$ is positive for the investigated
case, the positivity of the expression (\ref{N9}) is equivalent to the
positivity of the nominator for $Z_{1}$, written as 
\begin{equation}
2v_{1}v_{2}>v_{1}^{2}-v_{2}^{2}\text{.}  \label{N13}
\end{equation}%
This inequality can be decomposed as 
\begin{equation}
\left[ v_{1}-\left( 1+\sqrt[.]{2}\right) v_{2}\right] .\left[ v_{1}-\left( 1-%
\sqrt[.]{2}\right) v_{2}\right] <0  \label{N14}
\end{equation}%
and it is easily verified that it is fulfilled for 
\begin{equation}
\left( 1-\sqrt[.]{2}\right) v_{2}<v_{1}<\left( 1+\sqrt[.]{2}\right) v_{2}%
\text{ .}  \label{N15}
\end{equation}%
Consequently, it has to be proved that for all generalized velocities $\mid 
\frac{\partial \overrightarrow{\mathbf{r}}_{1}}{\partial F_{1}}\mid $ and $%
\mid \frac{\partial \overrightarrow{\mathbf{r}}_{2}}{\partial F_{2}}\mid $ 
\begin{equation}
\left( 1-\sqrt[.]{2}\right) \mid \frac{\partial \overrightarrow{\mathbf{r}}%
_{2}}{\partial F_{2}}\mid <\mid \frac{\partial \overrightarrow{\mathbf{r}}%
_{1}}{\partial F_{1}}\mid <\left( 1+\sqrt[.]{2}\right) \mid \frac{\partial 
\overrightarrow{\mathbf{r}}_{2}}{\partial F_{2}}\mid \text{.}  \label{N16}
\end{equation}%
Now noting that $1-\sqrt[.]{2}$ is a negative number, the initial inequality 
$\mid \frac{\partial \overrightarrow{\mathbf{r}}_{2}}{\partial F_{2}}\mid
<\mid \frac{\partial \overrightarrow{\mathbf{r}}_{1}}{\partial F_{1}}\mid $
after multiplying by $1-\sqrt[.]{2}$ gives the following chain of
inequalities 
\begin{equation*}
\left( 1-\sqrt[.]{2}\right) \mid \frac{\partial \overrightarrow{\mathbf{r}}%
_{1}}{\partial F_{1}}\mid <\left( 1-\sqrt[.]{2}\right) \mid \frac{\partial 
\overrightarrow{\mathbf{r}}_{2}}{\partial F_{2}}\mid
\end{equation*}%
\begin{equation}
<\left( 1+\sqrt[.]{2}\right) \mid \frac{\partial \overrightarrow{\mathbf{r}}%
_{2}}{\partial F_{2}}\mid <\left( 1+\sqrt[.]{2}\right) \mid \frac{\partial 
\overrightarrow{\mathbf{r}}_{1}}{\partial F_{1}}\mid \text{.}  \label{N17}
\end{equation}%
Now we take the derived inequality from (\ref{N17})%
\begin{equation}
\left( 1-\sqrt[.]{2}\right) \mid \frac{\partial \overrightarrow{\mathbf{r}}%
_{2}}{\partial F_{2}}\mid <\left( 1+\sqrt[.]{2}\right) \mid \frac{\partial 
\overrightarrow{\mathbf{r}}_{1}}{\partial F_{1}}\mid  \label{N18}
\end{equation}%
together with the other inequality, which can be derived also from (\ref{N17}%
) 
\begin{equation}
\left( 1-\sqrt[.]{2}\right) \mid \frac{\partial \overrightarrow{\mathbf{r}}%
_{1}}{\partial F_{1}}\mid <\left( 1+\sqrt[.]{2}\right) \mid \frac{\partial 
\overrightarrow{\mathbf{r}}_{2}}{\partial F_{2}}\mid \text{.}  \label{N19}
\end{equation}%
Since $\left( 1-\sqrt[.]{2}\right) $ is a negative number, this proves
immediately the left inequality in (\ref{N16}) 
\begin{equation}
\left( 1-\sqrt[.]{2}\right) \mid \frac{\partial \overrightarrow{\mathbf{r}}%
_{2}}{\partial F_{2}}\mid <\mid \frac{\partial \overrightarrow{\mathbf{r}}%
_{1}}{\partial F_{1}}\mid \text{ .}  \label{N20}
\end{equation}%
The second inequality in (\ref{N16}) \ $\mid \frac{\partial \overrightarrow{%
\mathbf{r}}_{1}}{\partial F_{1}}\mid <\left( 1+\sqrt[.]{2}\right) \mid \frac{%
\partial \overrightarrow{\mathbf{r}}_{2}}{\partial F_{2}}\mid $ is also
fulfilled because we are investigating the case $\mid \frac{\partial 
\overrightarrow{\mathbf{r}}_{2}}{\partial F_{2}}\mid <\mid \frac{\partial 
\overrightarrow{\mathbf{r}}_{1}}{\partial F_{1}}\mid $, consequently always $%
\mid \frac{\partial \overrightarrow{\mathbf{r}}_{2}}{\partial F_{2}}\mid
<\left( 1+\sqrt[.]{2}\right) \mid \frac{\partial \overrightarrow{\mathbf{r}}%
_{2}}{\partial F_{2}}\mid $ will be valid.
\subsection{Positivity of the geodesic distance for the case with the other
coefficient function}
\label{sec:PositGeodDistSecCoefFunct}
\bigskip Now formulaes (\ref{N6}) and (\ref{N7}) will be investigated for the case
when instead of the coefficient function $k=k_{1}$ in (\ref{N4}), the
coefficient function $k_{2}=-\frac{1}{k}$ shall be used.
\bigskip Formulae (\ref{N6}) for the geodesic distance is transformed as 
\begin{equation}
\widehat{R}_{AB}^{2}=\frac{c}{2\mid \frac{\partial \overrightarrow{\mathbf{r}%
}_{2}}{\partial F_{2}}\mid }\int \frac{D_{1}S_{1}\frac{\partial T_{1}}{%
\partial F_{1}}\left[ k\mid \frac{\partial \overrightarrow{\mathbf{r}}_{1}}{%
\partial F_{1}}\mid +\mid \frac{\partial \overrightarrow{\mathbf{r}}_{2}}{%
\partial F_{2}}\mid \right] }{k\mid \frac{\partial \overrightarrow{\mathbf{r}%
}_{1}}{\partial F_{1}}\mid }dF_{1}\text{.}  \label{N21}
\end{equation}%
Respectively, the second formulae (\ref{N7}) is transformed into the same
expression as (\ref{N21}), but with a minus sign in front 
\begin{equation}
\widehat{R}_{AB}^{2}=-\frac{c}{2\mid \frac{\partial \overrightarrow{\mathbf{r%
}}_{2}}{\partial F_{2}}\mid }\int \frac{D_{1}S_{1}\frac{\partial T_{1}}{%
\partial F_{1}}\left[ k\mid \frac{\partial \overrightarrow{\mathbf{r}}_{1}}{%
\partial F_{1}}\mid +\mid \frac{\partial \overrightarrow{\mathbf{r}}_{2}}{%
\partial F_{2}}\mid \right] }{k\mid \frac{\partial \overrightarrow{\mathbf{r}%
}_{1}}{\partial F_{1}}\mid }dF_{1}\text{.}  \label{N22}
\end{equation}%
The positivity of expression (\ref{N21}) (respectively then the other
geodesic distance (\ref{N22}) will be negative) will depend on the
coefficient function$\frac{\left[ k\mid \frac{\partial \overrightarrow{%
\mathbf{r}}_{1}}{\partial F_{1}}\mid +\mid \frac{\partial \overrightarrow{%
\mathbf{r}}_{2}}{\partial F_{2}}\mid \right] }{k}$, which can be written as
the ratio 
\begin{equation}
\frac{\left[ k\mid \frac{\partial \overrightarrow{\mathbf{r}}_{1}}{\partial
F_{1}}\mid +\mid \frac{\partial \overrightarrow{\mathbf{r}}_{2}}{\partial
F_{2}}\mid \right] }{k}=\frac{W_{1}}{W_{2}}\text{ ,}  \label{N23}
\end{equation}%
where $W_{1}$ and $W_{2}$ are determined as 
\begin{equation}
W_{1}:=2\mid \frac{\partial \overrightarrow{\mathbf{r}}_{1}}{\partial F_{1}}%
\mid .\mid \frac{\partial \overrightarrow{\mathbf{r}}_{2}}{\partial F_{2}}%
\mid +\mid \frac{\partial \overrightarrow{\mathbf{r}}_{1}}{\partial F_{1}}%
\mid ^{2}-\mid \frac{\partial \overrightarrow{\mathbf{r}}_{2}}{\partial F_{2}%
}\mid ^{2}\text{,}  \label{N24}
\end{equation}%
\begin{equation}
W_{2}:=\left[ \mid \frac{\partial \overrightarrow{\mathbf{r}}_{1}}{\partial
F_{1}}\mid ^{.}+\mid \frac{\partial \overrightarrow{\mathbf{r}}_{2}}{%
\partial F_{2}}\mid \right] \text{.}^{.}  \label{N25}
\end{equation}
\bigskip In this paper we shall not present the other proofs for the positivity of
the third geodesic distance (\ref{N21}), obtained after transformation of
the geodesic equation (\ref{N6}) and also the case about the positivity of
the fourth geodesic equation (\ref{N22}). All the detailed proofs will be
given in the forthcoming paper \cite{Dimitrov2026}.
\bigskip Below we shall only summarize the main new findings in this paper and
especially the four different types of geodesics.
\section{Four different expressions for the geodesic distance for the
general case of satellites, moving on arbitrary space trajectories}
\label{sec:FourDifferGeodesic}
\bigskip Here we shall present briefly the results on the different expressions for
the geodesic distance. The first formulae for the geodesic distance is expression (\ref{N6}) 
\begin{equation*}
\widehat{R}_{AB}^{2}=\frac{c}{2\mid \frac{\partial \overrightarrow{\mathbf{r}%
}_{2}}{\partial F_{2}}\mid }\int \frac{D_{1}S_{1}\frac{\partial T_{1}}{%
\partial F_{1}}\left[ \mid \frac{\partial \overrightarrow{\mathbf{r}}_{1}}{%
\partial F_{1}}\mid -k\mid \frac{\partial \overrightarrow{\mathbf{r}}_{2}}{%
\partial F_{2}}\mid \right] }{\mid \frac{\partial \overrightarrow{\mathbf{r}}%
_{1}}{\partial F_{1}}\mid }dF_{1}\text{.}
\end{equation*}%
The second expression for the geodesic distance is (\ref{N7}) 
\begin{equation*}
\widehat{R}_{AB}^{2}=\frac{c}{2\mid \frac{\partial \overrightarrow{\mathbf{r}%
}_{2}}{\partial F_{2}}\mid }\int \frac{D_{1}S_{1}\frac{\partial T_{1}}{%
\partial F_{1}}\left[ k\mid \frac{\partial \overrightarrow{\mathbf{r}}_{2}}{%
\partial F_{2}}\mid -\mid \frac{\partial \overrightarrow{\mathbf{r}}_{1}}{%
\partial F_{1}}\mid \right] }{\mid \frac{\partial \overrightarrow{\mathbf{r}}%
_{1}}{\partial F_{1}}\mid }dF_{1}\text{,}
\end{equation*}%
which is the same as (\ref{N6}), but with a minus sign in front. Note that
in both expressions there is asymmetry with respect to the interchange of
the indices $1\Longleftrightarrow 2$, which is a confirmation of the fact
that the geodesic distance $\widehat{R}_{AB}^{2}$ in the forward direction
will be different from the geodesic distance $\widehat{R}_{BA}^{2}$ in the
backward direction.
\bigskip Further after performing the transformation $k\rightarrow -\frac{1}{k}$ in
the transformation (\ref{N4}), another two expressions for the geodesic
distance have been derived: expression (\ref{N21})
\begin{equation*}
\widehat{R}_{AB}^{2}=\frac{c}{2\mid \frac{\partial \overrightarrow{\mathbf{r}%
}_{2}}{\partial F_{2}}\mid }\int \frac{D_{1}S_{1}\frac{\partial T_{1}}{%
\partial F_{1}}\left[ k\mid \frac{\partial \overrightarrow{\mathbf{r}}_{1}}{%
\partial F_{1}}\mid +\mid \frac{\partial \overrightarrow{\mathbf{r}}_{2}}{%
\partial F_{2}}\mid \right] }{k\mid \frac{\partial \overrightarrow{\mathbf{r}%
}_{1}}{\partial F_{1}}\mid }dF_{1}
\end{equation*}%
and also expression (\ref{N22}) 
\begin{equation*}
\widehat{R}_{AB}^{2}=-\frac{c}{2\mid \frac{\partial \overrightarrow{\mathbf{r%
}}_{2}}{\partial F_{2}}\mid }\int \frac{D_{1}S_{1}\frac{\partial T_{1}}{%
\partial F_{1}}\left[ k\mid \frac{\partial \overrightarrow{\mathbf{r}}_{1}}{%
\partial F_{1}}\mid +\mid \frac{\partial \overrightarrow{\mathbf{r}}_{2}}{%
\partial F_{2}}\mid \right] }{k\mid \frac{\partial \overrightarrow{\mathbf{r}%
}_{1}}{\partial F_{1}}\mid }dF_{1}\text{,}
\end{equation*}%
which is the same as the previous one, but with a minus sign in front.
\subsection{First case of positivity of the geodesic distance}
\label{sec:FirstCasePositiv}
\bigskip Earlier it was proved that the geodesic distance (\ref{N6}) is positive when 
$\mid \frac{\partial \overrightarrow{\mathbf{r}}_{2}}{\partial F_{2}}\mid
>\mid \frac{\partial \overrightarrow{\mathbf{r}}_{1}}{\partial F_{1}}\mid $,
but in the interval 
\begin{equation}
(\sqrt[.]{2}-1)\mid \frac{\partial \overrightarrow{\mathbf{r}}_{2}}{\partial
F_{2}}\mid <\mid \frac{\partial \overrightarrow{\mathbf{r}}_{1}}{\partial
F_{1}}\mid <\mid \frac{\partial \overrightarrow{\mathbf{r}}_{2}}{\partial
F_{2}}\mid  \label{N29}
\end{equation}%
both geodesic distance equations (\ref{N6}) and (\ref{N21}) are valid.
\subsection{Second case of positivity of the geodesic distance}
\label{sec:SecondCasePosit}
\bigskip For the case when $\mid \frac{\partial \overrightarrow{\mathbf{r}}_{1}}{%
\partial F_{1}}\mid >\mid \frac{\partial \overrightarrow{\mathbf{r}}_{2}}{%
\partial F_{2}}\mid $ expression (\ref{N7}) is valid and also expression (%
\ref{N21}).
\subsection{Third case of positivity of the geodesic distance}
\label{sec:ThirdCasePosit}
\bigskip When inequality $\mid \frac{\partial \overrightarrow{\mathbf{r}}_{1}}{%
\partial F_{1}}\mid <(\sqrt{2}-1)\mid \frac{\partial \overrightarrow{\mathbf{%
r}}_{2}}{\partial F_{2}}\mid $ holds, expression (\ref{N22}) for the
geodesic distance holds.
\subsection{Fourth case of positivity of the geodesic distance}
\label{sec:FourthCasePosit}
\bigskip This case is particularly interesting and needs further investigation. If $(%
\sqrt{2}-1)\mid \frac{\partial \overrightarrow{\mathbf{r}}_{2}}{\partial
F_{2}}\mid <\mid \frac{\partial \overrightarrow{\mathbf{r}}_{1}}{\partial
F_{1}}\mid <\mid \frac{\partial \overrightarrow{\mathbf{r}}_{2}}{\partial
F_{2}}\mid $ is fulfilled, then the geodesic distance (\ref{N6}) will be
positive. If the interval $((\sqrt{2}-1)$,$1)$ is extended a little so that
inequality 
\begin{equation}
(\sqrt{2}-1)\mid \frac{\partial \overrightarrow{\mathbf{r}}_{2}}{\partial
F_{2}}\mid <\mid \frac{\partial \overrightarrow{\mathbf{r}}_{1}}{\partial
F_{1}}\mid <\mid \frac{\partial \overrightarrow{\mathbf{r}}_{2}}{\partial
F_{2}}\mid <\sqrt{2}\mid \frac{\partial \overrightarrow{\mathbf{r}}_{2}}{%
\partial F_{2}}\mid  \label{N36}
\end{equation}%
is fulfilled, then additionally the formulae (\ref{N21}) will hold.
\subsection{Fifth case of positivity of the geodesic distance}
\label{sec:FifthCasePosit}
\bigskip If the inequality $\frac{\sqrt{2}}{2}\mid \frac{\partial \overrightarrow{%
\mathbf{r}}_{2}}{\partial F_{2}}\mid <\mid \frac{\partial \overrightarrow{%
\mathbf{r}}_{1}}{\partial F_{1}}\mid $ holds, then geodesic distance
formulae (\ref{N21}) will be valid. If the interval is extended 
\begin{equation}
\frac{\sqrt{2}}{2}\mid \frac{\partial \overrightarrow{\mathbf{r}}_{2}}{%
\partial F_{2}}\mid <\mid \frac{\partial \overrightarrow{\mathbf{r}}_{1}}{%
\partial F_{1}}\mid <\mid \frac{\partial \overrightarrow{\mathbf{r}}_{2}}{%
\partial F_{2}}\mid \text{,}  \label{N37}
\end{equation}%
then additionally geodesic distance (\ref{N6}) will be positive and thus
will be valid. Also, in the interval $\mid \frac{\partial \overrightarrow{%
\mathbf{r}}_{1}}{\partial F_{1}}\mid <\frac{\sqrt{2}}{2}\mid \frac{\partial 
\overrightarrow{\mathbf{r}}_{2}}{\partial F_{2}}\mid $ (\ref{N6}) will \
hold.
\section{The case of equality of the two space-time distances}
\label{sec:VarSpTimeDistEqualSpTime}
\bigskip The aim in this section will be to derive from the solution $d\widetilde{R}%
_{AB}^{2}(1\rightarrow 2)$ of the quadratic equation (\ref{M16}) the
space-time distance $\ \widetilde{R}_{AB}^{2}(1\rightarrow 2)$, then to
define the differential $d\widetilde{R}_{BA}^{2}(2\rightarrow 1)$ of the
space-time distance in the "backwards" direction. After finding the
"backwards" space-time distance $\widetilde{R}_{BA}^{2}(2\rightarrow 1)$, it
will be postulated that the two space-time distances $\widetilde{R}%
_{AB}^{2}(1\rightarrow 2)$ and $\widetilde{R}_{BA}^{2}(2\rightarrow 1)$ will
have to be equal and thus, the space-time distance will be invariantly
formulated. In finding the "backward" space-time distance $\widetilde{R}%
_{BA}^{2}(2\rightarrow 1)$, the dependence of the functions (\ref{M7A}) $%
D_{1}(F_{1},F_{2},\overrightarrow{\mathbf{r}}_{1},\overrightarrow{\mathbf{r}}%
_{2})$ and (\ref{M7B}) $D_{2}(F_{1},F_{2},\overrightarrow{\mathbf{r}}_{1},%
\overrightarrow{\mathbf{r}}_{2})$ on the space coordinates $\overrightarrow{%
\mathbf{r}}_{1}$ and $\overrightarrow{\mathbf{r}}_{2}$ will be taken into
account. There was no need to take into account this dependence when
deriving the space distance $\widetilde{R}_{AB}^{2}(1\rightarrow 2)$.
\bigskip The most interesting result from the invariant formulation of the space-time
distance is that it will not lead to a symmetry under the interchange of the
indices $1\Leftrightarrow 2$. After finding the geodesic distance from the
space-time distance, the formulae will turn out to depend on the first
generalized derivative $\mid \frac{\partial \overrightarrow{\mathbf{r}}_{1}}{%
\partial F_{1}}\mid $(generalized velocity) of the space vector coordinate $%
\overrightarrow{\mathbf{r}}_{1}$ of the first satellite, on the first $\frac{%
\partial \overrightarrow{\mathbf{r}}_{2}}{\partial F_{2}}$ and on second
derivative $\frac{\partial ^{2}\left( \overrightarrow{\mathbf{r}}%
_{2}^{.}\right) }{\partial \left( F_{2}\right) ^{2}}$ of the second vector
coordinate $\overrightarrow{\mathbf{r}}_{2}$ of the second satellite. This
will become evident when calculating expression (\ref{M69}) for $\widetilde{R%
}_{BA}^{2}(2\rightarrow 1)$ . Thus the asymmetry in the expressions for $%
\widetilde{R}_{AB}^{2}(1\rightarrow 2)$ and $\widetilde{R}%
_{BA}^{2}(2\rightarrow 1)$ with respect to the interchange of the indices $%
1\Leftrightarrow 2$ will represent a particular and important feature, a
consequence also of the derived relation (\ref{N2}) 
\begin{equation*}
D_{2}(dT_{2})S_{2}=k_{1,2}D_{1}(dT_{1})S_{1}\text{ .}
\end{equation*}%
The physical meaning of this relation and its analogy with earlier derived
by H. Bondi \cite{Bondi1}, \cite{Bondi2} formulae in Einstein's  
General Relativity have been clarified in the Introduction.
\subsection{The expression for the first space-time distance $\protect%
\widetilde{R}_{AB}^{2}(1\rightarrow 2)$}
\label{sec:FirstSpTimeDist}
\bigskip From the first equation (\ref{M21}) for the differential of the
space-time distance%
\begin{equation*}
d\widetilde{R}_{AB}^{2}(1\rightarrow 2)=\frac{cD_{1}(dT_{1})S_{1}(r_{1})}{%
\mid \frac{\partial \overrightarrow{\mathbf{r}}_{1}}{\partial F_{1}}\mid }%
\pm \frac{cD_{2}(dT_{2})S_{2}(r_{2})}{\mid \frac{\partial \overrightarrow{%
\mathbf{r}}_{1}}{\partial F_{1}}\mid })\text{,}
\end{equation*}%
the following differential equation in full derivatives can be obtained 
\begin{equation*}
\frac{\partial \widetilde{R}_{AB}^{2}(1\rightarrow 2)}{\partial F_{1}}dF_{1}+%
\frac{\partial \widetilde{R}_{AB}^{2}(1\rightarrow 2)}{\partial F_{2}}dF_{2}
\end{equation*}%
\begin{equation}
=\frac{cD_{1}(\frac{\partial T_{1}}{\partial F_{1}})S_{1}(r_{1})}{\mid \frac{%
\partial \overrightarrow{\mathbf{r}}_{1}}{\partial F_{1}}\mid }dF_{1}\pm 
\frac{cD_{2}(\frac{\partial T_{2}}{\partial F_{2}})S_{2}(r_{2})}{\mid \frac{%
\partial \overrightarrow{\mathbf{r}}_{1}}{\partial F_{1}}\mid }dF_{2}\text{.}
\label{M56}
\end{equation}%
The first equation, which can be obtained after integration is 
\begin{equation}
\widetilde{R}_{AB}^{2}(1\rightarrow 2)=c\int \frac{D_{1}(\frac{\partial
T_{1}}{\partial F_{1}})S_{1}(r_{1})}{\mid \frac{\partial \overrightarrow{%
\mathbf{r}}_{1}}{\partial F_{1}}\mid }dF_{1}+\Psi (F_{2})\text{.}
\label{M57}
\end{equation}%
After differentiation by $F_{2}$ and taking into account the second equation 
\begin{equation}
\frac{\partial \widetilde{R}_{AB}^{2}(1\rightarrow 2)}{\partial F_{2}}=\pm 
\frac{cD_{2}(\frac{\partial T_{2}}{\partial F_{2}})S_{2}(r_{2})}{\mid \frac{%
\partial \overrightarrow{\mathbf{r}}_{1}}{\partial F_{1}}\mid }\text{ ,}
\label{M58}
\end{equation}%
the equation for the derivative $\frac{\partial \Psi (F_{2})}{\partial F_{2}}
$ of the function $\Psi (F_{2})$ from expression (\ref{M57}) can be found to
be 
\begin{equation*}
\frac{\partial \widetilde{R}_{AB}^{2}(1\rightarrow 2)}{\partial F_{2}}%
=c\int \frac{\frac{\partial D_{1}}{\partial F_{2}}(\frac{\partial T_{1}}{%
\partial F_{1}})S_{1}(r_{1})}{\mid \frac{\partial \overrightarrow{\mathbf{r}}%
_{1}}{\partial F_{1}}\mid }dF_{1}
\end{equation*}%
\begin{equation}
+\frac{\partial \Psi (F_{2})}{\partial F_{2}}=\pm \frac{cD_{2}(\frac{%
\partial T_{2}}{\partial F_{2}})S_{2}(r_{2})}{\mid \frac{\partial 
\overrightarrow{\mathbf{r}}_{1}}{\partial F_{1}}\mid }\text{ .}  \label{M59}
\end{equation}%
After integration by $F_{2}$, the function $\Psi (F_{2})$ is obtained in the
form 
\begin{equation}
\Psi (F_{2})=\pm c\int \frac{D_{2}(\frac{\partial T_{2}}{\partial F_{2}}%
)S_{2}(r_{2})}{\mid \frac{\partial \overrightarrow{\mathbf{r}}_{1}}{\partial
F_{1}}\mid }dF_{2}-c\int \frac{\frac{\partial D_{1}}{\partial F_{2}}(\frac{%
\partial T_{1}}{\partial F_{1}})S_{1}(r_{1})}{\mid \frac{\partial 
\overrightarrow{\mathbf{r}}_{1}}{\partial F_{1}}\mid }dF_{1}dF_{2}\text{.}
\label{M60}
\end{equation}%
Taking into account that $\int \frac{\partial D_{1}}{\partial F_{2}}%
dF_{2}=D_{1}$ and substituting (\ref{M60}) into (\ref{M57}) for $\widetilde{R%
}_{AB}^{2}(1\rightarrow 2)$, one can obtain 
\begin{equation}
\widetilde{R}_{AB}^{2}(1\rightarrow 2)=\pm c\int \frac{D_{2}(\frac{\partial
T_{2}}{\partial F_{2}})S_{2}(r_{2})}{\mid \frac{\partial \overrightarrow{%
\mathbf{r}}_{1}}{\partial F_{1}}\mid }dF_{2}\text{.}  \label{M61}
\end{equation}
\subsection{The expression for the second ("inverse") space-time distance $%
\protect\widetilde{R}_{AB}^{2}(2\rightarrow 1)$}
\label{sec:SecondSpTimeDist}
\bigskip Following the approach, applied in the preceding paragraph, equality (\ref%
{M25}) can be written as 
\begin{equation*}
\frac{\partial \widetilde{R}_{BA}^{2}(2\rightarrow 1)}{\partial F_{1}}dF_{1}+%
\frac{\partial \widetilde{R}_{BA}^{2}(2\rightarrow 1)}{\partial F_{2}}dF_{2}
\end{equation*}%
\begin{equation}
=\pm \frac{cD_{1}(\frac{\partial T_{1}}{\partial F_{1}})S_{1}(r_{1})}{\mid 
\frac{\partial \overrightarrow{\mathbf{r}}_{2}}{\partial F_{2}}\mid }dF_{1}+%
\frac{cD_{2}(\frac{\partial T_{2}}{\partial F_{2}})S_{2}(r_{2})}{\mid \frac{%
\partial \overrightarrow{\mathbf{r}}_{2}}{\partial F_{2}}\mid }dF_{2}\text{.}
\label{M62}
\end{equation}%
Setting up equal the expressions in front of $dF_{1}$ and subsequently
integrating, one can obtain, adding also the unknown function $\varphi
(F_{2})$ (to be determined afterwards) 
\begin{equation}
\widetilde{R}_{BA}^{2}(2\rightarrow 1)=\pm c\int \frac{D_{1}(\frac{\partial
T_{1}}{\partial F_{1}})S_{1}(r_{1})}{\mid \frac{\partial \overrightarrow{%
\mathbf{r}}_{2}}{\partial F_{2}}\mid }dF_{1}+\varphi (F_{2})\text{.}
\label{M63}
\end{equation}%
Differentiating by $F_{2}$, taking into account expression (\ref{M7A}) for $%
D_{1}$ (which depends on both parameters $F_{1}$ and $F_{2}$) and also the
second equation (\ref{M62}) in total derivatives 
\begin{equation}
\frac{\partial \widetilde{R}_{BA}^{2}(2\rightarrow 1)}{\partial F_{2}}=\frac{%
cD_{2}(\frac{\partial T_{2}}{\partial F_{2}})S_{2}(r_{2})}{\mid \frac{%
\partial \overrightarrow{\mathbf{r}}_{2}}{\partial F_{2}}\mid }\text{,}
\label{M64}
\end{equation}%
the following equation is obtained for $\frac{\partial \widetilde{R}%
_{BA}^{2}(2\rightarrow 1)}{\partial F_{2}}$, in which the derivative $\frac{%
\partial \varphi (F_{2})}{\partial F_{2}}$ participates 
\begin{equation*}
\frac{\partial \widetilde{R}_{BA}^{2}(2\rightarrow 1)}{\partial F_{2}}=\pm
c\int \frac{\frac{\partial D_{1}}{\partial F_{2}}(\frac{\partial T_{1}}{%
\partial F_{1}})S_{1}(r_{1})}{\mid \frac{\partial \overrightarrow{\mathbf{r}}%
_{2}}{\partial F_{2}}\mid }dF_{1}
\end{equation*}%
\begin{equation}
\mp c\int \frac{D_{1}(\frac{\partial T_{1}}{\partial F_{1}})S_{1}(r_{1})}{%
\left( \mid \frac{\partial \overrightarrow{\mathbf{r}}_{2}}{\partial F_{2}}%
\mid \right) ^{2}}\frac{\partial }{\partial F_{2}}\left( \mid \frac{\partial 
\overrightarrow{\mathbf{r}}_{2}}{\partial F_{2}}\mid \right) dF_{1}+\frac{%
\partial \varphi (F_{2})}{\partial F_{2}}\text{.}  \label{M65}
\end{equation}%
Taking into account that 
\begin{equation}
\frac{\partial }{\partial F_{2}}\left( \mid \frac{\partial \overrightarrow{%
\mathbf{r}}_{2}}{\partial F_{2}}\mid \right) =\frac{\frac{\partial x_{2}}{%
\partial F_{2}}\frac{\partial ^{2}x_{2}}{\partial F_{2}^{2}}+\frac{\partial
y_{2}}{\partial F_{2}}\frac{\partial ^{2}y_{2}}{\partial F_{2}^{2}}+\frac{%
\partial z_{2}}{\partial F_{2}}\frac{\partial ^{2}z_{2}}{\partial F_{2}^{2}}%
}{\mid \frac{\partial \overrightarrow{\mathbf{r}}_{2}}{\partial F_{2}}\mid }%
\text{ }  \label{M66}
\end{equation}%
\begin{equation}
=\frac{\frac{\partial \overrightarrow{\mathbf{r}}_{2}}{\partial F_{2}}.\frac{%
\partial ^{2}\left( \overrightarrow{\mathbf{r}}_{2}^{.}\right) }{\partial
\left( F_{2}\right) ^{2}}}{\mid \frac{\partial \overrightarrow{\mathbf{r}}%
_{2}}{\partial F_{2}}\mid }\text{ }  \label{M67}
\end{equation}%
and performing the integration in (\ref{M65}), the expression for $\varphi
(F_{2})$ is obtained as 
\begin{equation*}
\varphi (F_{2})=c\int \frac{D_{2}(\frac{\partial T_{2}}{\partial F_{2}}%
)S_{2}(r_{2})}{\mid \frac{\partial \overrightarrow{\mathbf{r}}_{2}}{\partial
F_{2}}\mid }dF_{2}\mp c\int \frac{\frac{\partial D_{1}}{\partial F_{2}}%
S_{1}(\frac{\partial T_{1}}{\partial F_{1}})}{\mid \frac{\partial 
\overrightarrow{\mathbf{r}}_{2}}{\partial F_{2}}\mid }dF_{1}dF_{2}
\end{equation*}%
\begin{equation}
\pm c\int \frac{D_{1}(\frac{\partial T_{1}}{\partial F_{1}}%
)S_{1}dF_{1}\left( \frac{\partial \overrightarrow{\mathbf{r}}_{2}}{\partial
F_{2}}.\frac{\partial ^{2}\left( \overrightarrow{\mathbf{r}}_{2}^{.}\right) 
}{\partial \left( F_{2}\right) ^{2}}\right) }{\mid \frac{\partial 
\overrightarrow{\mathbf{r}}_{2}}{\partial F_{2}}\mid ^{3}}dF_{2}\text{ .}
\label{M68}
\end{equation}%
This expression should be added to formulae (\ref{M63}) for $\widetilde{R}%
_{BA}^{2}(2\rightarrow 1)$, which acquires the form 
\begin{equation*}
\widetilde{R}_{BA}^{2}(2\rightarrow 1)=\pm c\int \frac{D_{1}(\frac{\partial
T_{1}}{\partial F_{1}})S_{1}}{\mid \frac{\partial \overrightarrow{\mathbf{r}}%
_{2}}{\partial F_{2}}\mid }dF_{1}+c\int \frac{D_{2}(\frac{\partial T_{2}}{%
\partial F_{2}})S_{2}}{\mid \frac{\partial \overrightarrow{\mathbf{r}}_{2}}{%
\partial F_{2}}\mid }dF_{2}
\end{equation*}%
\begin{equation*}
\mp c\int \frac{\frac{\partial D_{1}}{\partial F_{2}}S_{1}(\frac{\partial
T_{1}}{\partial F_{1}})}{\mid \frac{\partial \overrightarrow{\mathbf{r}}_{2}%
}{\partial F_{2}}\mid }dF_{1}dF_{2}
\end{equation*}%
\begin{equation}
\pm c\int \frac{D_{1}(\frac{\partial T_{1}}{\partial F_{1}}%
)S_{1}dF_{1}\left( \frac{\partial \overrightarrow{\mathbf{r}}_{2}}{\partial
F_{2}}.\frac{\partial ^{2}\left( \overrightarrow{\mathbf{r}}_{2}^{.}\right) 
}{\partial \left( F_{2}\right) ^{2}}\right) }{\mid \frac{\partial 
\overrightarrow{\mathbf{r}}_{2}}{\partial F_{2}}\mid ^{3}}dF_{2}\text{.}
\label{M69}
\end{equation}
\section{ Equality between the two space-time distances $\protect\widetilde{R%
}_{AB}^{2}(1\rightarrow 2)$ and $\protect\widetilde{R}_{AB}^{2}(2\rightarrow
1)$ and derivation of the modified compatibility condition}
\label{sec:EqualTwoSpTimeDistances}
\bigskip In order to determine the space-time distance uniquely, let us set up equal
the two expressions (\ref{M61}) for $\widetilde{R}_{AB}^{2}(1\rightarrow 2)$
from the previous section and expression (\ref{M69}) for $\widetilde{R}%
_{AB}^{2}(2\rightarrow 1)$. Since there are terms with signs $\mp $ and $\pm 
$, two cases shall be investigated.
\subsection{1st Case: The case with the lower sign}
\label{sec:ConLowerSigns}
\bigskip The corresponding equation is 
\begin{equation*}
\int \frac{D_{2}(\frac{\partial T_{2}}{\partial F_{2}})S_{2}}{\mid \frac{%
\partial \overrightarrow{\mathbf{r}}_{1}}{\partial F_{1}}\mid }dF_{2}=\int 
\frac{D_{1}(\frac{\partial T_{1}}{\partial F_{1}})S_{1}}{\mid \frac{\partial 
\overrightarrow{\mathbf{r}}_{2}}{\partial F_{2}}\mid }dF_{2}
\end{equation*}%
\begin{equation*}
-\int \frac{\frac{\partial D_{1}}{\partial F_{2}}S_{1}(\frac{\partial T_{1}%
}{\partial F_{1}})}{\mid \frac{\partial \overrightarrow{\mathbf{r}}_{2}}{%
\partial F_{2}}\mid }dF_{1}dF_{2}-\int \frac{D_{2}(\frac{\partial T_{2}}{%
\partial F_{2}})S_{2}}{\mid \frac{\partial \overrightarrow{\mathbf{r}}_{2}}{%
\partial F_{2}}\mid }dF_{2}
\end{equation*}%
\begin{equation}
+\int \frac{D_{1}(\frac{\partial T_{1}}{\partial F_{1}})S_{1}dF_{1}\left( 
\frac{\partial \overrightarrow{\mathbf{r}}_{2}}{\partial F_{2}}.\frac{%
\partial ^{2}\left( \overrightarrow{\mathbf{r}}_{2}^{.}\right) }{\partial
\left( F_{2}\right) ^{2}}\right) }{\mid \frac{\partial \overrightarrow{%
\mathbf{r}}_{2}}{\partial F_{2}}\mid ^{3}}dF_{2}\text{.}  \label{M70}
\end{equation}%
However, previously from the condition $d\widetilde{R}_{AB}^{2}(1\rightarrow
2)=d\widetilde{R}_{BA}^{2}(2$ $\rightarrow 1)$ for the equality between the
two differentials, the following relation (\ref{N2}) has been found%
\begin{equation*}
D_{2}(dT_{2})S_{2}=k_{1,2}(\mid \frac{\partial \overrightarrow{\mathbf{r}}%
_{1}}{\partial F_{1}}\mid ,\mid \frac{\partial \overrightarrow{\mathbf{r}}%
_{2}}{\partial F_{2}}\mid )D_{1}(dT_{1})S_{1}\text{,}
\end{equation*}%
where $k_{1,2}(\mid \frac{\partial \overrightarrow{\mathbf{r}}_{1}}{\partial
F_{1}}\mid ,\mid \frac{\partial \overrightarrow{\mathbf{r}}_{2}}{\partial
F_{2}}\mid )$ is defined as the ratio (\ref{N3}) 
\begin{equation*}
k_{1,2}(\mid \frac{\partial \overrightarrow{\mathbf{r}}_{1}}{\partial F_{1}}%
\mid ,\mid \frac{\partial \overrightarrow{\mathbf{r}}_{2}}{\partial F_{2}}%
\mid ):=\frac{\mid \frac{\partial \overrightarrow{\mathbf{r}}_{2}}{\partial
F_{2}}\mid \mp \mid \frac{\partial \overrightarrow{\mathbf{r}}_{1}}{\partial
F_{1}}\mid }{\mid \frac{\partial \overrightarrow{\mathbf{r}}_{1}}{\partial
F_{1}}\mid \pm \mid \frac{\partial \overrightarrow{\mathbf{r}}_{2}}{\partial
F_{2}}\mid }\text{.}
\end{equation*}%
Substituting relation (\ref{N2}) into the compatibility condition (\ref{M36}%
) 
\begin{equation*}
\int \frac{D_{2}(\frac{\partial T_{2}}{\partial F_{2}})S_{2}(r_{2})}{\mid 
\frac{\partial \overrightarrow{\mathbf{r}}_{1}}{\partial F_{1}}\mid }%
dF_{2}=\int \frac{D_{1}(\frac{\partial T_{1}}{\partial F_{1}})S_{1}(r_{1})}{%
\mid \frac{\partial \overrightarrow{\mathbf{r}}_{2}}{\partial F_{2}}\mid }%
dF_{1}\text{,}
\end{equation*}%
the following relation can be obtained 
\begin{equation}
\int D_{1}\frac{\partial T_{1}}{\partial F_{1}}S_{1}\left[ \frac{k_{1,2}}{%
\mid \frac{\partial \overrightarrow{\mathbf{r}}_{1}}{\partial F_{1}}\mid }-%
\frac{1}{\mid \frac{\partial \overrightarrow{\mathbf{r}}_{2}}{\partial F_{2}}%
\mid }\right] dF_{1}=0\text{ .}  \label{M72}
\end{equation}
\subsection{First condition for the fulfillment of the compatibility relation%
}
\label{sec:ConditFulfCompatCondit}
\bigskip The first condition is related to the equality to zero of the expression in
the brackets, which is fulfilled when 
\begin{equation}
k_{1,2}=\frac{\mid \frac{\partial \overrightarrow{\mathbf{r}}_{1}}{\partial
F_{1}}\mid }{\mid \frac{\partial \overrightarrow{\mathbf{r}}_{2}}{\partial
F_{2}}\mid }\text{ .}  \label{M73}
\end{equation}%
Taking into account that $k_{1,2}$ according to (\ref{N3}) can be taken with
the lower and with the upper sign, this will correspond to two cases:
\subsubsection{First subcase when $k_{1,2}=k_{1}$}
\bigskip The first subcase is realized when $k_{1}=\frac{\mid \frac{\partial 
\overrightarrow{\mathbf{r}}_{2}}{\partial F_{2}}\mid +\mid \frac{\partial 
\overrightarrow{\mathbf{r}}_{1}}{\partial F_{1}}\mid }{\mid \frac{\partial 
\overrightarrow{\mathbf{r}}_{1}}{\partial F_{1}}\mid -\mid \frac{\partial 
\overrightarrow{\mathbf{r}}_{2}}{\partial F_{2}}\mid }$. Then equality (\ref%
{M73}) assumes the form of a quadratic relation between the generalized
velocities 
\begin{equation}
\left( \mid \frac{\partial \overrightarrow{\mathbf{r}}_{1}}{\partial F_{1}}%
\mid \right) ^{2}-\left( \mid \frac{\partial \overrightarrow{\mathbf{r}}_{2}%
}{\partial F_{2}}\mid \right) ^{2}-2\mid \frac{\partial \overrightarrow{%
\mathbf{r}}_{1}}{\partial F_{1}}\mid \mid \frac{\partial \overrightarrow{%
\mathbf{r}}_{2}}{\partial F_{2}}\mid =0\text{ .}  \label{M74}
\end{equation}%
This equality corresponds to 
\begin{equation}
\mid \frac{\partial \overrightarrow{\mathbf{r}}_{2}}{\partial F_{2}}\mid =%
\frac{\epsilon }{\sqrt{2}}.\left[ \mid \frac{\partial \overrightarrow{%
\mathbf{r}}_{1}}{\partial F_{1}}\mid -\mid \frac{\partial \overrightarrow{%
\mathbf{r}}_{2}}{\partial F_{2}}\mid \right] \text{ ,}  \label{M75}
\end{equation}%
where $\epsilon =\pm 1$. The case $\epsilon =-1$ cannot be realized, because 
$\mid \frac{\partial \overrightarrow{\mathbf{r}}_{2}}{\partial F_{2}}\mid
=(-1-\sqrt{2}).\mid \frac{\partial \overrightarrow{\mathbf{r}}_{1}}{\partial
F_{1}}\mid $ and this will contradict to the requirement that both $\mid 
\frac{\partial \overrightarrow{\mathbf{r}}_{1}}{\partial F_{1}}\mid $ and $%
\mid \frac{\partial \overrightarrow{\mathbf{r}}_{2}}{\partial F_{2}}\mid $
have to be positive. The other case $\epsilon =+1$ leads to 
\begin{equation}
\mid \frac{\partial \overrightarrow{\mathbf{r}}_{2}}{\partial F_{2}}\mid =(%
\sqrt{2}-1).\mid \frac{\partial \overrightarrow{\mathbf{r}}_{1}}{\partial
F_{1}}\mid  \label{M76}
\end{equation}%
and since $\sqrt{2}-1>0$, the requirement for positivity of both $\mid \frac{%
\partial \overrightarrow{\mathbf{r}}_{1}}{\partial F_{1}}\mid $ and $\mid 
\frac{\partial \overrightarrow{\mathbf{r}}_{2}}{\partial F_{2}}\mid $ is
fulfilled. Since $(\sqrt{2}-1).\mid \frac{\partial \overrightarrow{\mathbf{r}%
}_{1}}{\partial F_{1}}\mid <(\sqrt{2}+1).\mid \frac{\partial \overrightarrow{%
\mathbf{r}}_{1}}{\partial F_{1}}\mid $ , equality (\ref{M76}) seems to
contradict the previously established inequality (\ref{N29}), written as 
\begin{equation}
\mid \frac{\partial \overrightarrow{\mathbf{r}}_{2}}{\partial F_{2}}\mid <%
\frac{1}{\sqrt{2}-1}.\mid \frac{\partial \overrightarrow{\mathbf{r}}_{1}}{%
\partial F_{1}}\mid =\left( \sqrt{2}+1\right) .\mid \frac{\partial 
\overrightarrow{\mathbf{r}}_{1}}{\partial F_{1}}\mid \text{.}  \label{M77}
\end{equation}%
However, the fulfillment of this inequality, necessary for the positivity of
the geodesic distances (\ref{N6}) and (\ref{N21}) for the present case is
not obligatory. The equality (\ref{M76}) in fact means that $\mid \frac{%
\partial \overrightarrow{\mathbf{r}}_{2}}{\partial F_{2}}\mid <$ $\mid \frac{%
\partial \overrightarrow{\mathbf{r}}_{1}}{\partial F_{1}}\mid $. This
inequality was discussed earlier in relation to the positivity of the
geodesic distances (\ref{N7}) and (\ref{N21}). It can be concluded that
equality (\ref{M76}) leads to no contradictions.
\subsubsection{Second subcase when $k_{1,2}=k_{2}$}
\bigskip The second case is realized for 
\begin{equation}
k_{2}=\frac{\mid \frac{\partial \overrightarrow{\mathbf{r}}_{2}}{\partial
F_{2}}\mid -\mid \frac{\partial \overrightarrow{\mathbf{r}}_{1}}{\partial
F_{1}}\mid }{\mid \frac{\partial \overrightarrow{\mathbf{r}}_{1}}{\partial
F_{1}}\mid +\mid \frac{\partial \overrightarrow{\mathbf{r}}_{2}}{\partial
F_{2}}\mid }=\frac{\mid \frac{\partial \overrightarrow{\mathbf{r}}_{1}}{%
\partial F_{1}}\mid }{\mid \frac{\partial \overrightarrow{\mathbf{r}}_{2}}{%
\partial F_{2}}\mid }\text{ .}  \label{M78}
\end{equation}%
This leads to the equality 
\begin{equation}
\left( \mid \frac{\partial \overrightarrow{\mathbf{r}}_{1}}{\partial F_{1}}%
\mid \right) ^{2}-\left( \mid \frac{\partial \overrightarrow{\mathbf{r}}_{2}%
}{\partial F_{2}}\mid \right) ^{2}+2\mid \frac{\partial \overrightarrow{%
\mathbf{r}}_{1}}{\partial F_{1}}\mid \mid \frac{\partial \overrightarrow{%
\mathbf{r}}_{2}}{\partial F_{2}}\mid =0\text{ ,}  \label{M79}
\end{equation}%
from where it can be found 
\begin{equation}
\mid \frac{\partial \overrightarrow{\mathbf{r}}_{1}}{\partial F_{1}}\mid =(%
\sqrt{2}\epsilon -1).\mid \frac{\partial \overrightarrow{\mathbf{r}}_{2}}{%
\partial F_{2}}\mid \text{.}  \label{M80}
\end{equation}%
Again, the case $\epsilon =-1$ cannot be realized, because the velocity $%
v_{1}=\mid \frac{\partial \overrightarrow{\mathbf{r}}_{1}}{\partial F_{1}}%
\mid $ becomes negative, but the case $\epsilon =+1$ leads to 
\begin{equation}
\mid \frac{\partial \overrightarrow{\mathbf{r}}_{2}}{\partial F_{2}}\mid =%
\frac{1}{(\sqrt{2}-1)}.\mid \frac{\partial \overrightarrow{\mathbf{r}}_{1}}{%
\partial F_{1}}\mid =(\sqrt{2}+1).\mid \frac{\partial \overrightarrow{%
\mathbf{r}}_{1}}{\partial F_{1}}\mid \text{.}  \label{M81}
\end{equation}
This equality is in contradiction with the inequality $\mid \frac{\partial 
\overrightarrow{\mathbf{r}}_{2}}{\partial F_{2}}\mid <\frac{1}{(\sqrt{2}-1)}%
.\mid \frac{\partial \overrightarrow{\mathbf{r}}_{1}}{\partial F_{1}}\mid $,
which is the first part of the inequalities chain (\ref{N36}) 
\begin{equation*}
(\sqrt{2}-1)\mid \frac{\partial \overrightarrow{\mathbf{r}}_{2}}{\partial
F_{2}}\mid <\mid \frac{\partial \overrightarrow{\mathbf{r}}_{1}}{\partial
F_{1}}\mid <\mid \frac{\partial \overrightarrow{\mathbf{r}}_{2}}{\partial
F_{2}}\mid <\sqrt{2}\mid \frac{\partial \overrightarrow{\mathbf{r}}_{2}}{%
\partial F_{2}}\mid \text{,}
\end{equation*}%
due to which the geodesic distance formulae (\ref{N6}) is not valid.
\bigskip The second part of inequality (\ref{N36}) can be written as 
\begin{equation}
\mid \frac{\partial \overrightarrow{\mathbf{r}}_{1}}{\partial F_{1}}\mid <%
\sqrt{2}(\sqrt{2}+1)\mid \frac{\partial \overrightarrow{\mathbf{r}}_{2}}{%
\partial F_{2}}\mid =(\sqrt{2}+2)\mid \frac{\partial \overrightarrow{\mathbf{%
r}}_{2}}{\partial F_{2}}\mid \text{ .}  \label{M82}
\end{equation}%
The geodesic distance formulae (\ref{N21}) will hold only when the
inequality 
\begin{equation}
(\sqrt{2}+1)\mid \frac{\partial \overrightarrow{\mathbf{r}}_{2}}{\partial
F_{2}}\mid =\mid \frac{\partial \overrightarrow{\mathbf{r}}_{1}}{\partial
F_{1}}\mid <(\sqrt{2}+2)\mid \frac{\partial \overrightarrow{\mathbf{r}}_{2}}{%
\partial F_{2}}\mid  \label{M83}
\end{equation}%
is valid. When 
\begin{equation}
\mid \frac{\partial \overrightarrow{\mathbf{r}}_{1}}{\partial F_{1}}\mid >(%
\sqrt{2}+1)\mid \frac{\partial \overrightarrow{\mathbf{r}}_{2}}{\partial
F_{2}}\mid \text{,}  \label{M84}
\end{equation}%
formulae (\ref{N21}) will not be valid.
\subsection{Second condition for the fulfillment of the compatibility
relation and derivation of an integral equation with respect to the
generalized velocity $\mid \frac{\partial \protect\overrightarrow{\mathbf{r}}%
_{1}}{\partial F_{1}}\mid $}
\label{sec:SecConditFulfCompatCondit}
\bigskip The compatibility condition (\ref{M72}) 
\begin{equation*}
\int D_{1}\frac{\partial T_{1}}{\partial F_{1}}S_{1}\left[ \frac{k_{1,2}}{%
\mid \frac{\partial \overrightarrow{\mathbf{r}}_{1}}{\partial F_{1}}\mid }-%
\frac{1}{\mid \frac{\partial \overrightarrow{\mathbf{r}}_{2}}{\partial F_{2}}%
\mid }\right] dF_{1}=0\text{ }
\end{equation*}%
will be satisfied when the following condition is fulfilled 
\begin{equation}
\mid \frac{\partial \overrightarrow{\mathbf{r}}_{2}}{\partial F_{2}}\mid =%
\frac{\int D_{1}\frac{\partial T_{1}}{\partial F_{1}}S_{1}dF_{1}}{\int
D_{1}\frac{\partial T_{1}}{\partial F_{1}}S_{1}\frac{k_{1,2}}{\mid \frac{%
\partial \overrightarrow{\mathbf{r}}_{1}}{\partial F_{1}}\mid }dF_{1}}\text{
.}  \label{M90}
\end{equation}%
This is a more general condition in comparison with the previously
investigated conditions (\ref{M74}) and (\ref{M78}), which can be considered
only as partial conditions.
\bigskip Relation (\ref{M90}) shall now be investigated for the first value (\ref{N4})
\begin{equation*}
k_{1}=k=\frac{\mid \frac{\partial \overrightarrow{\mathbf{r}}_{2}}{\partial
F_{2}}\mid +\mid \frac{\partial \overrightarrow{\mathbf{r}}_{1}}{\partial
F_{1}}\mid }{\mid \frac{\partial \overrightarrow{\mathbf{r}}_{1}}{\partial
F_{1}}\mid -\mid \frac{\partial \overrightarrow{\mathbf{r}}_{2}}{\partial
F_{2}}\mid }\text{ , }
\end{equation*}%
obtained from the equality (\ref{N3}) by taking the lower signs in the
nominator and in the denominator. Using the previous notations (\ref{N12})
for 
\begin{equation*}
v_{1}=\mid \frac{\partial \overrightarrow{\mathbf{r}}_{1}}{\partial F_{1}}%
\mid \text{, }v_{2}=\mid \frac{\partial \overrightarrow{\mathbf{r}}_{2}}{%
\partial F_{2}}\mid \text{,}
\end{equation*}%
relation (\ref{M90}) can be written in the form 
\begin{equation}
y=\frac{\int D_{1}\frac{\partial T_{1}}{\partial F_{1}}S_{1}dF_{1}}{\int
D_{1}\frac{\partial T_{1}}{\partial F_{1}}S_{1}\frac{(v_{1}+v_{2})}{%
v_{1}(v_{1}-v_{2})}dF_{1}}\text{ .}  \label{M91}
\end{equation}%
Denoting the expression in the denominator by $N(F_{1},v_{1},v_{2})$ and
using the simple equality 
\begin{equation}
\frac{1}{v_{1}(v_{1}-v_{2})}=\frac{1}{v_{2}}.\left( \frac{1}{v_{1}-v_{2}}-%
\frac{1}{v_{1}}\right) \text{ ,}  \label{M92}
\end{equation}%
it can be derived 
\begin{equation}
Nv_{2}=\int D_{1}\frac{\partial T_{1}}{\partial F_{1}}S_{1}dF_{1}\text{ ,}
\label{M93}
\end{equation}%
where 
\begin{equation}
N(F_{1},v_{1},v_{2}):=2v_{2}\int \frac{D_{1}\frac{\partial T_{1}}{\partial
F_{1}}S_{1}}{v_{1}-v_{2}}dF_{1}-y_{2}\int \frac{D_{1}\frac{\partial T_{1}}{%
\partial F_{1}}S_{1}}{v_{1}}dF_{1}\text{ .}  \label{M94}
\end{equation}%
Now introducing the notation 
\begin{equation}
f(F_{1})=D_{1}\frac{\partial T_{1}}{\partial F_{1}}S_{1}\text{ ,}
\label{M95}
\end{equation}%
equation (\ref{M93}) with account of the notation (\ref{M94}) for $%
N(F_{1},v_{1},v_{2})$ can be written in the form of an integral equation
with respect to the variable $v_{1}=\mid \frac{\partial \overrightarrow{%
\mathbf{r}}_{1}}{\partial F_{1}}\mid $ 
\begin{equation}
2\int \frac{f}{v_{1}-v_{2}}dF_{1}-\int \frac{f}{v_{1}}dF_{1}=\frac{1}{v_{2}%
}\int fdF_{1}\text{ .}  \label{M96}
\end{equation}
\subsection{Definition \protect\bigskip of the function $f$ $\ $ by means of
the derivatives of the Euclidean distance and the first radius-vector}
\label{sec:DerivRadius}
\bigskip Let us remind also that the function $D_{1}$ was defined by the formulae (%
\ref{M7A})
\begin{equation*}
D_{1}(F_{1},F_{2},\overrightarrow{\mathbf{r}}_{1},\overrightarrow{\mathbf{r}}%
_{2}):=\overrightarrow{\mathbf{r}}_{1}.\frac{\partial \overrightarrow{%
\mathbf{r}}_{1}}{\partial F_{1}}-\overrightarrow{\mathbf{r}}_{2}.\frac{%
\partial \overrightarrow{\mathbf{r}}_{1}}{\partial F_{1}}\text{. }
\end{equation*}%
Taking into account this formulae and also formulae (\ref{M10}) 
\begin{equation*}
\frac{\partial T_{1}}{\partial F_{1}}=\frac{1}{c}.\frac{1}{%
S_{1}(r_{1})}.\mid \frac{\partial \overrightarrow{\mathbf{r}}_{1}}{\partial
F_{1}}\mid \text{,}
\end{equation*}%
the function $f(F_{1})$ in (\ref{M95}) can be represented as%
\begin{equation}
f(F_{1})=D_{1}\frac{\partial T_{1}}{\partial F_{1}}S_{1}=\frac{1}{c}.\left( 
\overrightarrow{\mathbf{r}}_{1}.\frac{\partial \overrightarrow{\mathbf{r}}%
_{1}}{\partial F_{1}}-\overrightarrow{\mathbf{r}}_{2}.\frac{\partial 
\overrightarrow{\mathbf{r}}_{1}}{\partial F_{1}}\right) .\mid \frac{\partial 
\overrightarrow{\mathbf{r}}_{1}}{\partial F_{1}}\mid \text{ }  \label{M96A}
\end{equation}%
\begin{equation}
=\frac{1}{2c}.\left[ \frac{\partial }{\partial F_{1}}\left( \overrightarrow{%
\mathbf{r}}_{1}-\overrightarrow{\mathbf{r}}_{2}\right) ^{2}\right] .\mid 
\frac{\partial \overrightarrow{\mathbf{r}}_{1}}{\partial F_{1}}\mid =\frac{1%
}{2c}.\left[ \frac{\partial }{\partial F_{1}}R_{AB}^{2}\right] .\mid \frac{%
\partial \overrightarrow{\mathbf{r}}_{1}}{\partial F_{1}}\mid \text{,}
\label{M96B}
\end{equation}%
where 
\begin{equation}
R_{AB}^{2}=\left( \overrightarrow{\mathbf{r}}_{1}-\overrightarrow{\mathbf{r}}%
_{2}\right) ^{2}=(x_{1}-x_{2})^{2}+(y_{1}-y_{2})^{2}+(z_{1}-z_{2})^{2}
\label{M96C}
\end{equation}%
is the usual Euclidean distance between the two space points $%
(x_{1},y_{1},z_{1})$ and $(x_{2},y_{2},z_{2})$. Note also that since
according to our initial assumptions the two radius-vectors $\overrightarrow{%
\mathbf{r}}_{1}=\overrightarrow{\mathbf{r}}_{1}(F_{1})$ and $\overrightarrow{%
\mathbf{r}}_{2}=\overrightarrow{\mathbf{r}}_{2}(F_{2})$ depend only on the
generalized parameters $F_{1}$ and $F_{2}$ and thus the function $D_{1}$ in (%
\ref{M96A}) is equal to 
\begin{equation}
D_{1}=\left( \overrightarrow{\mathbf{r}}_{1}.\frac{\partial \overrightarrow{%
\mathbf{r}}_{1}}{\partial F_{1}}-\overrightarrow{\mathbf{r}}_{2}.\frac{%
\partial \overrightarrow{\mathbf{r}}_{1}}{\partial F_{1}}\right) =\frac{%
\partial }{\partial F_{1}}\left( \overrightarrow{\mathbf{r}}_{1}-%
\overrightarrow{\mathbf{r}}_{2}\right) ^{2}\text{,}  \label{M96D}
\end{equation}%
because the derivative $\frac{\partial }{\partial F_{1}}$ is acting only on
the radius-vector $\overrightarrow{\mathbf{r}}_{1}=\overrightarrow{\mathbf{r}%
}_{1}(F_{1})$. However, if the process of sending the signal by the
signal-emitting satellite and its reception by the second, signal-receiving
satellite depends on the mutual dependence of the two generalized parameters 
$F_{1}$ and $F_{2}$ (as previously clarified, for the partial case of motion
of the satellites on two elliptic orbits they may be defined as the
eccentric anomaly angles $E_{1}$ and $E_{2}$ for plane elliptic orbits or
the true anomaly angles $f_{1}$ and $f_{2}$ for space-oriented orbits), then
it is possible that the dependence of the radius-vectors will be more
complicated, i.e. $\overrightarrow{\mathbf{r}}_{1}=\overrightarrow{\mathbf{r}%
}_{1}(F_{1},F_{2})$ and $\overrightarrow{\mathbf{r}}_{2}=\overrightarrow{%
\mathbf{r}}_{2}(F_{1},F_{2})$. In a future publication it will be shown that
such a dependence will appear if a signal has to be transmitted along a
chain of $k-$satellites.
\bigskip Also, the expressions (\ref{M96A}) and (\ref{M96B}) do not depend on the
functions $S_{1}(r_{1})$ and $S_{2}(r_{2})$, defined by formulae (\ref{M11})
as 
\begin{equation*}
S_{1}(r_{1})=\sqrt[.]{\frac{c^{2}+2V_{1}}{c^{2}-2V_{1}}}\text{, }%
S_{2}(r_{2})=\sqrt[.]{\frac{c^{2}+2V_{2}}{c^{2}-2V_{2}}}\text{\ .}
\end{equation*}
\subsection{A partial solution of the integral equation}
\label{sec:PartSolut}
\bigskip A partial solution of the equation (\ref{M96}) can be found if the
expression in the brackets is zero for every choice of the function $%
f=f(F_{1})$ 
\begin{equation}
\int f.\left[ \frac{2}{v_{1}-v_{2}}-\frac{1}{v_{1}}-\frac{1}{v_{2}}\right]
dF_{1}=0\text{ . }  \label{M97}
\end{equation}%
We recover the already known partial solution (\ref{M76}) $v_{1}=(\sqrt[.]{2}%
+1)v_{2}$ (or $\mid \frac{\partial \overrightarrow{\mathbf{r}}_{2}}{\partial
F_{2}}\mid =(\sqrt{2}-1).\mid \frac{\partial \overrightarrow{\mathbf{r}}_{1}%
}{\partial F_{1}}\mid $). For this solution it was established that since$%
\mid \frac{\partial \overrightarrow{\mathbf{r}}_{1}}{\partial F_{1}}\mid
>\mid \frac{\partial \overrightarrow{\mathbf{r}}_{2}}{\partial F_{2}}\mid $,
the geodesic distance equation (\ref{N6}) cannot be valid, but instead,
geodesic equation (\ref{N21}) was proved to be properly defined (in the
meaning of being positive). Later, equation (\ref{N21}) can be transformed to
the equation  
\begin{equation*}
\widehat{R}_{AB}^{2}=\frac{c}{2\mid \frac{\partial \overrightarrow{\mathbf{r}%
}_{2}}{\partial F_{2}}\mid }.\int \frac{D_{1}S_{1}\frac{\partial T_{1}}{%
\partial F_{1}}\left[ 2\mid \frac{\partial \overrightarrow{\mathbf{r}}_{1}}{%
\partial F_{1}}\mid \mid \frac{\partial \overrightarrow{\mathbf{r}}_{2}}{%
\partial F_{2}}\mid +\mid \frac{\partial \overrightarrow{\mathbf{r}}_{1}}{%
\partial F_{1}}\mid ^{2}-\mid \frac{\partial \overrightarrow{\mathbf{r}}_{2}%
}{\partial F_{2}}\mid ^{2}\right] }{\left( \mid \frac{\partial 
\overrightarrow{\mathbf{r}}_{1}}{\partial F_{1}}\mid +\mid \frac{\partial 
\overrightarrow{\mathbf{r}}_{2}}{\partial F_{2}}\mid \right) }dF_{1}\text{ .} \label{M87}
\end{equation*}%
It is important to note that the integration in the last integral can be
changed to an integration along $v_{1}=\mid \frac{\partial \overrightarrow{%
\mathbf{r}}_{1}}{\partial F_{1}}\mid $ by using the simple formulae 
\begin{equation}
dF_{1}=\frac{\partial F_{1}}{\partial v_{1}}dv_{1}=\frac{1}{\frac{\partial
v_{1}}{\partial F_{1}}}dv_{1}\text{ ,}  \label{M98}
\end{equation}%
where 
\begin{equation*}
\frac{\partial v_{1}}{\partial F_{1}}=\frac{\partial }{\partial F_{1}}\sqrt[.%
]{\left( \frac{\partial x_{1}}{\partial F_{1}}\right) ^{2}+\left( \frac{%
\partial y_{1}}{\partial F_{1}}\right) ^{2}+\left( \frac{\partial z_{1}}{%
\partial F_{1}}\right) ^{2}}
\end{equation*}%
\begin{equation}
=\frac{\frac{\partial \overrightarrow{\mathbf{r}}_{1}}{\partial F_{1}}.\frac{%
\partial ^{2}\overrightarrow{\mathbf{r}}_{1}}{\partial F_{1}^{2}}}{\mid 
\frac{\partial \overrightarrow{\mathbf{r}}_{1}}{\partial F_{1}}\mid }\text{ .%
}  \label{M99}
\end{equation}%
Thus, we obtain the formulae for the change of the integration variable 
\begin{equation}
dF_{1}=\frac{\mid \frac{\partial \overrightarrow{\mathbf{r}}_{1}}{\partial
F_{1}}\mid }{\frac{\partial \overrightarrow{\mathbf{r}}_{1}}{\partial F_{1}}.%
\frac{\partial ^{2}\overrightarrow{\mathbf{r}}_{1}}{\partial F_{1}^{2}}}%
d\left( \mid \frac{\partial \overrightarrow{\mathbf{r}}_{1}}{\partial F_{1}}%
\mid \right) =\frac{v_{1}}{\frac{\partial \overrightarrow{\mathbf{r}}_{1}}{%
\partial F_{1}}.\frac{\partial ^{2}\overrightarrow{\mathbf{r}}_{1}}{\partial
F_{1}^{2}}}.dv_{1}\text{ .}  \label{M99A}
\end{equation}%
Clearly, the integral in (\ref{M87}) can be conveniently represented in
terms of a function, depending only on $\mid \frac{\partial \overrightarrow{%
\mathbf{r}}_{1}}{\partial F_{1}}\mid $ and $\mid \frac{\partial 
\overrightarrow{\mathbf{r}}_{2}}{\partial F_{2}}\mid $, if the function (\ref%
{M95}) $f(F_{1})=D_{1}\frac{\partial T_{1}}{\partial F_{1}}S_{1}$ can be
expressed also in terms of $\mid \frac{\partial \overrightarrow{\mathbf{r}}%
_{1}}{\partial F_{1}}\mid $ and $\mid \frac{\partial \overrightarrow{\mathbf{%
r}}_{2}}{\partial F_{2}}\mid $. Further it will be proved that this will be
possible to be performed if the integral equation (\ref{M97}) $\int f.\left[
\frac{2}{v_{1}-v_{2}}-\frac{1}{v_{1}}-\frac{1}{v_{2}}\right] dF_{1}=0$ is
solved with respect to the function $f(F_{1})$.
\subsection{A general solution of the integral equation}
\label{sec:GeneralSolut}
\bigskip Let us differentiate twice by the generalized variable $F_{1}$ the integral
equation (\ref{M97}). Furher the prime sign "$^{\prime \text{"}}$ will
denote the derivative with respect to the generalized dynamical parameter $%
F_{1}$: 
\begin{equation}
2\frac{f^{\prime }}{(v_{1}-v_{2})}-\frac{2f}{(v_{1}-v_{2})}.\frac{\partial
v_{1}}{\partial F_{1}}-\frac{f^{\prime }}{v_{1}}+\frac{f}{v_{1}^{2}}.\frac{%
\partial v_{1}}{\partial F_{1}}=\frac{1}{v_{2}}f^{\prime }\text{ .}
\label{M100}
\end{equation}%
This is a simple differential equation with respect to the function $%
f(F_{1}) $ 
\begin{equation}
L_{1}(v_{1},v_{2}).f^{\prime }+L_{2}(v_{1},v_{2},\frac{\partial v_{1}}{%
\partial F_{1}}).f=0\text{ ,}  \label{M101}
\end{equation}%
where the functions $L_{1}(v_{1},v_{2})$ and $L_{2}(v_{1},v_{2},\frac{%
\partial v_{1}}{\partial F_{1}})$ are defined as 
\begin{equation}
L_{1}(v_{1},v_{2}):=\frac{1}{v_{1}-v_{2}}-(\frac{1}{v_{1}}+\frac{1}{v_{2}})%
\text{ ,}  \label{M102}
\end{equation}%
\begin{equation}
L_{2}(v_{1},v_{2},\frac{\partial v_{1}}{\partial F_{1}}):=\frac{\partial
v_{1}}{\partial F_{1}}.\left( \frac{1}{v_{1}^{2}}-\frac{2}{v_{1}-v_{2}}%
\right) \text{ .}  \label{M103}
\end{equation}%
The solution of the differential equation (\ref{M101}) is given by the
following simple formulae 
\begin{equation}
f(v_{1},v_{2},\frac{\partial v_{1}}{\partial F_{1}})=\int \exp \left( -%
\frac{\partial v_{1}}{\partial F_{1}}.\frac{v_{2}}{v_{1}}.\frac{%
(v_{1}-v_{2}-2v_{1}^{2})}{(v_{1}v_{2}-v_{1}^{2}+v_{2}^{2})}\right) dF_{1}%
\text{ .}  \label{M104}
\end{equation}%
If formulaes (\ref{M99}) for $\frac{\partial v_{1}}{\partial F_{1}}=\frac{%
\frac{\partial \overrightarrow{\mathbf{r}}_{1}}{\partial F_{1}}.\frac{%
\partial ^{2}\overrightarrow{\mathbf{r}}_{1}}{\partial F_{1}^{2}}}{\mid 
\frac{\partial \overrightarrow{\mathbf{r}}_{1}}{\partial F_{1}}\mid }=\frac{%
\frac{\partial \overrightarrow{\mathbf{r}}_{1}}{\partial F_{1}}.\frac{%
\partial ^{2}\overrightarrow{\mathbf{r}}_{1}}{\partial F_{1}^{2}}}{v_{1}}$%
and (\ref{M99A}) for $dF_{1}=\frac{v_{1}}{\frac{\partial \overrightarrow{%
\mathbf{r}}_{1}}{\partial F_{1}}.\frac{\partial ^{2}\overrightarrow{\mathbf{r%
}}_{1}}{\partial F_{1}^{2}}}.dv_{1}$ are taken into account, expression (\ref%
{M104}) can be rewritten as 
\begin{equation}
f(v_{1},v_{2},\frac{\partial \overrightarrow{\mathbf{r}}_{1}}{\partial F_{1}}%
.\frac{\partial ^{2}\overrightarrow{\mathbf{r}}_{1}}{\partial F_{1}^{2}}%
)=\int \exp \left[ -K_{F_{1}}.\frac{v_{2}}{v_{1}^{2}}.\frac{%
(v_{1}-v_{2}-2v_{1}^{2})}{(v_{1}v_{2}-v_{1}^{2}+v_{2}^{2})}\right] \frac{%
v_{1}}{K_{F_{1}}}.dv_{1}\text{ ,}  \label{M105}
\end{equation}%
where $K_{F_{1}}$ denotes the function (scalar product of $\frac{\partial 
\overrightarrow{\mathbf{r}}_{1}}{\partial F_{1}}$ and $\frac{\partial ^{2}%
\overrightarrow{\mathbf{r}}_{1}}{\partial F_{1}^{2}}$) 
\begin{equation}
K_{F_{1}}\equiv \frac{\partial \overrightarrow{\mathbf{r}_{1}}}{\partial
F_{1}}.\frac{\partial ^{2}\overrightarrow{\mathbf{r}_{1}}\text{ }}{\partial
F_{1}^{2}}\text{ .}  \label{M106}
\end{equation}
In fact, $K_{F_{1}}$ represents the scalar product of the components of the
velocity vector $\frac{\partial \overrightarrow{\mathbf{r}}_{1}}{\partial
F_{1}}$ and the acceleration vector $\frac{\partial ^{2}\overrightarrow{%
\mathbf{r}}_{1}}{\partial F_{1}^{2}}$ and since the two vectors are
orthogonal, the expression for $K_{F_{1}}$ in (\ref{M106}) should be set up
to be equal to zero. However, this is true only if the parameter $F_{1}$ is
a natural parameter for the curve $\overrightarrow{\mathbf{r}_{1}}=%
\overrightarrow{\mathbf{r}_{1}}(F_{1})$.
\bigskip In the subsequent section a singular integral equation will be derived, and
the singularity will depend not only on $K_{F_{1}}$, but also on
polynomials, which include the generalized velocities of the first satellite
and of the second satellite, defined by the expressions in formulae (\ref%
{N12}) 
\begin{equation*}
v_{1}=\mid \frac{\partial \overrightarrow{\mathbf{r}}_{1}}{\partial F_{1}}%
\mid \text{, }v_{2}=\mid \frac{\partial \overrightarrow{\mathbf{r}}_{2}}{%
\partial F_{2}}\mid \text{ }
\end{equation*}%
This is the reason why the expression for $K_{F_{1}}$ will not be set up
equal to zero.
\subsection{Examples for plane elliptic and space-oriented orbits, when the
expression for $K_{F_{1}}$ is different from zero. Physical meaning of the
function $K_{F_{1}}$}
\label{sec:KnotZero}
\bigskip In fact, $K$ will be equal to zero only in case if the parameter $F_{1}$ is
a natural parameter for the curve $\frac{\partial \overrightarrow{\mathbf{r}}%
_{1}}{\partial F_{1}}$. Then 
\begin{equation}
\left( \frac{\partial \overrightarrow{\mathbf{r}}_{1}}{\partial F_{1}}%
\right) ^{2}=const\text{ (or is equal to }1\text{)}  \label{M106A1}
\end{equation}%
and after differentiation it will give $K=\frac{\partial \overrightarrow{%
\mathbf{r}_{1}}}{\partial F_{1}}.\frac{\partial ^{2}\overrightarrow{\mathbf{r%
}_{1}}\text{ }}{\partial F_{1}^{2}}=0$. Since we are creating a theory for
inter-satellite communications between satellites, moving on generally
determined space orbits, such a theory should be correctly defined also for
the partial cases of plane elliptic orbits and space-oriented orbits.
\bigskip This means that for both these cases it should be found when the expression
for $K_{F_{1}}$ will be different from zero or equal to zero.
\bigskip If the generalized parameter $F_{1}$ is the eccentric anomaly angle $E$ and
the plane $x-y$ orbital coordinates are parametrized as \ (\ref{ABC3}) \ (we
are omitting the indice $1$) 
\begin{equation*}
x_{.}=a_{.}(\cos E_{.}-e_{.})\text{, }y_{.}=a_{.}\sqrt[.]{1-e_{.}^{2}}\sin
E_{.}\text{,}
\end{equation*}%
then the generalized velocity $v_{E}=\sqrt[.]{\left( \frac{\partial x}{%
\partial E}\right) ^{2}+\left( \frac{\partial x}{\partial E}\right) ^{2}}$
has been calculated in the papers \cite{Dimitrov2017} and \cite{Dimitrov2025}
to be 
\begin{equation}
v_{E}=\frac{na}{\sqrt[.]{1-e\cos E}}\sqrt[.]{1-e^{2}\cos ^{2}E}\text{. }
\label{M106A2}
\end{equation}%
Consequently, the eccentric anomaly angle $E$ in the parametric
representation of the curve (\ref{ABC3}) is not a natural parameter.
\bigskip Now we shall find a simple relation between the generalized velocity (not
only in the form of $v_{E}$ (\ref{M106A2})) and the function $K_{F_{1}}$,
after representing it in the form 
\begin{equation}
K_{F_{1}}=\frac{\partial \overrightarrow{\mathbf{r}_{1}}}{\partial F_{1}}.%
\frac{\partial ^{2}\overrightarrow{\mathbf{r}_{1}}\text{ }}{\partial
F_{1}^{2}}=\frac{1}{2}\frac{\partial }{\partial F_{1}}\left( \frac{\partial 
\overrightarrow{\mathbf{r}_{1}}}{\partial F_{1}}\right) ^{2}  \label{M106AA3}
\end{equation}%
\begin{equation}
=\frac{1}{2}\frac{\partial }{\partial F_{1}}\left[ \left( \frac{\partial
x_{1}}{\partial F_{1}}\right) ^{2}+\left( \frac{\partial y_{1}}{\partial
F_{1}}\right) ^{2}+\left( \frac{\partial z_{1}}{\partial F_{1}}\right) ^{2}%
\right] =\frac{1}{2}\frac{\partial v_{F_{1}}^{2}}{\partial F_{1}}\text{ .}
\label{M106AA4}
\end{equation}%
Thus, the function $K_{F_{1}}$ is equal to one half of the change of the
square of the velocity with respect to the chosen dynamical parameter $F_{1}$%
. Consequently, if the square of the velocity $v_{F_{1}}^{2}$ is constant,
the function $K_{F_{1}}$ according to the above formulae will be equal to
zero.
\section{Transforming the geodesic distance equation and finding a relation
between the geodesic distance and the Euclidean distance}
\label{sec:RelatGeodEuclid}
\bigskip The purpose of this section will be to show that because of the relation
between the Euclidean distance $R_{AB}^{2}$ and the function $f(F_{1})$ in (%
\ref{M96B}) 
\begin{equation*}
f(F_{1})=D_{1}S_{1}\frac{\partial T_{1}}{\partial F_{1}}=\frac{1}{2c}.\left[ 
\frac{\partial }{\partial F_{1}}\left( \overrightarrow{\mathbf{r}}_{1}-%
\overrightarrow{\mathbf{r}}_{2}\right) ^{2}\right] .\mid \frac{\partial 
\overrightarrow{\mathbf{r}}_{1}}{\partial F_{1}}\mid
\end{equation*}%
\begin{equation*}
=\frac{1}{2c}.\left[ \frac{\partial }{\partial F_{1}}R_{AB}^{2}\right] .\mid 
\frac{\partial \overrightarrow{\mathbf{r}}_{1}}{\partial F_{1}}\mid \text{,}
\end{equation*}%
the geodesic distance equation (\ref{N21}) 
\begin{equation*}
\widehat{R}_{AB}^{2}=\frac{c}{2\mid \frac{\partial \overrightarrow{\mathbf{r}%
}_{2}}{\partial F_{2}}\mid }\int \frac{D_{1}S_{1}\frac{\partial T_{1}}{%
\partial F_{1}}.\left[ 2\mid \frac{\partial \overrightarrow{\mathbf{r}}_{1}}{%
\partial F_{1}}\mid \mid \frac{\partial \overrightarrow{\mathbf{r}}_{2}}{%
\partial F_{2}}\mid +\mid \frac{\partial \overrightarrow{\mathbf{r}}_{1}}{%
\partial F_{1}}\mid ^{2}-\mid \frac{\partial \overrightarrow{\mathbf{r}}_{2}%
}{\partial F_{2}}\mid ^{2}\right] }{\left( \mid \frac{\partial 
\overrightarrow{\mathbf{r}}_{1}}{\partial F_{1}}\mid +\mid \frac{\partial 
\overrightarrow{\mathbf{r}}_{2}}{\partial F_{2}}\mid \right) }dF_{1}\text{ }
\end{equation*}%
will give the relation between the Euclidean distance $R_{AB}^{2}$ and the
geodesic distance $\widehat{R}_{AB}^{2}$.
\bigskip Making use of formulae (\ref{M96B}), introducing the previous notations $%
\mid \frac{\partial \overrightarrow{\mathbf{r}}_{1}}{\partial F_{1}}\mid
=v_{1}$ and $\mid \frac{\partial \overrightarrow{\mathbf{r}}_{2}}{\partial
F_{2}}\mid =v_{2}$ and also defining the function 
\begin{equation*}
\widetilde{K}(v_{1},v_{2}):=\frac{\left[ 2\mid \frac{\partial 
\overrightarrow{\mathbf{r}}_{1}}{\partial F_{1}}\mid \mid \frac{\partial 
\overrightarrow{\mathbf{r}}_{2}}{\partial F_{2}}\mid +\mid \frac{\partial 
\overrightarrow{\mathbf{r}}_{1}}{\partial F_{1}}\mid ^{2}-\mid \frac{%
\partial \overrightarrow{\mathbf{r}}_{2}}{\partial F_{2}}\mid ^{2}\right] }{%
\left( \mid \frac{\partial \overrightarrow{\mathbf{r}}_{1}}{\partial F_{1}}%
\mid +\mid \frac{\partial \overrightarrow{\mathbf{r}}_{2}}{\partial F_{2}}%
\mid \right) }
\end{equation*}%
\begin{equation}
=\frac{2v_{1}v_{2}+v_{1}^{2}-v_{2}^{2}}{v_{1}+v_{2}}\text{ ,}  \label{M107A}
\end{equation}%
the geodesic distance (\ref{M87}) can be written in a form, not dependent
on the velocity of light $c$ 
\begin{equation}
\widehat{R}_{AB}^{2}=\frac{1}{4.v_{2}}\int \left[ \frac{\partial }{\partial
F_{1}}(R_{AB}^{2})\right] .v_{1}.\widetilde{K}(v_{1},v_{2})dF_{1}\text{ .}
\label{M108}
\end{equation}%
Making use of the relation (\ref{M99A}) $dF_{1}=\frac{v_{1}}{\frac{\partial 
\overrightarrow{\mathbf{r}}_{1}}{\partial F_{1}}.\frac{\partial ^{2}%
\overrightarrow{\mathbf{r}}_{1}}{\partial F_{1}^{2}}}.dv_{1}=\frac{v_{1}}{%
K_{F_{1}}}dv_{1}$, the above relation can be represented also as 
\begin{equation}
\widehat{R}_{AB}^{2}=\frac{1}{4v_{2}}\int \left[ \frac{\partial }{\partial
F_{1}}(R_{AB}^{2})\right] .v_{1}^{2}.\frac{\widetilde{K}(v_{1},v_{2})}{%
K_{F_{1}}}dv_{1}\text{ .}  \label{M109}
\end{equation}%
In this form, the geodesic distance $\widehat{R}_{AB}^{2}$ is related to the
non-inertiality of the reference system, attached to the first
signal-emitting satellite. This non-inertiality is evident from the presence
of the function $K_{F_{1}}$(\ref{M106AA4}). Note however that a function $%
K_{F_{2}}=\frac{1}{2}\frac{\partial v_{F_{2}}^{2}}{\partial F_{2}}$ is not
present in formulae (\ref{M109}) and moreover, the function $\widetilde{K}%
(v_{1},v_{2})$ in (\ref{M107A}) is not symmetric with respect to the
interchange $v_{1}\Leftrightarrow v_{2}$ of the velocities of the two
satellites. Consequently 
\begin{equation}
\widetilde{K}(v_{2},v_{1})=\frac{2v_{1}v_{2}+v_{2}^{2}-v_{1}^{2}}{v_{1}+v_{2}%
}\neq \widetilde{K}(v_{1},v_{2})\text{ }  \label{M107B}
\end{equation}%
and the defined geodesic distance $\widehat{R}_{BA}^{2}$ in the "backward"
direction ($B\rightarrow A$) 
\begin{equation}
\widehat{R}_{BA}^{2}=\frac{1}{4v_{1}}\int \left[ \frac{\partial }{\partial
F_{2}}(R_{AB}^{2})\right] .v_{2}^{2}.\frac{\widetilde{K}(v_{2},v_{1})}{%
K_{F_{2}}}dv_{2}\text{ }  \label{M109B}
\end{equation}%
will be entirely different from the geodesic distance $\widehat{R}_{AB}^{2}$
(\ref{M109}) in the "forward" direction $A\rightarrow B$. Formulae (\ref{M87}%
) for the geodesic distance $\widehat{R}_{AB}^{2}$ can be written also in a
form, where it depends only on the derivative $\frac{\partial T_{1}}{%
\partial F_{1}}$ of the first propagation time and also on $D_{1}S_{1}$ 
\begin{equation}
\widehat{R}_{AB}^{2}=\frac{c}{2v_{2}}\int D_{1}S_{1}\frac{\partial T_{1}}{%
\partial F_{1}}\left[ \frac{2v_{1}v_{2}+v_{1}^{2}-v_{2}^{2}}{v_{1}+v_{2}}%
\right] dF_{1}\text{ . }  \label{M109C}
\end{equation}%
Thus, if the derivative $\frac{\partial T_{1}}{\partial F_{1}}$ is known and
also the two velocities $v_{1}$ and $v_{2}$ at each point on the two space
trajectories, the \ geodesic distance $\widehat{R}_{AB}^{2}$ in the
"forward" direction can be calculated without knowing the acceleration $%
\frac{\partial v_{1}}{\partial F_{1}}$. Thus the two equivalent expressions (%
\ref{M109}) and (\ref{M109C}) suggest the physical interpretation that the
derivative $\frac{\partial T_{1}}{\partial F_{1}}$ is related to the
non-inertiality of the first reference system, attached to the first
satellite.
\section{Appendix A: Coefficient functions $\protect\widetilde{A}_{1}$, $%
\protect\widetilde{A}_{2}$ and $\protect\widetilde{A}_{3\text{ }}$in the
expression for the differential of the second propagation time as a function
of the differential of the first propagation time}
\label{sec:AppA}
\bigskip The coefficient function $\widetilde{A}_{1}$ in the expression (\ref{DOPC7}) 
$dT_{2}=\frac{1}{c}\sqrt[.]{\widetilde{A}_{1}(dT_{1})^{2}+\widetilde{A}%
_{2}(dT_{1})+\widetilde{A}_{3}}$ \ for the differential $dT_{2}$ is of the
following form 
\begin{equation*}
\widetilde{A}_{1}:=c^{2}\left( \frac{\widetilde{R}_{2}}{\widetilde{R}_{1}}%
\right) ^{2}.\left( \frac{\widetilde{S}_{1}}{\widetilde{S}_{2}}\right) ^{2}
\end{equation*}%
\begin{equation*}
+\left( \frac{c^{2}-2V_{2}}{c^{2}+2V_{2}}\right) \left( \frac{\partial z_{2}%
}{\partial F_{1}}\right) ^{2}\left( \frac{1}{\widetilde{R}_{1}}\right) ^{2}
\end{equation*}%
\begin{equation}
+2\left( \frac{c^{2}-2V_{2}}{c^{2}+2V_{2}}\right) \left( \frac{\partial z_{2}%
}{\partial F_{1}}\right) ^{2}\left( \frac{\partial z_{2}}{\partial F_{2}}%
\right) ^{2}\left( \frac{\widetilde{S}_{1}}{\widetilde{S}_{2}}\right) \left( 
\frac{1}{\widetilde{R}_{1}}\right) ^{2}\text{.}  \label{GGGG1}
\end{equation}%
We remind that the expressions for $\widetilde{S}_{i}$ $(i=1,2)$ are defined
by the formulae (\ref{DOPC5}) 
\begin{equation*}
\widetilde{S}_{i}:=\left( \frac{\partial z_{i}}{\partial F_{i}}\right) +%
\frac{(x_{1}-x_{2})}{(z_{1}-z_{2})}\frac{\partial x_{i}}{\partial F_{i}}
\end{equation*}%
\begin{equation*}
+\frac{(y_{1}-y_{2})}{(z_{1}-z_{2})}\frac{\partial y_{i}}{\partial F_{i}}%
\text{.}
\end{equation*}%
The expression for $\widetilde{R}_{i}$ $(i=1,2)$ is defined by the formulae (%
\ref{DOPC6}) 
\begin{equation*}
\widetilde{R}_{i}:=\frac{1}{c}\left( \frac{c^{2}-2V_{i}}{c^{2}+2V_{i}}%
\right) \left[ \left( \frac{\partial x_{i}}{\partial F_{i}}\right)
^{2}+\left( \frac{\partial y_{i}}{\partial F_{i}}\right) ^{2}+\left( \frac{%
\partial z_{i}}{\partial F_{i}}\right) ^{2}\right] \text{.}
\end{equation*}%
The dependence of the coordinate $z_{2}$ on both the generalized parameters 
$F_{1}$ and $F_{2}$ can be seen, if from the hyperplane equation (\ref{ABC2A22}%
) $dz_{2}$ is expressed as 
\begin{equation}
dz_{2}=dz_{1}+\frac{F}{2(z_{1}-z_{2})}\text{,}  \label{GGGG2}
\end{equation}%
where expression $F$ is 
\begin{equation*}
F:=2(x_{1}-x_{2})dx_{1}-2(x_{1}-x_{2})dx_{2}
\end{equation*}%
\begin{equation}
+2(y_{1}-y_{2})dy_{1}-2(y_{1}-y_{2})dy_{2}-dR_{AB}^{2}\text{.}  \label{GGGG3}
\end{equation}%
The coefficient function $\widetilde{A}_{2}$ is given by the expression 
\begin{equation*}
\widetilde{A}_{2}:=-c^{2}\left( \frac{\widetilde{R}_{2}}{\widetilde{S}_{2}}%
\right) ^{2}.\left( \frac{\widetilde{S}_{1}}{\widetilde{R}_{1}(z_{1}-z_{2})}%
\right) .dR_{AB}^{2}
\end{equation*}%
\begin{equation}
-\left( \frac{c^{2}-2V_{2}}{c^{2}+2V_{2}}\right) .\left( \frac{\partial z_{2}%
}{\partial F_{1}}\right) .\left( \frac{\partial z_{2}}{\partial F_{2}}%
\right) \frac{dR_{AB}^{2}}{\widetilde{R}_{1}\widetilde{S}_{2}}\text{.}
\label{GGGG4}
\end{equation}%
The coefficient function $\widetilde{A}_{3}$ is given by 
\begin{equation}
\widetilde{A}_{3}:=c^{2}\left( \frac{\widetilde{R}_{2}}{\widetilde{S}_{2}}%
\right) ^{2}.\frac{\left( dR_{AB}^{2}\right) ^{2}}{4(z_{1}-z_{2})^{2}}\text{.%
}  \label{GGGG5}
\end{equation}%
The coefficient functions $Q_{1}(E_{1},E_{2})$, $Q_{2}(E_{1},E_{2})$, $%
Q_{3}(E_{1},E_{2})$ and $Q_{4}(E_{1},E_{2})$ in the cubic equation (\ref{ABC19}) have
the following form 
\begin{equation}
Q_{1}(E_{1},E_{2}):=\frac{P_{1}(E_{1})}{P_{2}(E_{2})}\frac{\partial
T_{1}(E_{1})}{\partial E_{1}}\text{,}  \label{BBB1}
\end{equation}%
\begin{equation}
Q_{2}(E_{1},E_{2}):=\left( \frac{\partial T_{2}(E_{1},E_{2})}{\partial E_{2}}%
\right) ^{2}  \notag \\
\end{equation}%
\begin{equation}
\\
+2\frac{P_{1}(E_{1})}{P_{2}(E_{2})}\frac{\partial T_{1}(E_{1})}{\partial
E_{1}}\frac{\partial T_{2}(E_{1},E_{2})}{\partial E_{2}}  \notag \\
\end{equation}%
%
%
%
%
%
%
%
\begin{equation}
\\
-\frac{1}{P_{2}(E_{2})}\frac{\partial R_{AB}^{2}}{\partial E_{2}}-\frac{%
S_{2}(E_{1},E_{2})}{P_{2}(E_{2})}\text{ ,}  \label{BBB2}
\end{equation}%
\begin{equation*}
Q_{3}(E_{1},E_{2}):=\frac{P_{1}(E_{1})}{P_{2}(E_{2})}\left( \frac{\partial
T_{1}(E_{1})}{\partial E_{1}}\right) ^{2}
\end{equation*}%
\begin{equation}
-\frac{1}{P_{2}(E_{2})}\frac{\partial R_{AB}^{2}}{\partial E_{1}}-\frac{%
S_{1}(E_{1},E_{2})}{P_{2}(E_{2})}\text{,}  \label{BBB3}
\end{equation}%
%
%
%
%
%
%
%
\begin{equation}
Q_{4}(E_{1},E_{2}):=\left( \frac{\partial T_{2}(E_{1},E_{2})}{\partial E_{1}}%
\right) ^{2}\text{.}  \label{BBB4}
\end{equation}
\bigskip The coefficient functions $G_{1}(E_{1}.E_{2})$ and $G_{2}(E_{1},E_{2})$ in (%
\ref{ABC22}) have the following form 
\begin{equation*}
G_{1}(E_{1}.E_{2}):=\frac{1}{2}\frac{S_{2}(E_{1},E_{2})}{P_{2}(E_{2})}
\end{equation*}%
\begin{equation}
+\left( \frac{P_{1}(E_{1})}{2P_{2}(E_{2})}\right) ^{2}\left( \frac{\partial
T_{1}(E_{1})}{\partial E_{1}}\right) ^{2}-\frac{P_{1}(E_{1})}{2P_{2}(E_{2})}%
\frac{\partial T_{1}(E_{1})}{\partial E_{1}}\text{,}  \label{BBB5}
\end{equation}%
\begin{equation*}
G_{2}(E_{1},E_{2}):=\frac{1}{2}\frac{S_{1}(E_{1},E_{2})}{P_{2}(E_{2})}
\end{equation*}%
\begin{equation}
+\frac{1}{P_{2}(E_{2})}\frac{\partial R_{AB}^{2}}{\partial E_{1}}-\frac{%
P_{1}(E_{1})}{2P_{2}(E_{2})}\left( \frac{\partial T_{1}(E_{1})}{\partial
E_{1}}\right) ^{2}\text{.}  \label{BBB6}
\end{equation}%
The function $\overline{G}_{1}(E_{1}.E_{2})$ in (\ref{ABC24}) is obtained
from (\ref{BBB5}) by interchanging the indices $1$ and $2$ 
\begin{equation*}
\overline{G}_{1}(E_{1}.E_{2}):=\frac{1}{2}\frac{S_{1}(E_{1},E_{2})}{%
P_{1}(E_{1})}
\end{equation*}%
\begin{equation*}
+\left( \frac{P_{2}(E_{2})}{2P_{1}(E_{1})}\right) ^{2}\left( \frac{\partial
T_{2}(E_{1},E_{2})}{\partial E_{2}}\right) ^{2}
\end{equation*}%
\begin{equation}
-\frac{P_{2}(E_{2})}{2P_{1}(E_{1})}\frac{\partial T_{2}(E_{1},E_{2})}{%
\partial E_{2}}\text{.}  \label{BBB7}
\end{equation}%
The function $\overline{G}_{2}(E_{1}.E_{2})$ can be obtained in an analogous
way from (\ref{BBB6}).
\bigskip The function $N$ in (\ref{ABC25}) can be written as follows 
\begin{equation*}
N:=\frac{1}{4}\left( \frac{dE_{1}}{dE_{2}}\right) ^{2}\left( 1+4\left( \frac{%
\partial T_{1}}{\partial E_{2}}\right) \right) ^{2}
\end{equation*}%
\begin{equation*}
+\frac{S_{1}(E_{1},E_{2})}{P_{2}(E_{2})}\left( \frac{dE_{1}}{dE_{2}}\right)
^{2}+\frac{S_{2}(E_{1},E_{2})}{P_{2}(E_{2})}
\end{equation*}%
\begin{equation}
+\frac{1}{P_{2}(E_{2})}\frac{\partial R_{AB}^{2}}{\partial E_{2}}-\frac{%
2P_{1}(E_{1})}{P_{2}(E_{2})}\frac{dE_{1}}{dE_{2}}\left( \frac{\partial
T_{1}(E_{1})}{\partial E_{1}}\right) ^{2}\text{.}  \label{BBB8}
\end{equation}
\section{Appendix B: Derivation of the expression for the space-time
distance $\protect\widetilde{R}_{AB}^{2}(1\rightarrow 2)$}
\bigskip Setting up the expression (\ref{M31}) equal to the second equation for $%
\frac{\partial \widetilde{R}_{AB}^{2}(1\rightarrow 2)}{\partial F_{2}}$ in (%
\ref{M30}), the following expression for $\frac{\partial \Psi (F_{2})}{%
\partial F_{2}}$ can be obtained%
\begin{equation}
\frac{\partial \Psi (F_{2})}{\partial F_{2}}=\pm \frac{cD_{2}(\frac{\partial
T_{2}}{\partial F_{2}})S_{2}(r_{2})}{\mid \frac{\partial \overrightarrow{%
\mathbf{r}}_{1}}{\partial F_{1}}\mid }-c\int \frac{\frac{\partial D_{1}}{%
\partial F_{2}}(\frac{\partial T_{1}}{\partial F_{1}})S_{1}(r_{1})}{\mid 
\frac{\partial \overrightarrow{\mathbf{r}}_{1}}{\partial F_{1}}\mid }dF_{1}%
\text{.\ \ }  \label{MM1}
\end{equation}%
After integrating, the following expression is obtained for the function $%
\Psi (F_{2})$ 
\begin{equation}
\Psi (F_{2})=\pm c\int \frac{D_{2}(\frac{\partial T_{2}}{\partial F_{2}}%
)S_{2}(r_{2})}{\mid \frac{\partial \overrightarrow{\mathbf{r}}_{1}}{\partial
F_{1}}\mid }dF_{2}-c\iint \frac{\frac{\partial D_{1}}{\partial F_{2}}(\frac{%
\partial T_{1}}{\partial F_{1}})S_{1}(r_{1})}{\mid \frac{\partial 
\overrightarrow{\mathbf{r}}_{1}}{\partial F_{1}}\mid }dF_{1}dF_{2}\text{.}
\label{MM2}
\end{equation}%
The first term in the above equality can be transformed into the following
sum after performing an integration by parts 
\begin{equation}
\pm c\frac{D_{2}S_{2}(r_{2})T_{2}}{\mid \frac{\partial \overrightarrow{%
\mathbf{r}}_{1}}{\partial F_{1}}\mid }\mp \frac{c}{\mid \frac{\partial 
\overrightarrow{\mathbf{r}}_{1}}{\partial F_{1}}\mid }\int T_{2}\frac{%
\partial }{\partial F_{2}}\left[ S_{2}(r_{2})D_{2}\right] dF_{2}\text{.}
\label{MM3}
\end{equation}%
The second term in (\ref{MM2}) after integration by parts with respect to $%
F_{2}$ and $F_{1}$ can be transformed to the sum%
\begin{equation}
-c\frac{D_{1}S_{1}(r_{1})T_{1}}{\mid \frac{\partial \overrightarrow{\mathbf{r%
}}_{1}}{\partial F_{1}}\mid }+c\int T_{1}\frac{\partial }{\partial F_{1}}%
\left[ \frac{S_{1}(r_{1})D_{2}}{\mid \frac{\partial \overrightarrow{\mathbf{r%
}}_{1}}{\partial F_{1}}\mid }\right] dF_{1}\text{ .}  \label{MM4}
\end{equation}%
Summing up all the four terms in (\ref{MM3}) and (\ref{MM4}), one can obtain
the function $\Psi (F_{2})$. After integrating the first equation in (\ref%
{M30}), performing integration by parts with respect to the variable $F_{1}$
and adding the four terms in the expression for $\Psi (F_{2})$, one can
obtain the following expression for $\widetilde{R}_{AB}^{2}$ \ 
\begin{equation}
\widetilde{R}_{AB}^{2}=c\frac{D_{1}S_{1}(r_{1})T_{1}}{\mid \frac{\partial 
\overrightarrow{\mathbf{r}}_{1}}{\partial F_{1}}\mid }-c\int T_{1}\frac{%
\partial }{\partial F_{1}}\left[ \frac{S_{1}(r_{1})D_{2}}{\mid \frac{%
\partial \overrightarrow{\mathbf{r}}_{1}}{\partial F_{1}}\mid }\right]
dF_{1}+\Psi (F_{2})  \label{MM5}
\end{equation}
\begin{equation}
=c\frac{D_{1}S_{1}(r_{1})T_{1}}{\mid \frac{\partial \overrightarrow{\mathbf{r%
}}_{1}}{\partial F_{1}}\mid }-c\int T_{1}\frac{\partial }{\partial F_{1}}%
\left[ \frac{S_{1}(r_{1})D_{1}}{\mid \frac{\partial \overrightarrow{\mathbf{r%
}}_{1}}{\partial F_{1}}\mid }\right] dF_{1}  \label{MM6}
\end{equation}%
\begin{equation}
\pm c\frac{D_{2}S_{2}(r_{2})T_{2}}{\mid \frac{\partial \overrightarrow{%
\mathbf{r}}_{1}}{\partial F_{1}}\mid }\mp c\int T_{2}\frac{\partial }{%
\partial F_{2}}\left[ \frac{S_{2}(r_{2})D_{2}}{\mid \frac{\partial 
\overrightarrow{\mathbf{r}}_{1}}{\partial F_{1}}\mid }\right] dF_{2}
\label{MM7}
\end{equation}%
\begin{equation}
-c\frac{D_{1}S_{1}(r_{1})T_{1}}{\mid \frac{\partial \overrightarrow{\mathbf{r%
}}_{1}}{\partial F_{1}}\mid }+c\int T_{1}\frac{\partial }{\partial F_{1}}%
\left[ \frac{S_{1}(r_{1})D_{1}}{\mid \frac{\partial \overrightarrow{\mathbf{r%
}}_{1}}{\partial F_{1}}\mid }\right] dF_{1}\text{.}  \label{MM8}
\end{equation}%
All the terms in the decomposition for $\widetilde{R}_{AB}^{2}$ cancel with
the exception of the third and the fourth in (\ref{MM7}). After performing
the integration by parts in the fourth term, the following expression is
obtained for the space-time distance$\widetilde{\text{ }R}_{AB}^{2}$ 
\begin{equation}
\widetilde{R}_{AB}^{2}(1\rightarrow 2)=\pm c\int S_{2}(r_{2})\frac{D_{2}}{%
\mid \frac{\partial \overrightarrow{\mathbf{r}}_{1}}{\partial F_{1}}\mid }%
\frac{\partial T_{2}}{\partial F_{2}}dF_{2}\text{.\ \ }  \label{MM9}
\end{equation}%
This expression is consistently derived because it can be obtained also
after integration by parts of the first integral equation (\ref{M37}).
Analogously, after integrating by parts of the second integral equation (\ref%
{M38}), one can obtain a similar equality for the space-time distance $%
\widetilde{R}_{AB}^{2}(1\rightarrow 2)$ with interchanged indices $%
1\Leftrightarrow 2$ 
\begin{equation}
\widetilde{R}_{BA}^{2}(2\rightarrow 1)=\pm c\int S_{1}(r_{1})\frac{D_{1}}{%
\mid \frac{\partial \overrightarrow{\mathbf{r}}_{2}}{\partial F_{2}}\mid }%
\frac{\partial T_{1}}{\partial F_{1}}dF_{1}\text{.\ \ }  \label{MM10}
\end{equation}%
Now compare the right-hand sides of (\ref{MM9}) and (\ref{MM10}), which are
equal according to the condition for inter-satellite communications (\ref{M36}%
) $\int \frac{D_{2}(\frac{\partial T_{2}}{\partial F_{2}})S_{2}(r_{2})}{%
\mid \frac{\partial \overrightarrow{\mathbf{r}}_{1}}{\partial F_{1}}\mid }%
dF_{2}=\int \frac{D_{1}(\frac{\partial T_{1}}{\partial F_{1}})S_{1}(r_{1})}{%
\mid \frac{\partial \overrightarrow{\mathbf{r}}_{2}}{\partial F_{2}}\mid }%
dF_{1}$. Consequently, the condition for inter-satellite communications
expresses the equality between the two space-time distances $\widetilde{R}%
_{AB}^{2}(1\rightarrow 2)$ and \ $\widetilde{R}_{BA}^{2}(2\rightarrow 1)$.
\section{Discussion}
\bigskip The main goal of this paper is to present a new theoretical formalism for
propagation of signals (electromagnetic, radio or laser) between satellites,
moving on different space trajectories. It should be clarified that
propagation of signals between satellites means that a signal is being send
by a satellite and is received by another satellite. This means that this is
not just a propagation of the signal, as this has been investigated in
previous papers \cite{Dimitrov2017}, \cite{Dimitrov2022}, \cite{Dimitrov2025}%
, but in a sense is a more complicated process, when the movement of the
satellites has to be taken into account and also the curving of the
trajectory of the signal, so that the signal will "hit" the receiver of the second 
satellite at the proper moment.
\bigskip The development of the theoretical formalism has started in the previous
publications \cite{Dimitrov2017}, \cite{Dimitrov2019}, \cite{Dimitrov2023}, 
\cite{Dimitrov2025b}, where the new method of intersecting null cones with
the hyperplane equation for the differential of the Euclidean distance has
been proposed. The most simple case of transferring a signal from a moving
satellite on the elliptic orbit with parameters $(E_{1},a_{1},e_{1})$ to a
second satellite, moving on an elliptic orbit with parameters $%
(E_{2},a_{2},e_{2})$ has been considered. For the first time, the physical
notions about space-time distance, condition for inter-satellite
communications and geodesic distance have been introduced and in a strict
mathematical way it has been proved that they possess specific properties:
the space-time distance as a General Relativity notion can be negative,
equal to zero or positive, while the geodesic distance (also a General Relativity 
notion) as the real distance travelled by the signal can be positive and greater 
than the Euclidean distance.
\bigskip Now two important problems arise: 1. Can the same notions be introduced for
the case of other metrics? If this is the case, then such a model can be
applied to the case, when a satellite in a near-Earth orbit emits a signal,
which is received by a satellite in an orbit around the Moon.2. Should the
satellites necessarily move on plane orbits or on space-oriented orbits? \
The problem here appears in reference to the current and future 6G and 7G
satellite network systems. The necessity for quick communication (i.e. of "ultra low latency") between the
satellites in the network will require the satellites to be very "densely situated". This
means that when a satellite moves along an arbitrary space trajectory, it
may not be necessarily characterized by the standard $6$ Kepler parameters $%
(f,a,e,\Omega ,\omega ,I)$. There might be various disturbing forces due to
which the satellite$^{\prime }$s motion displays considerable deviations
from its trajectory. That is why in celestial mechanics very long time ago
the method of disturbing forces has been developed and it may be found not
only in the cited monographs \cite{Brauer}, \cite{KopEfroimKaplan}, but also
in many other monographs. Also, when a satellite moves in a network, because
of the large number of other satellites, it might be more appropriate to
characterize its motion by introducing its space trajectory. In such a
theoretical model there may be one dynamical parameter, characterizing the
motion of the satellite - this may be the eccentric anomaly angle $E$ for
plane elliptic orbits, but also the true anomaly angle $f$ \ in the
transformations (\ref{K11}), (\ref{K22}), (\ref{K33}) for the space-oriented
Kepler orbits. But as explained in the monograph \cite{Betz}, there may be
more than $6$ parameters, related to the motion of the satellite. In this
theoretical model, we have assumed that there is only one dynamical
parameter, which will be denoted by $F$. In future the model will be 
considerably extended. 
\bigskip In this paper the first main achievement is that it has been proved that the
notions of space-time distance, condition for inter-satellite communications
and geodesic distance can be introduced also for the more general case of
transmission of signals between satellites, moving on arbitrary space
trajectories, each one characterized by its corresponding dynamical
parameter $F_{1}$ and $F_{2}$. Moreover, these physical notions have the
same physical meaning as for the case of satellites, moving on plane elliptic 
orbits and exchanging signals. In more
details the calculations and the conclusions of this paper have been
presented in the review paper \cite{Dimitrov2026}.
\bigskip The second achievement is related to clarifying the notion of
irreversibility with respect to the signals, propagating in the
gravitational field. Since in a satellite network the most optimal way for
sending a signal presupposes also propagating the signal in the "backward"
direction, this subject is of key importance for the notion of "ultra-low
latency", introduced with respect to 6G and 7G satellite networks. Achieving
ultra-low latency means a minimal time for propagation of a signal to reach
a certain destination and then to return back. Again, the definition of
latency \ does not take into account the action of the gravitational field
on the propagation of the signal, but the formulae (\ref{AA19}) for the
Shapiro delay is considered to be obvious and undisputable.
\bigskip In this paper "irreversibility" means that the propagation time $T_{2}-T_{1}$
in the "forward" direction (i.e. from satellite $1$ with propagation time $%
T_{1}$) to satellite $2$ (with propagation time $T_{2}$) is different from
the propagation time $T_{1}-T_{2}$ for propagation of the signal in the
"backward" direction $(2\rightarrow 1)$. The fulfillment of the causality
condition $T_{2}>T_{1}$ in the "forward" direction and $T_{1}>T_{2}$ in the
"backward" direction, if this defined for the Euclidean space, is related to
the s.c. range equation $c(T_{2}-T_{1})-R_{mn}=0$ in the "forward" direction
and $c(T_{1}-T_{2})-R_{nm}=0$ in the backward direction. In the above two
range equations $R_{mn}$ and \ $R_{nm}$ are Euclidean distances and $m$ and $%
n$ are three-dimensional indices. The range equations are in fact
consequences of the null-cone equations for the Euclidean metric \cite%
{Einstein} and they have a simple physical meaning - the speed of light $c$,
multiplied by the difference in the propagation times, gives the Euclidean
distance $R_{mn}$ or $R_{nm}$, travelled by the signal. The two
distances are the same in the two directions.
\bigskip Naturally, the problem arises how to generalize the range equation in a
curved space-time, which leaves its impact on the signal communications. As
noted in \cite{CommunCurved}, the space-time curvature causes $0.10$ $radian$
phase-shift for communication between the Earth and the International Space
Station. Since in a curved space-time signals travel along geodesic
lines, it will be natural to accept the geodesic distance as the distance,
which will be present in the range equation. In principle, the distance
along the geodesic line, which is given by the solutions of the geodesic
equation $\frac{d^{2}x^{\nu }}{d\lambda ^{2}}+\Gamma _{\alpha \beta }^{\nu }%
\frac{dx^{\alpha }}{d\lambda }\frac{dx^{\beta }}{d\lambda }=0$ \ \ \ $\alpha
,\beta ,\nu =0,1,2,3$ should be taken into account. However, in this paper
and in its more extended version \cite{Dimitrov2026}, as well as in the
previous papers \cite{Dimitrov2017}, \cite{Dimitrov2019}, \cite{Dimitrov2023}%
, \cite{Dimitrov2025}, the "geodesic distance" $\widehat{R}_{AB}^{.}$\ is
the one, obtained from the space-time distance and the condition for
inter-satellite communications. In the review paper \cite{Dimitrov2025}
arguments were given that the found geodesic distance $\widehat{R}_{AB}^{.}$ 
satisfies all the requirements to represent the path, travelled by the signal.
\bigskip That is why, in this paper it was proposed to accept the range equations $%
c(T_{2}-T_{1})-\widehat{R}_{AB}^{.}=0$ in the "forward" direction and $%
c(T_{1}-T_{2})-\widehat{R}_{BA}^{.}=0$ in the backward direction as
generalizations for the case of a curved space-time. Now it is easily seen
that if the two geodesic distances $\widehat{R}_{AB}^{.}$ and $\widehat{R}%
_{BA}^{.}$ are equal, this will lead to a contradiction with the causality
principle, because the for the first equation the causality $T_{2}>T_{1}$
will be fulfilled. However,  for the second case $T_{1}>T_{2}$ will not be
fulfilled because both $\widehat{R}_{AB}^{.}$ and $\widehat{R}_{BA}^{.}$ as
geodesic distances are positive.
\bigskip Consequently, the second achievement in this paper is the strict
mathematical proof that for the general case of satellites on arbitrary
space trajectories, $\widehat{R}_{AB}^{.}(1\rightarrow 2)$ and $\widehat{R}%
_{BA}^{.}(2\rightarrow 1)$ are different. In fact, it turned out that the
two space-time distances $\widetilde{R}_{AB}^{.}(1\rightarrow 2)$ and $%
\widetilde{R}_{BA}^{.}(2\rightarrow 1)$ are not equal, but in order to
compute the "forward" and the "backward" geodesic distances, yet it was
necessary to assume that the differentials of the squares of the two
space-time distances are equal, i.e. $d\widetilde{R}_{AB}^{2}(1\rightarrow
2)=d\widetilde{R}_{BA}^{2}(2\rightarrow 1)$. This assumption may seem to be
an artificial one, but an additional calculation has been performed for the
case, when $\widetilde{R}_{AB}^{2}(1\rightarrow 2)=\widetilde{R}%
_{BA}^{2}(2\rightarrow 1)=const$. The result was that the two radius vectors
of the satellites should coincide, i.e. $\overrightarrow{\mathbf{r}}_{1}=%
\overrightarrow{\mathbf{r}}_{2}$. In such a case, no inter-satellite
communications will be possible. It may seem that this is an analogy with
the case of plane elliptic orbits when the orbital parameters $%
(E_{1},a_{1},e_{1})$ of the first satellite coincide with the parameters $%
(E_{2},a_{2},e_{2})$. In fact, there is no analogy between the two cases,
because the two eccentric anomaly angles are determined up to some angle and
they enter the transformation formulaes (\ref{Kep}) in the trigonometric
functions $\cos E$ and $\sin E$. In other words, each of the parameters $%
E_{1}$ and $E_{2}$ was one among the total three parameters (the others
being $(a_{1},e_{1})$ or $(a_{2},e_{2})$), determining the curves $x_{1}=$
That is why it was possible to assert that the two orbits can be rotated at
some angle in one and the same plane.
\bigskip Thus, the calculation of the different space-time distances $\widetilde{R}%
_{AB}^{.}(1\rightarrow 2)$ and $\widetilde{R}_{BA}^{.}(2\rightarrow 1)$
naturally leads to the idea about the different geodesic distances $\widehat{%
R}_{AB}^{.}(1\rightarrow 2)$ and $\widehat{R}_{BA}^{.}(2\rightarrow 1)$,
which was confirmed by the calculation in formulae (\ref{M109}) 
\begin{equation*}
\widehat{R}_{AB}^{2}=\frac{1}{4v_{2}}\int \left[ \frac{\partial }{\partial
F_{1}}(R_{AB}^{2})\right] .v_{1}^{2}.\frac{\widetilde{K}(v_{1},v_{2})}{%
K_{F_{1}}}dv_{1}
\end{equation*}%
and (\ref{M109B}) for $\widehat{R}_{BA}^{2}$.
\bigskip Now it may be noted that in the denominator of formulae (\ref{M109}) for the
geodesic distance participates the function (\ref{M106AA3}), which is
related with the derivative of the square of the velocity $v_{F_{1}}^{2}$ 
\begin{equation*}
K_{F_{1}}=\frac{\partial \overrightarrow{\mathbf{r}_{1}}}{\partial F_{1}}.%
\frac{\partial ^{2}\overrightarrow{\mathbf{r}_{1}}\text{ }}{\partial
F_{1}^{2}}=\frac{1}{2}\frac{\partial }{\partial F_{1}}\left( \frac{\partial 
\overrightarrow{\mathbf{r}_{1}}}{\partial F_{1}}\right) ^{2}=\frac{1}{2}%
\frac{\partial v_{F_{1}}^{2}}{\partial F_{1}}\text{ .}
\end{equation*}%
Note also that the function (\ref{M107B}) $\widetilde{K}(v_{2},v_{1})=\frac{%
2v_{1}v_{2}+v_{2}^{2}-v_{1}^{2}}{v_{1}+v_{2}}\neq \widetilde{K}(v_{1},v_{2})$
does not contain the function $K_{F_{1}}$, the expression (\ref{M109}) for $%
\widehat{R}_{AB}^{2}$ does not contain the function $K_{F_{2}}$. Previously,
it was mentioned that the equivalent to (\ref{M109}) expression (\ref{M109C}%
) 
\begin{equation*}
\widehat{R}_{AB}^{2}=\frac{c}{2v_{2}}\int D_{1}S_{1}\frac{\partial T_{1}}{%
\partial F_{1}}\left[ \frac{2v_{1}v_{2}+v_{1}^{2}-v_{2}^{2}}{v_{1}+v_{2}}%
\right] dF_{1}\text{ }
\end{equation*}%
is also related to the non-inertiality of reference system, attached to the
first satellite, but this is through the derivative $\frac{\partial T_{1}}{%
\partial F_{1}}$ of the first propagation time $T_{1}$.
\bigskip The third achievement in this paper is in clarifying the quantitative
expression of non-inertiality in the two reference systems, attached to the
signal-emitting and the signal-receiving satellite. More concretely, if for
the first system the quantity $D_{1}(dT_{1})S_{1}$ is defined, then the
quantity $D_{2}(dT_{2})S_{2}$ is related through the quantity $%
D_{1}(dT_{1})S_{1}$ by the \ transformation (\ref{N2}) 
\begin{equation*}
D_{2}(dT_{2})S_{2}=k_{1,2}(\mid \frac{\partial \overrightarrow{\mathbf{r}}%
_{1}}{\partial F_{1}}\mid ,\mid \frac{\partial \overrightarrow{\mathbf{r}}%
_{2}}{\partial F_{2}}\mid )D_{1}(dT_{1})S_{1}\text{,}
\end{equation*}
where the coefficient (\ref{N3}) $k_{1,2}(\mid \frac{\partial 
\overrightarrow{\mathbf{r}}_{1}}{\partial F_{1}}\mid ,\mid \frac{\partial 
\overrightarrow{\mathbf{r}}_{2}}{\partial F_{2}}\mid ):=\frac{v_{2}\mp v_{1}%
}{v_{1}\pm v_{2}}$ is expressed as a ratio of the combination of velocities $%
v_{1}$ and $v_{2}$. In the introduction it was shown that formulae (\ref{N2}%
) is analogous to the simple formulae (\ref{NN3}) $v=\frac{k^{2}-1}{k^{2}+1}$
in the monographs \cite{Bondi1} and \cite{Bondi2}, where $k$ is defined as
the ratio of the time intervals of emission and reception.
\bigskip In the present case, $k_{1,2}$ is no longer a "time" ratio due to two
reasons: 1. The velocities $v_{1}$ and $v_{2}$ depend on the space
coordinates. 2. The coefficient functions $D_{1}$ and $D_{2}$, which are to
be defined further, also are functions of both set of coordinates $%
(x_{1},y_{1},z_{1})$ and $(x_{2},y_{2},z_{2})$.
\bigskip If in formulae (\ref{N2}) $k_{1}$ $=\overline{k}_{1}$ and then $k_{2}=-\frac{%
1}{\overline{k}_{1}}$ (obtained with a minus sign in the process of
calculation), then the quantities $D_{1}(dT_{1})S_{1}$ and $%
D_{2}(dT_{2})S_{2}$ may be interpreted as some kind of a "space-time" clock,
defined as depending on the coordinate system. This is a "space-time"
definition, different from the definition by Weyl in \cite{Weyl} in terms
only of time. Consequently, there is no longer any "universal"
time, independent of the coordinate system, as stated in the monograph \cite%
{Bergmann}.
\bigskip Formulae (\ref{N2}) admits an interesting generalization, if one accepts
that there are $n-$satellites, which transmit a signal from the first
satellite to the second one, then from the second to the third one and so
finally the signal is transmitted to the $n-$th satellite. Then it can
easily be calculated from (\ref{N2}) that 
\begin{equation}
(D_{n}dT_{n})S_{n}=k_{(n-1),n}.k_{(n-2),(n-1)}....k_{l-1,l}...k_{1,2}D_{1}(dT_{1})S_{1}%
\text{ , }  \label{M110}
\end{equation}%
where the coefficient functions $k_{l-1,l}$ are defined as 
\begin{equation}
k_{l-1,l}=\frac{v_{l}\mp v_{l-1}}{v_{l-1}\pm v_{l}}\text{ ,\ }  \label{M111}
\end{equation}%
correspondingly $k_{l-1}$ will be with the upper sign and $k_{l}$ will be
with the lower sign 
\begin{equation}
k_{l-1}=\frac{v_{l}-v_{l-1}}{v_{l-1}+v_{l}}\text{ , }k_{l}=\frac{%
v_{l}+v_{l-1}}{v_{l-1}-v_{l}}=-\text{ }\frac{1}{k_{l-1}}\text{.}
\label{M112}
\end{equation}%
Thus in the "forward" direction of propagation of the signal along the chain
of satellites 
\begin{equation}
D_{n}(dT_{n})S_{n}=k_{n-1}k_{n-2}.....k_{l}...k_{1}D_{1}(dT_{1})S_{1}
\label{M113}
\end{equation}%
and in the "backwards" propagation of the signal 
\begin{equation}
D_{1}(dT_{1})S_{1}=(-1)^{n-1}k_{1}k_{2}.....k_{l}...k_{n-1}D_{n}(dT_{n})S_{n}%
\text{ \ .}  \label{M114}
\end{equation}%
It may seem that for $n$ an odd number (of satellites) one may have $%
(-1)^{n-1}=+1$ and thus there will be reversibility of propagation of the
signal in the "forward" direction (from the first to the $n-$th satellite)
and in the "backward" direction (from the $n-$th satellite to the first one).
\bigskip However, such a conclusion will be entirely wrong and further these two
formulaes will be rewritten in the correct way, thus proving the
irreversibility in the propagation of the signal, i.e. different "forward"
propagation of the signal in comparison with the "backward" propagation.
\bigskip The subtle moment here is that the two set of coordinates $%
(x_{1},y_{1},z_{1})$ and $(x_{2},y_{2},z_{2})$ are defined at the moment of
emission of the signal by the first satellite. By the time the signal has
travelled from the first to the second satellite, the coordinates of the
second satellite have changed to $(x_{2}^{(1)},y_{2}^{(1)},z_{2}^{(1)})$.
This means that when the second satellite sends back the signal in the
"backward" direction, the initial coordinates will be $%
(x_{2}^{(1)},y_{2}^{(1)},z_{2}^{(1)})$, respectively the initial velocity of
the second satellite will be $%
v_{2}^{(1)}(x_{2}^{(1)},y_{2}^{(1)},z_{2}^{(1)})\neq
v_{2}(x_{2},y_{2},z_{2}) $. Also, by the time when the "backward" signal is
emitted from the second satellite in the direction of the first one, the
coordinates of the first satellite due to his motion during the time of
propagation of the signal (from satellite $1$ towards satellite $2$) have
also changed from $(x_{1},y_{1},z_{1})$ to $%
(x_{1}^{(1)},y_{1}^{(1)},z_{1}^{(1)})$, respectively the velocity of the
first satellite will change from $v_{1}^{.}(x_{1},y_{1},z_{1})$ to $%
v_{1}^{(1)}(x_{1}^{(1)},y_{1}^{(1)},z_{1}^{(1)})$.
\bigskip It may be concluded that formula (\ref{N2A2}) $D_{1}(dT_{1})S_{1}=-%
\widetilde{k}_{1}D_{2}(dT_{2})S_{2}$ for the propagation of the signal in
the "backward" direction may still be applied, but instead of $k_{1}$ here
we have written $\widetilde{k}_{1}$. Instead of the defining formulae (\ref%
{N2A0}) for $k_{1}=\frac{k^{2}-1}{1+k^{2}}$, where $k^{2}=\frac{%
v_{1}(x_{1},y_{1},z_{1})}{v_{2}(x_{2},y_{2},z_{2})}$, the coefficient
function $\widetilde{k}_{1}$ will be defined as 
\begin{equation}
\widetilde{k}_{1}^{(1)}=\frac{\widetilde{k}_{(1)}^{2}-1}{1+\widetilde{k}%
_{(1)}^{2}}=\frac{%
v_{2}^{(1)}(x_{2}^{(1)},y_{2}^{(1)},z_{2}^{(1)})+v_{1}^{(1)}(x_{1}^{(1)},y_{1}^{(1)},z_{1}^{(1)})%
}{%
v_{2}^{(1)}(x_{2}^{(1)},y_{2}^{(1)},z_{2}^{(1)})-v_{1}^{(1)}(x_{1}^{(1)},y_{1}^{(1)},z_{1}^{(1)})%
}\text{ , }  \label{M115}
\end{equation}%
\begin{equation}
\widetilde{k}_{(1)}^{.}=\sqrt[.]{\text{\ }\frac{%
v_{2}^{(1)}(x_{2}^{(1)},y_{2}^{(1)},z_{2}^{(1)})}{%
v_{1}^{(1)}(x_{1}^{(1)},y_{1}^{(1)},z_{1}^{(1)})}}\text{ .}  \label{M116}
\end{equation}%
Further we may call the set of coordinates $%
(x_{1}^{(1)},y_{1}^{(1)},z_{1}^{(1)})$ and $%
(x_{2}^{(1)},y_{2}^{(1)},z_{2}^{(1)})$ \ "first iteration" of the initial
coordinates $(x_{1},y_{1},z_{1})$ and $(x_{2},y_{2},z_{2})$. We have put
also the indice $1$ in round brackets to denote that the corresponding
expressions for $\widetilde{k}_{1}^{(1)}$ and $\widetilde{k}_{(1)}^{2}$ in (%
\ref{M115}) and (\ref{M116}) are associated with the first coordinate
iterations $(x_{1}^{(1)},y_{1}^{(1)},z_{1}^{(1)})$ and $%
(x_{2}^{(1)},y_{2}^{(1)},z_{2}^{(1)})$.
\bigskip Let us clarify first what is the meaning of these "iterations", which are
related with the corresponding positions of the second, signal-receiving
satellite at the moment of reception of the signal and the moment of
reception of the signal by the first satellite, after it has been send
"backwards" by the second satellite in the direction of the first satellite.
Evidently, the first iteration \ coordinates $%
(x_{2}^{(1)},y_{2}^{(1)},z_{2}^{(1)})$ are the coordinates, after the
initial position of the second satellite has changed during the time of
travel $(T_{2}-T_{1})$ from position $(x_{2},y_{2},z_{2})$ to position $%
(x_{2}^{(1)},y_{2}^{(1)},z_{2}^{(1)})$. Thus we have the equality 
\begin{equation*}
x_{2}^{(1)}=x_{2}+\left[
t_{.}^{(2)}(x_{2}^{(1)},y_{2}^{(1)},z_{2}^{(1)})-t_{.}^{(2)}(x_{2},y_{2}^{(1)},z_{2}^{(1)})%
\right] \times 
\end{equation*}%
\begin{equation}
\times \frac{\partial \overrightarrow{\mathbf{r}}%
_{2}(F_{2},x_{2}^{(1)},y_{2}^{(1)},z_{2}^{(1)})}{\partial F_{2}}\mid
_{x_{2}^{(1)}}\frac{\partial F_{2}}{\partial t^{(2)}}\mid
_{t^{(t_{2})}=t(x_{2}^{(1)},y_{2}^{(1)},z_{2}^{(1)})}\text{ ,}  \label{M118}
\end{equation}%
where 
\begin{equation}
\frac{\partial \overrightarrow{\mathbf{r}}%
_{2}(F_{2},x_{2}^{(1)}(F_{2}),y_{2}^{(1)}(F_{2}),z_{2}^{(1)}(F_{2}))}{%
\partial t^{(2)}}\mid _{x_{2}^{(1)},y_{2}^{(1)},z_{2}^{(1)}}  \label{M118A}
\end{equation}%
is the $x_{2}^{(1)}$ component of the tangent vector  to the space curve $%
\overrightarrow{\mathbf{r}}%
_{2}(F_{2},x_{2}^{(1)}(F_{2}),y_{2}^{(1)}(F_{2}),z_{2}^{(1)}(F_{2}))$ of
motion of the second satellite.  
\bigskip In (\ref{M118}) \ the function $%
t^{(2)}(x_{2}^{(1)},y_{2}^{(1)},z_{2}^{(1)})$ \ is the celestial time of
motion of the second satellite at the moment of reception of the signal,
when the second satellite$^{\prime }$s coordinates are $%
(x_{2}^{(1)},y_{2}^{(1)},z_{2}^{(1)})$ and further in (\ref{M119}) $%
t_{.}^{(2)}(x_{2},y_{2}^{.},z_{2}^{.})$ will be the celestial time of the
second satellite at the space point $(x_{2},y_{2}^{.},z_{2}^{.})$,
corresponding to the initial moment, at which the signal was send from the
first satellite. At this initial moment the coordinates of the first
satellite were  $(x_{1},y_{1},z_{1})$. Since the difference between the two
celestial times, corresponding to the space points $%
(x_{2}^{(1)},y_{2}^{(1)},z_{2}^{(1)})$ and $(x_{2},y_{2},z_{2})$ should
correspond to the difference in the propagation time of the signal $%
T_{2}(x_{2}^{(1)},y_{2}^{(1)},z_{2}^{(1)})$ and $T_{1}(x_{1},y_{1},z_{1})$,
equation (\ref{M118}) should be supplemented by the equation 
\begin{equation}
T_{2}(x_{2}^{(1)},y_{2}^{(1)},z_{2}^{(1)})-T_{1}(x_{1},y_{1},z_{1})=t_{.}^{(2)}(x_{2}^{(1)},y_{2}^{(1)},z_{2}^{(1)})-t_{.}^{(2)}(x_{2},y_{2}^{.},z_{2}^{.})%
\text{ .}  \label{M119}
\end{equation}%
\ The derivative $\frac{\partial \overrightarrow{\mathbf{r}}%
_{2}(F_{2},x_{2}^{(1)},y_{2}^{(1)},z_{2}^{(1)})}{\partial F_{2}}$\ in (\ref%
{M118}) is in fact the velocity $v_{2}(x_{2}^{(1)},y_{2}^{(1)},z_{2}^{(1)})$%
, which represents the tangent vector to the space curve. It is taken at the
moment, when the second satellite is at the point $%
(x_{2}^{(1)},y_{2}^{(1)},z_{2}^{(1)})$, when it has received the signal from
the first satellite. Note also that this point is not fixed as an initial
point, it can be an arbitrary point on the trajectory of the second
satellite, but with account of the first iterated coordinate.  Analogous
formulaes to (\ref{M118}) can be written also for the components $y_{2}^{(1)}
$ and \ $z_{2}^{(1)}$.
\bigskip Now it remains to rewrite formulaes (\ref{M113}) and (\ref{M114}) for the
"forward" and "backward" propagation of the signal through a chain of
satellites in the correct way by accounting of the number of different
"iterations" of the coordinates and thus it will be proved that the
chain-propagation of the signal in the "backward" direction is different
from the "forward" propagation of the signal. The number of iterations will
be placed above each coefficient function in brackets. For the "forward"
chain-propagation of the signal, the correct formulae (instead of (\ref{M113}%
)) is 
\begin{equation}
D_{n}(dT_{n})S_{n}=\widetilde{k}_{n-1}^{(n-1)}\widetilde{k}%
_{n-2}^{(n-2)}.....\widetilde{k}_{l}^{(l)}...\widetilde{k}%
_{1}^{(1)}D_{1}(dT_{1})S_{1}  \label{M120}
\end{equation}%
and in the "backwards" propagation of the signal (instead of (\ref{M114}) 
\begin{equation}
D_{1}(dT_{1})S_{1}=(-1)^{n-1}\widetilde{k}_{1}^{(n-1)}\widetilde{k}%
_{2}^{(n-2)}.....\widetilde{k}_{l}^{(n-l)}...\widetilde{k}%
_{n-1}^{(1)}D_{n}(dT_{n})S_{n}\text{ .}  \label{M121}
\end{equation}%
It becomes clear what is the the reason for the irreversibility - in the "forward"
chain propagation formulae (\ref{M120}) the $l$-th coefficient function $%
\widetilde{k}_{l}^{(l)}$ is of the $l-$th \ iteration, while in the
"backward" chain propagation formulae (\ref{M121}) the $l-$th coefficient
function is $\widetilde{k}_{l}^{(n-l)}$ and it is of the $(n-l)-$th
iteration. The $l-$th iteration $\widetilde{k}_{l}^{(l)}$ in analogy with
the defining formulae (\ref{M115}) for  $\widetilde{k}_{1}^{(1)}$ will be   
\begin{equation}
\widetilde{k}_{l}^{(l)}=\frac{\widetilde{k}_{(l)}^{2}-1}{1+\widetilde{k}%
_{(l)}^{2}}=\frac{%
v_{l+1}^{(l)}(x_{l+1}^{(l)},y_{l+1}^{(l)},z_{l+1}^{(l)})+v_{l}^{(l)}(x_{l}^{(l)},y_{l}^{(l)},z_{l}^{(l)})%
}{%
v_{l+1}^{(l)}(x_{l+1}^{(l)},y_{l+1}^{(l)},z_{l+1}^{(l)})-v_{l}^{(l)}(x_{l}^{(l)},y_{l}^{(l)},z_{l}^{(1)})%
}\text{ , }  \label{M122}
\end{equation}%
where $\widetilde{k}_{(l)}^{.}$ is defined in analogy with (\ref{M116}) as
an $l-$th iteration  
\begin{equation}
\widetilde{k}_{(l)}^{.}=\sqrt[.]{\text{\ }\frac{%
v_{l+1}^{(l)}(x_{l+1}^{(l)},y_{l+1}^{(l)},z_{l+1}^{(l)})}{%
v_{l}^{(l)}(x_{l}^{(l)},y_{l}^{(l)},z_{l}^{(l)})}}\text{ .}  \label{M123}
\end{equation}%
It should be stressed that the number of the set of coordinates, for example
$(x_{l+1}^{.},y_{l+1}^{.},z_{l+1}^{.})$, is different from the number of the
iterations, which is written above the coordinate in the round bracket $%
(x_{l+1}^{(l)},y_{l+1}^{(l)},z_{l+1}^{(l)})$. Similarly, the coefficient
function $\widetilde{k}_{l}^{(n-l)}$ in formulae (\ref{M121}) can be written
by using the same formulaes as (\ref{M122}) and (\ref{M123}), but everywhere
the $(l)-$th iterations should be replaced by the  $(n-l)$-th iteration. The 
$l-$th number of the set of coordinates  $(x_{l}^{.},y_{l}^{.},z_{l}^{.})$
remains unchanged. In such way, it becomes clear that formulaes (\ref{M120})
and (\ref{M121}) give the expressions for the "forward" and "backward \
propagation of the signal, which is different in the different directions. 
\bigskip Now it remains to present the analogous to (\ref{M118}) and (\ref{M119}) for
the "backwards" propagation of the signal from satellite $2$ towards
satellite $1$. These formulaes, as well the ones, which will be presented
below, naturally can be generalized for the case of a "chain" of $n$
satellites, in which the first one sends a signal along the chain to the $n-$%
the satellite and then the $n-$th satellite sends the signal along the chain
in the "backwards" direction to the first satellite. 
\begin{equation*}
x_{1}^{(1)}=x_{1}+\left[
t_{.}^{(1)}(x_{1}^{(1)},y_{1}^{(1)},z_{1}^{(1)})-t_{.}^{(1)}(x_{1},y_{1}^{(1)},z_{1}^{(1)})%
\right] \times 
\end{equation*}%
\begin{equation}
\times \frac{\partial \overrightarrow{\mathbf{r}}%
_{1}(F_{1},x_{1}^{(1)},y_{1}^{(1)},z_{1}^{(1)})}{\partial F_{1}}\mid
_{x_{1}^{(1)}}\frac{\partial F_{1}}{\partial t^{(1)}}\mid
_{t^{(1)}=t^{(1)}(x_{1}^{(1)},y_{1}^{(1)},z_{1}^{(1)})}\text{ ,}
\label{M125}
\end{equation}
where  $\frac{\partial \overrightarrow{\mathbf{r}}%
_{1}(F_{1},x_{1}^{(1)},y_{1}^{(1)},z_{1}^{(1)})}{\partial F_{1}}%
=v_{1}(F_{1},x_{1}^{(1)},y_{1}^{(1)},z_{1}^{(1)})$ is the velocity of the
first satellite, taken at the moment when the "backwards" signal, send from
satellite $2$ is received back from satellite $1$.  The functions  $%
t_{.}^{(1)}(x_{1}^{(1)},y_{1}^{(1)},z_{1}^{(1)})$ and $%
t_{.}^{(1)}(x_{1},y_{1}^{(1)},z_{1}^{(1)})$ are the celestial times of the
first satellite at the corresponding points. The motion of the first
satellite is more complicated in comparison with the second one, because at
the time of the "backward" reception of the signal, the difference \ between
the celestial times is equal to the sum of the propagation time $\left[
T_{2}(x_{2}^{(1)},y_{2}^{(1)},z_{2}^{(1)})-T_{1}(x_{1},y_{1},z_{1})\right] $
from satellite $1$ to satellite $2$ and also the "backward" propagation time 
$\left[ \widetilde{T}_{1}(x_{1}^{(1)},y_{1}^{(1)},z_{1}^{(1)})-\widetilde{T}%
_{2}(x_{2}^{(1)},y_{2}^{(1)},z_{2}^{(1)})\right] $ from satellite $2$
towards satellite $1$. Thus the analogue of equation (\ref{M119}) for the
"backward" propagation of the signal is 
\begin{equation*}
\left[ T_{2}(x_{2}^{(1)},y_{2}^{(1)},z_{2}^{(1)})-T_{1}(x_{1},y_{1},z_{1})%
\right] +\left[ \widetilde{T}_{1}(x_{1}^{(1)},y_{1}^{(1)},z_{1}^{(1)})-%
\widetilde{T}_{2}(x_{2}^{(1)},y_{2}^{(1)},z_{2}^{(1)})\right] 
\end{equation*}%
\begin{equation}
=t_{.}^{(1)}(x_{1}^{(1)},y_{1}^{(1)},z_{1}^{(1)})-t_{.}^{(1)}(x_{1},y_{1}^{.},z_{1}^{.})%
\text{ .}  \label{M126}
\end{equation}%
\newline
It should be noted that the first term  $%
T_{2}(x_{2}^{(1)},y_{2}^{(1)},z_{2}^{(1)})$ and the last one $\widetilde{T}%
_{2}(x_{2}^{(1)},y_{2}^{(1)},z_{2}^{(1)})$ do not cancel. This is related
with the complex nature of the physical phenomena of "forward" and
"backward" propagation of the signal, on one side, and on the other side, it
is related to the motion of the satellites during the propagation of the
signal in both directions, which leads to the change of the coordinates. So
these are two different phenomena, which may be studied separately. For
example, what will happen if the velocity of the second satellite is zero
(and is in a steady position, when it emits the "backwards" signal), but
only the first satellite is moving, or vice versa, the first satellite is
not moving, but the second is moving. 
\bigskip The system of four equations (\ref{M118}), (\ref{M119}), (\ref{M125}) and (%
\ref{M126}) is a system of "iterative " equations, determining the first-
iterated coordinates $(x_{2}^{(1)},y_{2}^{(1)},z_{2}^{(1)})$ and $%
(x_{1}^{(1)},y_{1}^{(1)},z_{1}^{(1)})$ on the initial coordinates  $%
(x_{1},y_{1}^{.},z_{1}^{.})$ and $(x_{2},y_{2}^{.},z_{2}^{.})$ and also on
the propagation times $T_{1}$ and $T_{2}$, defined on the two sets of the
initial coordinates and on  the two sets of "first iterated " coordinates.
But in fact, there will be four additional equations, because similarly to
the equation (\ref{M118}) for the "iterated" coordinate $x_{2}^{(1)}$,
two other equations have to be written for the components \ $y_{2}^{(1)}$
and \ $z_{2}^{(1)}$. Analogous to equation (\ref{M125}) for the component $%
x_{1}^{(1)}$, two additional  equations have to be written for the
components \ $y_{1}^{(1)}$ and \ $z_{1}^{(1)}$. Thus, the total number of
equations becomes eight. 
\bigskip By differentiating each of the eight equations with respect to the two set
of initial coordinates $(x_{1},y_{1}^{.},z_{1}^{.})$ and $%
(x_{2},y_{2}^{.},z_{2}^{.})$ and then with respect to the two sets of
"iterated" coordinates $(x_{1}^{(1)},y_{1}^{(1)},z_{1}^{(1)})$ \ and $%
(x_{2}^{(1)},y_{2}^{(1)},z_{2}^{(1)})$, this system of $8$ equations will be
transformed to an equivalent system of $45$  nonlinear differential
equations, which will describe the dynamic process of propagation of the
signal between two satellites in the "forward" and in the "backward"
direction. This number is computed as follows, taking into account that the
number of the differential equations for each of the equations (\ref{M118}),
(\ref{M119}), (\ref{M125}) and (\ref{M126}) is equal to the number of
coordinates, on which it depends. This is so because each one of these
equations is subsequently differentiated along each of the coordinates.
Thus, the first equation (\ref{M118}) depends on four coordinates,
consequently after differentiation along each of the coordinates it will give
four differential equations. But since analogous to (\ref{M118})
differential equations have to be written for $y_{2}^{(1)}$ and  $z_{2}^{(1)}
$, the total number of differential equations will be $4\times 3=12$. This
is true under the assumption that $x_{1}^{(1)}=x_{1}^{(1)}(x_{1})$, also $%
y_{1}^{(1)}=y_{1}^{(1)}(y_{1})$ and $z_{1}^{(1)}=z_{1}^{(1)}(z_{1})$. 
\bigskip The second equation (\ref{M119}) depends on $9$ coordinates, consequently
after differentiation it will give $9$ new differential equations. 
\bigskip The third equation (\ref{M125}) depends on $4$ coordinates, but equations
have to be written also for the $y_{1}^{(1)}$ and $z_{1}^{(1)}$ coordinates.
The number of differential equations obtained from (\ref{M125}) will be $%
4\times 3=12$. 
\bigskip The last (fourth) equation (\ref{M126}) depends on $9$ coordinates,
therefore after differentiation it will give $9$ equations. 
\bigskip The total number of nonlinear differential equations, obtained from  (\ref%
{M118}), (\ref{M119}), (\ref{M125}) and (\ref{M126}) will be $12+9+12+9=42$
equations. 
\bigskip However, one may assume also another "scheme" for the dependence of the
coordinates, for example 
\begin{equation}
x_{1}^{(1)}=x_{1}^{(1)}(x_{1},y_{1},z_{1})\text{ , }%
y_{1}^{(1)}=y_{1}^{(1)}(x_{1},y_{1},z_{1})\text{ \ , }%
z_{1}^{(1)}=z_{1}^{(1)}(x_{1},y_{1},z_{1})\text{ ,}  \label{M127}
\end{equation}
\begin{equation}
x_{2}^{(1)}=x_{2}^{(1)}(x_{2},y_{2},z_{2})\text{ , }%
y_{2}^{(1)}=y_{2}^{(1)}(x_{2},y_{2},z_{2})\text{ , }%
z_{2}^{(1)}=z_{2}^{(1)}(x_{2},y_{2},z_{2})\text{ .}  \label{M128}
\end{equation}%
Then the number of equations will be larger. 
\section{Acknowledgements}
\bigskip The author is grateful to the organizers of the Eighteenth Hybrid Conference
of the Euro-American Consortium for Promoting the Application of Mathematics in
Technical and Natural Sciences (AMITaNS26) 27 June - 4 July 2026, Albena
resourt, Bulgaria and especially to Prof. Michail Todorov and to Assoc. Prof.
Meglena Lazarova for the opportunity to participate in this conference.


\begin{thebibliography}{99}
\bibitem{GenNetworks} Wang Qi, Chen X., Qiao Qi, Li M., Gerstacker W.
2025 \textit{Multi-Satellite Cooperative Information Communication and
Location Sensing in LEO Satellite Constellations}, arXiv: 2501.07220v1
[cs.IT]
\bibitem{Hoyhtya} Hoyhtya M. 2025 \textit{Satellite Communications and
Networks} (Cham:Springer Nature Switzerland AG)
\bibitem{Shapiro} Shapiro I 1964 \textit{Fourth Test of General
Relativity} Phys. Rev. Lett. \textbf{13} 789
\bibitem{Straumann} Straumann N. 2013 \textit{General Relativity}, second
edition (Dordrecht:Springer Science+Business Media)
\bibitem{Dimitrov2022} Dimitrov B G 2022 \textit{Elliptic Integrals for
Calculation of the Propagation Time of a Signal, Emitted and Received by
Satellites on One Orbit} AIP Conf. Proc. \textbf{2522} 978-0-7354-4361-7
(Preprint gen-ph/2201.13208v2)
\bibitem{Dimitrov2025} Dimitrov B G 2025 \textit{New Applications of
Elliptic Functions and Integrals in GPS Inter-Satellite Communications with
Account of General Relativity Theory}, MDPI Special Issue "\textit{Advances
in Elliptic Equations and Their Applications}" Mathematics \textbf{13(8)}
1286 Available at: https://www.mdpi.com/2227-7390/13/8/1286
\bibitem{Elbert} Elbert, B. 2008 \textit{Introduction to Satellite
Communications}, Third Ed. (Boston, London:Artech House, INC.)
\bibitem{PrasSol} Prasolov V and Solov'yev Y 1997 \textit{Elliptic Functions
and Elliptic Integrals. AMS Translations of Mathematical Monographs 170}
(Rhode Island: Providence) Russian original: Prasolov V and Solov'yev Y 1997 
\textit{Elliptic Functions and Algebraic Equations }(Moscow:Factorial
Publishing)
\bibitem{Smirnov3} Smirnov V 1964 \textit{A Course of Higher Mathematics:
vol. 3, part 2: Complex Variables Special Functions. International Series of
Monographs in Pure and Applied Mathematics: volume 60}\emph{\ } (Pergamon
Press)
\bibitem{Fichten2} Fichtenholz G 2003 \textit{Course in Differential and
Integral Calculus, vol.2, 8th ed.} (Moscow:Fizmatlit)
\bibitem{KopEfroimKaplan} Kopeikin S, Efroimsky M and Kaplan G 2011 \textit{%
Relativistic Celestial Mechanics of the Solar System} (New York:Wiley-VCH)
\bibitem{LaserCommun} NASA to Demonstrate Laser Communications from Space
Station,
https://www.nasa.gov/technology/nasa-to-demonstrate-laser-communications-from-space-station/
\bibitem{CrosVendor} \textit{U.S. military satellites achieve first
cross-vendor laser communications link},
https://spacenews.com/u-s-military-satellites-achieve-first-cross-vendor-laser-communications-link/
\bibitem{SmallSatel} \textit{Small Satellite Architectures Get New Boosts
From SDA, NRO}, Jan. 10, 2025,
https://www.airandspaceforces.com/sda-nro-small-satellite-architectures/
\bibitem{Wu6GMobile} Wu Y., Singh S., Taleb, T., Roy A., Dhilon, H.,
Kanagarathinam M., De A 2021 \textit{6G Mobile Wireless Networks} (Cham:
Springer Nature Switzerland AG) Switzerland
\bibitem{4G7G} Saurabh, Arya N., Aggarwal A., Sharma G \ 2025 \textit{%
Advancement in Mobile Network Technology: Evolution from 4G to 7G} \ IEEE
Xplore 2025 International Conference on Innovations and Emerging
Technologies in AI \& Communications Systems (IETACS) 1742\ 
\bibitem{6GBeyond} Rappaport T., Xing Y., Kanhera O., Ju S., Madanayake A.,
Mandal S., Alkhateeb A., Trichopoulos G 2019 \textit{Wireless Communications
and Applications Above }$100$\textit{\ }$GHz$\textit{:Opportunities and
Challenges for 6G and Beyond IEEE Access }\textbf{7 }78729
\bibitem{SharifPhD} Sharif S 2023 \textit{Multi-Objective Resource
Optimization in Space-Aerial-Ground-Sea Integrated Networks}, Thesis for the
fulfillment of the requirements for the degree Master of Science, Lakehead
University, Thunder Bay, Ontario, Canada available at
https://knowledgecommons.lakeheadu.ca/server/api/core/bitstreams/224dc4ef-88d5-4c08-bd3d-f688410b0a6a/content
\bibitem{Dimitrov2017} Dimitrov, B. 2017 \textit{Two Null Gravitational
Cones in the Theoryof GPS-Intersatellite Communications Between Two Moving
Satellites.I.Physical and Mathematical Theory of the Space-Time Interval and
the Geodesic Distance on Intersecting Null Cones} https://arxiv.org/abs/1712.01101v3)
\bibitem{Dimitrov2019} Dimitrov B. 2019 \textit{New Mathematical Models of
GPS Intersatellite Communications in the Gravitational Field of the Near-Earth Space} 
 AIP Conf. Proc. \textbf{2075, \ }040007 (Preprint gr-qc/1909.12753v1)
\bibitem{Dimitrov2023} Dimitrov B. 2023 \textit{Algebraic Geometry and
Elliptic Integrals Approach for Calculation of the Propagation Time Between
GPS Satellites with Account of the General Relativity Theory} J. Phys.:
Conf. Ser. \textbf{2675, \ }012026
\bibitem{Dimitrov2026} Dimitrov B., 2026 \textit{Space-Time Distance,
Geodesic Distance and Ranging Equation on Intersecting Null Cones in the
Algebraic Geometry Theory of GPS Inter-Satellite Communications}, to be
submitted to the MDPI journal Mathematics
\bibitem{Betz} Betz, J. 2016 \textit{Engineering Satellite-Based Navigation
and Timing. Global Navigation Satellite Systems, Signals and Receivers} (New
Delhi:Wiley India, Pvt)
\bibitem{Brauer} Brauer, D., Clemence G. 1961 \textit{Methods of Celestial
Mechanics} (New York, London:Academic Press)
\bibitem{IntegrSatel} Xiao, L, Jiang, T, Xiao, P. 2025 \textit{Satellite
Terrestrial Network Fundamentals For Mobile Communications} (Boston,
London:Artech House), UK
\bibitem{Grewal} Grewal M.,Weil L and Andrews A 2001 \textit{Global
Positioning Systems, Inertial Navigation and Integration} (New York,
Toronto:John Wiley \& Sons, Inc.),USA
\bibitem{Inertial} Zhou H, Zhou Y, Song H, and \ Wang S 2021 \textit{%
Inertial Navigation Technology. Navigation: Science and Technology. vol.7}
(Dalian, Singapore:National Defense Industry Press+Springer Nature Singapore
Pte Ltd.)
\bibitem{SINS} Holovach S, Nesterenko O, Hehelskyi O, and Pazdrii O \textit{%
Alternative Mems Navigation During GNSS/Starlink Outages}, Preprint
https://www.preprints.org/manuscript/202602.0937, 11 February 2026
\bibitem{HauschSpriHand} Hauschild A "\textit{Basic Observation Equations}",
in Teunissen P, and Montenbruck O (eds) \ 2017 \ Springer \textit{Handbook
of Global Navigation Systems} (Cham:Springer International Publishing),
Switzerland
\bibitem{Einstein} Einstein A. 1954 \textit{The Meaning of Relativity. The
Stafford Little Lectures of Princeton University, May 1921} (Princeton, New
Jersey: Princeton University Press), USA
\bibitem{Synge1} Synge J. 1960 \textit{Relativity:The General Theory (}%
Amsterdam\textit{:}North-Holland Publishing Company)\textit{, }Netherlands 
\textit{\ }
\bibitem{Synge2} Synge J. 1956 \textit{Relativity:The Special Theory (}%
Amsterdam:North-Holland Publishing Company), Netherlands\textit{\ } \textbf{%
\ }
\bibitem{Bahder1} Bahder T. 2001 \textit{Navigation in Curved Space-Time}
Amer. J. Phys\emph{.} \textbf{\ 69}\emph{, }315
\bibitem{Causal} Minguzzi E, Sanchez M 2008 \textit{The Causal Hierarchy of
Spacetimes, in Recent Developments of Pseudo-Riemannian Geometry} ESI Lect.
Math. Phys. (Zurich:Eur. Math. Soc. Publ. House), Switzerland
https://arxiv.org/pdf/gr-qc/0609119
\bibitem{Harris} Harris J 1992 \textit{Algebraic Geometry. A First Course.
GraduateTexts in Mathematics} (NewYork:Springer-Verlag Inc.), Germany
\bibitem{Weyl} Weyl H 1922 \textit{Space-Time-Matter} (London:Methuen S. CO.
LTD), UK https://www.gutenberg.org/files/43006/43006-pdf.pdf\textit{\ }
\bibitem{Bergmann} Bergmann P. 1942 \textit{Introduction to the Theory of
Relativity} (Bombay, Calcutta:Asia Publishing House), India
\bibitem{Bondi1} Bondi H. 1964 \textit{Relativity and Common Sense. A New
Approach to Einstein} (Garden City, New York:Anchor Books, Doubleday \&
Company, Inc.), USA
\bibitem{Bondi2} Bondi H. 1967 \textit{Assumption and Myth in Physical Theory%
} (Cambridge:Cambridge University Press), USA
\bibitem{Pauli} Pauli W 1958 \textit{Theory of Relativity} (London, New
York:Pergamon Press, Inc.), USA
\bibitem{Dimitrov2024} Dimitrov B 2024 \textit{Topological, Differential
Geometry Methods and Modified Variational Approach for Calculation of the
Propagation Time of a Signal, Emitted by a GPS-Satellite and Depending on
the Full Set of 6 Kepler Parameters} J. Phys.: Conf. Ser. \textbf{2910}
012001 open access article at
https://iopscience.iop.org/article/10.1088/1742-6596/2910/1/012001 (Preprint
https://arxiv.org/abs/2501.04043)\textit{\ }
\bibitem{KopKaplan} Kopeikin S., Kaplan G 2024 \textit{\ Lunar Time in
General Relativity} Phys. Rev. \textbf{D110 }084047
\bibitem{AshbyPatla} Ashby N., Patla B 2024 \textit{A Relativistic Framework
to Estimate Clock Rates on the Moon} ApJ \textbf{168} 112
\bibitem{TuryshAJ} Turyshev Sl 2026 \textit{High-Precision Relativistic Time
Scales for Cislunar Navigation} ApJ \textbf{997} 97
\bibitem{PrattAllnut} Pratt T, Allnut J \ 2020 \textit{Satellite
Communications}, Third Ed. (Chichester, West Sussex: John Wiley \& Sons)
\bibitem{Dimitrov2025b} Dimitrov B. 2025 \textit{Mathematical and Physical
Modelling of Communications Between GPS/GNSS Satellites With Acount of the
General Relativity Effects} S E S 2 0 2 5 Twenty-first International
Scientific Conference SPACE, ECOLOGY, SAFETY Sofia Bulgaria Space Research
and Technology Institute, Bulgarian Academy of Sciences Accessed at
http://space.bas.bg/SES/archive/SES\%202025\_DOKLADI/3\_Remote\%20Sensing/5%
\_Dimitrov.pdf
\bibitem{DiffEq} Fedoriuk M. 2003 \textit{Ordinary Differential Equations};
third edition (Sankt Peterburg:LAN Publishing House) \ Russia
\bibitem{IntegEq} Courant R., Hilbert D 1953 \textit{Methods of Mathematical
Physics, vol. 1 (}New York:Interscience Publishers, INC.) USA
\bibitem{CommunCurved} Exirifard Q., Culf E., Karimi E 2021 \textit{Towards
communication in a curved space-time geometry} Commun. in Physics \textbf{4}%
, 171
\end{thebibliography}
\end{document}